\documentclass{article}

\usepackage[letterpaper,top=2cm,bottom=2cm,left=2cm,right=1.8cm,marginparwidth=1.75cm]{geometry}

\usepackage{amssymb,amsmath,amsthm,bm}

\usepackage{authblk}
\usepackage{xcolor}
\usepackage{pifont}

\usepackage{graphicx}
\usepackage{subcaption}
\usepackage{float}
\usepackage{multirow}
\usepackage{rotating}
\usepackage{pdflscape}
\usepackage{longtable}

\usepackage{microtype}
\usepackage{comment}
\usepackage{soul}
\usepackage{todonotes}
\usepackage[normalem]{ulem}

\usepackage{hyperref}
\hypersetup{
    colorlinks=true,
    linkcolor=blue,
    citecolor=blue,
    urlcolor=blue
}

\usepackage{xcolor}
\usepackage{lmodern}

\title{
\Large \bfseries
\textcolor{black!35!blue}{\texttt{\LARGE{MARUT}}: An Exascale-Ready, GPU-Accelerated High-Order CFD Framework with AMR for High-Speed Flows and Finite-Rate Chemistry}
}
\author[]{Trishit Mondal}
\author[]{Ameya D. Jagtap\thanks{Corresponding author: Ameya D. Jagtap (ajagtap@wpi.edu, ameyadjagtap@gmail.com)}}

\affil[]{\textit{\small{Aerospace Engineering Department, Worcester Polytechnic Institute, Worcester, MA 01609, USA.}}}
\date{}
\begin{document}

\maketitle

\begin{abstract}
We present \texttt{MARUT}, a scalable multi-GPU computational fluid dynamics (CFD) framework designed for high-fidelity simulations of compressible flows spanning subsonic to hypersonic regimes, including chemically reacting nonequilibrium flows with finite-rate chemistry and adaptive mesh refinement (AMR). The framework addresses a central challenge in contemporary scientific computing: the development of numerically accurate and computationally scalable algorithms capable of resolving strongly nonlinear, multiscale flow physics on emerging heterogeneous supercomputing architectures. Built around a distributed-memory MPI-parallel infrastructure and implemented natively on NVIDIA GPUs, \texttt{MARUT} combines high-order spectral discontinuous Galerkin discretisations with strong-stability-preserving Runge--Kutta time integration to achieve low-dissipation and high-resolution representation of shocks, vortical structures and reactive interfaces. Dynamic AMR further enables efficient concentration of computational resources in localized regions of physical complexity, thereby substantially reducing computational cost while preserving solution fidelity. \texttt{MARUT} is designed to maintain strong parallel efficiency through GPU-resident computations and scalable MPI communication strategies, achieving near-linear strong scaling across multiple GPUs. The solver is validated against a broad suite of canonical benchmark problems involving inviscid, viscous, and reactive compressible flows, including subsonic, transonic, supersonic, and hypersonic configurations with multi-species nonequilibrium chemistry. The numerical predictions show close agreement with established reference solutions. Beyond its immediate performance characteristics, the framework reflects the broader transition of computational science towards modular, adaptive and AI-compatible simulation ecosystems. Owing to its implementation in \texttt{Julia}, \texttt{MARUT} can be naturally integrated with differentiable programming, machine-learning libraries and scientific AI workflows, enabling future developments in autonomous flow control, inverse modeling and data-driven scientific discovery. The combination of high-order accuracy, adaptive resolution, GPU acceleration and distributed scalability therefore establishes \texttt{MARUT} as a flexible platform for \textit{next-generation exascale simulations} of complex compressible-flow phenomena across high-speed reactive-flow applications. \texttt{MARUT}'s official website: \href{https://sites.google.com/view/marutcfd-com/home}{marutcfd.com}.

\end{abstract}

\maketitle
\vspace{0.2cm}

 \begin{small}Keywords:  \textit{High-Fidelity CFD}, \textit{Compressible Flows}, \textit{Multi-GPU}, \textit{MPI Parallelization}, \textit{Adaptive Mesh Refinement}, \textit{AI compatible CFD Solver};
\end{small}

\section{Introduction}
High-fidelity simulation of compressible flows over complex geometries remains a central challenge in computational fluid dynamics (CFD) \cite{moin1998direct}. Modern aerospace applications, including hypersonic vehicle design, supersonic propulsion systems, shock–boundary layer interactions, and high-altitude flight, require numerical methods capable of accurately resolving strong discontinuities, thin viscous layers, and a wide range of interacting multiscale flow phenomena \cite{gaitonde2015progress}. Achieving this level of fidelity within practical computational turnaround times is increasingly difficult on traditional CPU-based architectures, as high-resolution grids and high-order discretizations lead to prohibitive computational costs \cite{slotnick2014cfd}. This limitation motivates the development of GPU-accelerated solvers, which can leverage massive parallelism to enable efficient, high-order simulations of complex compressible flow physics at scale \cite{vermeire2017utility}.

High-order discontinuous Galerkin (DG) methods have emerged as a particularly attractive framework for these problems \cite{cockburn1989tvb,hesthaven2008nodal}. Their element-local formulation enables high arithmetic intensity and excellent parallel efficiency, while their spectral convergence properties significantly reduce the number of degrees of freedom required to achieve a given accuracy compared to low-order schemes \cite{klockner2009nodal}. In addition, their strong local conservation properties are essential for faithfully capturing shock-dominated and multi-scale flow physics in compressible regimes. DG methods are also naturally compatible with unstructured and non-conforming meshes, making them especially well-suited for adaptive mesh refinement (AMR) \cite{berger1989local}. This capability is important for efficiently resolving localized and dynamically evolving features such as shocks, shear layers, and reaction fronts in compressible and reacting flows without incurring prohibitive global mesh resolution costs. However, the combination of high-order accuracy, AMR, and realistic three-dimensional geometries leads to rapidly increasing computational workloads that exceed the capabilities of single-GPU or single-node execution. This motivates the development of scalable multi-GPU solvers that can exploit distributed memory parallelism through message passing interface (MPI) while fully leveraging intra-node GPU acceleration. Modern GPU clusters provide massive parallel throughput and high memory bandwidth, but achieving strong scaling requires careful algorithmic design to minimize communication overhead, balance dynamically adaptive workloads, and maintain high GPU occupancy across heterogeneous mesh resolutions.
In this context, adaptive high-order solvers for compressible flows with finite-rate chemistry introduce additional computational challenges due to stiff source terms, tightly coupled thermo-chemical processes, and stringent stability constraints at high Mach numbers. These complexities further amplify the need for a tightly integrated software framework that combines multi-GPU acceleration, efficient MPI-based domain decomposition, and dynamic AMR strategies.

In this work, we present \textbf{\texttt{MARUT}}, a scalable, multi-GPU, high-order CFD framework for compressible reacting and non-reacting flows with AMR, designed to fully exploit modern heterogeneous computing architectures. The framework is implemented in the Julia programming language using \texttt{Trixi.jl} \cite{schlottkelakemper2025trixi} as the underlying codebase, combining high computational performance with flexibility and extensibility. 
By integrating high-order numerical accuracy, geometric adaptivity, and zero-bottleneck GPU parallelization, \textbf{\texttt{MARUT}} enables robust, low-dissipation, and high-fidelity large-scale simulations across a wide range of shock-dominated and high-speed reactive flow regimes.

The main contributions of this work can be summarized as follows:

\begin{itemize}
   \item We developed a fully GPU-resident hyperbolic solver for the spectral DG discretization, in which each strong-stability-preserving Runge--Kutta (SSP-RK) substage is executed as a sequence of element-local kernels comprising entropy-stable split-form volume integration, shock-capturing indicator evaluation, conforming and non-conforming (mortar) interface flux computations, and a combined surface-integration and Jacobian-rescaling step. The entire time-stepping procedure remains on the GPU, eliminating host-device data transfers during hyperbolic evolution. In contrast to the existing GPU implementation within the Trixi ecosystem, which is limited to Cartesian \texttt{TreeMesh} discretizations, our framework provides an end-to-end GPU execution path for curvilinear, unstructured \texttt{P4estMesh} geometries, thereby enabling high-order entropy-stable shock capturing and non-conforming adaptive mesh coupling on complex domains. All hyperbolic boundary conditions are enforced through dedicated GPU kernels specialized by boundary type, including free-stream and subsonic far-field boundaries, supersonic inflow and outflow boundaries, as well as slip-wall and symmetry boundaries.

   \item  We evaluate the viscous (parabolic) terms entirely on the GPU using the Bassi--Rebay (BR1) formulation, with all viscous boundary conditions enforced through dedicated device-resident kernels, including no-slip adiabatic and isothermal walls, symmetry boundaries, and Dirichlet far-field conditions. This enables high-fidelity simulations of wall-bounded and boundary-layer-resolved flows while maintaining a fully GPU-resident execution path. Whereas \texttt{TrixiCUDA.jl} does not provide GPU acceleration for parabolic operators, \texttt{MARUT} implements the complete viscous operator and associated boundary-condition treatment on the GPU, enabling fully device-resident evaluation of viscous terms throughout the simulation.

    \item We introduce \textbf{\texttt{GPUForest}}, a GPU-resident forest-of-quadtree/octree data structure whose tree connectivity, Morton element ordering, per-element refinement levels, and curvilinear element geometry reside entirely in GPU memory, together with a GPU-resident conservative AMR strategy that dynamically resolves multiscale flow structures, including shocks, shear layers, boundary-layer interactions, and reaction fronts; while overcoming the primary computational bottleneck in adaptive multiscale simulations, thereby enabling substantially higher throughput and scalability on modern GPU architectures. In contrast, \texttt{TrixiCUDA.jl} offers no GPU based mesh adaptation, with refinement and coarsening handled on the host; \texttt{GPUForest} keeps the forest, its connectivity, and the adaptation kernels resident on the GPU throughout the simulation.

    \item  We integrate finite-rate chemistry and thermal non-equilibrium directly on the GPU through per-node source-term kernels that evaluate the species mass-production rates of a finite-rate mechanism and the Landau--Teller vibrational--translational energy exchange of a two-temperature model, enabling reactive and non-equilibrium hypersonic simulations within GPU. Such GPU-resident finite-rate chemistry and thermal-non-equilibrium source-term integration is not available in \texttt{TrixiCUDA.jl}. 

    \item  We establish efficient multi-GPU scalability with strong and weak scaling performance on NVIDIA GPUs, achieving high parallel efficiency for large-scale simulations, an important milestone toward enabling exascale-ready, production-scale, high-fidelity CFD and reactive-flow simulations on next-generation heterogeneous computing platforms.

    \item We demonstrate the robustness of the framework across a wide range of compressible-flow regimes, including subsonic, transonic, supersonic, and hypersonic flows with finite-rate chemistry and stiff reactive source terms; highlighting its applicability to next-generation aerospace systems.
\end{itemize}

The paper is organized as follows: Section~2 introduces the governing compressible Navier--Stokes equations. 
Section~3 describes the details of the GPU implementation, including the treatment of hyperbolic and parabolic terms, source term integration, and the AMR strategy. In Section~4, we present numerical results for a range of test cases spanning Euler and Navier-Stokes regimes, covering subsonic, transonic, supersonic, and hypersonic flows with finite-rate chemistry, along with a comparison of CPU and GPU performance. Section~5 is devoted to multi-GPU scalability, where both strong and weak scaling studies are reported. Finally, Section~6 summarizes the main findings and conclusions of the study.

\section{Governing Equations}
\label{sec:governing}
The compressible Navier-Stokes equations in conservation form describe the conservation of mass, momentum, and energy for a fluid. The general vector form is given by:

\begin{equation}\label{CNSe}
    \frac{\partial \mathbf{U}}{\partial t} + \frac{\partial \mathbf{F}_{\mathrm{inv}}}{\partial x} + \frac{\partial \mathbf{G}_{\mathrm{inv}}}{\partial y} + \frac{\partial \mathbf{H}_{\mathrm{inv}}}{\partial z} = \frac{\partial \mathbf{F}_v}{\partial x} + \frac{\partial \mathbf{G}_v}{\partial y} + \frac{\partial \mathbf{H}_v}{\partial z}, ~~ \mathbf{x} = \{x,y,z\} \in \Omega \subset \mathbb{R}^3,~ t \in  \mathbb{R}^+
\end{equation}
with appropriate boundary and initial conditions.

The conserved variable vector $\mathbf{U}$ and the inviscid fluxes $\mathbf{F}_{\mathrm{inv}}, \mathbf{G}_{\mathrm{inv}}, \mathbf{H}_{\mathrm{inv}}$ are defined as:
\begin{equation*}
    \mathbf{U} = \begin{bmatrix} \rho \\ \rho u \\ \rho v \\ \rho w \\ E \end{bmatrix} \quad
    \mathbf{F}_{\mathrm{inv}} = \begin{bmatrix} \rho u \\ \rho u^2 + p \\ \rho uv \\ \rho uw \\ (E+p)u \end{bmatrix}, \quad
    \mathbf{G}_{\mathrm{inv}} = \begin{bmatrix} \rho v \\ \rho uv \\ \rho v^2 + p \\ \rho vw \\ (E+p)v \end{bmatrix}, \quad
    \mathbf{H}_{\mathrm{inv}} = \begin{bmatrix} \rho w \\ \rho uw \\ \rho vw \\ \rho w^2 + p \\ (E+p)w \end{bmatrix}
\end{equation*}
where $\rho$ is density, $u, v, w$ are the velocity components, $E$ is the total energy per unit volume, and  $p$ is the static pressure.

The viscous flux vectors $\mathbf{F}_v, \mathbf{G}_v, \mathbf{H}_v$ account for the stress tensor $\tau_{ij}$ and heat flux $q_j$:
\begin{equation*}
    \mathbf{F}_v = \begin{bmatrix} 0 \\ \tau_{xx} \\ \tau_{xy} \\ \tau_{xz} \\ u\tau_{xx} + v\tau_{xy} + w\tau_{xz} - q_x \end{bmatrix}, \quad
    \mathbf{G}_v = \begin{bmatrix} 0 \\ \tau_{yx} \\ \tau_{yy} \\ \tau_{yz} \\ u\tau_{yx} + v\tau_{yy} + w\tau_{yz} - q_y \end{bmatrix}, \quad
    \mathbf{H}_v = \begin{bmatrix} 0 \\ \tau_{zx} \\ \tau_{zy} \\ \tau_{zz} \\ u\tau_{zx} + v\tau_{zy} + w\tau_{zz} - q_z \end{bmatrix}
\end{equation*}

The viscous stress tensor for a Newtonian fluid is:
\begin{equation*}
    \tau_{ij} = \mu \left( \frac{\partial u_i}{\partial x_j} + \frac{\partial u_j}{\partial x_i} \right) - \frac{2}{3} \mu \delta_{ij} \frac{\partial u_k}{\partial x_k}
\end{equation*}
The heat flux vector is defined by Fourier's law as $ q_j = -k \frac{\partial T}{\partial x_j},$ where $\mu$ is the dynamic viscosity and $k$ is the thermal conductivity. The system is closed by the equation of state, typically the ideal gas law: $p = (\gamma - 1)(E - \frac{1}{2}\rho(u^2+v^2+w^2))$, where $\gamma$ is the ratio of specific heats.

\subsection{Thermochemical non-equilibrium with finite-rate chemistry}
\label{ssec:tcneq}
At hypersonic velocities, the assumption of thermodynamic equilibrium
breaks down because of finite-rate chemical reactions and delayed
relaxation of internal energy modes~\cite{anderson1989hypersonic,park1989nonequilibrium}.
Under such conditions, the flow is modeled as a reacting,
multi-species mixture in thermochemical non-equilibrium, governed by
the compressible Navier--Stokes equations augmented with species
transport equations and a separate vibrational--electronic energy
equation. In the commonly employed two-temperature formulation, the
translational and rotational energy modes are assumed to remain in
equilibrium at a common temperature, while the vibrational and
electronic modes are described through an additional nonequilibrium
temperature~\cite{park1989nonequilibrium}. Such formulations are widely used in the
simulation of hypersonic chemically reacting flows and atmospheric
entry problems~\cite{anderson1989hypersonic,candler1991computation}.
The vector of conserved variables is defined as
\begin{equation}
\mathbf{U} =
\left[
\rho_1, \dots, \rho_{N_s},
\rho u, \rho v, \rho w,
E, E_v
\right]^T,
\end{equation}
where $\rho_s$ is the partial density of species $s$,
$\rho = \sum_s \rho_s$ is the mixture density, and
$E_v$ is the vibrational--electronic energy per unit volume. The species mass fractions are $Y_s = \rho_s/\rho$.

\vspace{0.1cm}
\noindent The inviscid flux in the $x$-direction is
\begin{equation}
\mathbf{F}_{\mathrm{inv}} =
\left[
\rho_1 u, \dots, \rho_{N_s} u,
\rho u^2 + p,
\rho u v,
\rho u w,
(E+p)u,
E_v u
\right]^T,
\end{equation}
with analogous expressions in $y$ and $z$ directions. The pressure is given by the mixture equation of state
\begin{equation*}
p = \rho R T,
\qquad
R = \sum_s Y_s R_s,
\end{equation*}
where $T$ is the translational temperature and $R_s$ is the specific
gas constant of species $s$.

\vspace{0.1cm}
\noindent Species diffusion is modeled using a mixture-averaged Fick law,
\begin{equation*}
J_{s,i} = -\rho D_s \frac{\partial Y_s}{\partial x_i},
\qquad
\sum_{s=1}^{N_s} J_{s,i} = 0,
\end{equation*}
where $D_s$ is the effective diffusion coefficient of species $s$.

\vspace{0.1cm}
\noindent
The translational and vibrational heat fluxes are modeled as
\begin{equation*}
q_i = -k_{tr} \frac{\partial T}{\partial x_i},
\qquad
q_{v,i} = -k_v \frac{\partial T_v}{\partial x_i},
\end{equation*}
where $k_{tr}$ and $k_v$ are translational and vibrational thermal
conductivities, and $T_v$ is the vibrational temperature.

\vspace{0.1cm}
\noindent The viscous flux in the $x$-direction is
\begin{equation}
\mathbf{F}_v =
\begin{bmatrix}
- J_{1,x} \\
\vdots \\
- J_{N_s,x} \\
\tau_{xx} \\
\tau_{xy} \\
\tau_{xz} \\
u \tau_{xx} + v \tau_{xy} + w \tau_{xz}
- q_x
- \sum_s h_s J_{s,x} \\
- q_{v,x}
- \sum_s e_{v,s} J_{s,x}
\end{bmatrix}.
\end{equation}

Here $h_s = h_s^0 + \int c_{p,s}(T)\,dT$ is the sensible enthalpy, $h_s^0$ is the formation enthalpy, $e_{v,s}(T_v)$ is the vibrational energy of species $s$.

\vspace{0.1cm}
\noindent The source vector is
\begin{equation}
\mathbf{S} =
\left[
\dot{\omega}_1, \dots, \dot{\omega}_{N_s},
0, 0, 0, 0,
Q_{T\text{-}v} + \sum_s \dot{\omega}_s e_{v,s}
\right]^T,
\end{equation}
where $\dot{\omega}_s$ is the net chemical production rate of species
$s$.

\vspace{0.1cm}
\noindent \textbf{Chemical kinetics}:
Finite-rate chemistry is modeled using Arrhenius laws,
\begin{equation*}
\dot{\omega}_s = W_s \sum_r (\nu''_{sr} - \nu'_{sr}) R_r,
\end{equation*}
where $W_s$ is the molecular weight of species $s$, and
$\nu'_{sr}$ and $\nu''_{sr}$ are the stoichiometric coefficients of
reactants and products, respectively.

The reaction rate for reaction $r$ is
\begin{equation*}
R_r = k_{f,r} \prod_s (\rho_s/W_s)^{\nu'_{sr}} - k_{b,r} \prod_s (\rho_s/W_s)^{\nu''_{sr}},
\end{equation*}
with forward rate coefficient
\begin{equation}
k_{f,r} = A_r T_a^{\beta_r} \exp\!\left(-\frac{E_r}{R T_a}\right),
\qquad
T_a = \sqrt{T T_v}.
\end{equation}
where $A_r$, $\beta_r$, and $E_r$ are the Arrhenius constants, and
$T_a$ is the effective two-temperature reaction temperature.

\vspace{0.1cm}
\noindent \textbf{Thermal nonequilibrium coupling}: The vibrational relaxation term follows a Landau--Teller model,
\begin{equation}
Q_{T\text{-}v}
=
\sum_{s \in \mathrm{mol}}
\rho_s
\frac{e_{v,s}^{eq}(T) - e_{v,s}(T_v)}{\tau_s},
\end{equation}
where $\tau_s$ is the vibrational relaxation time.

\vspace{0.1cm}
\noindent \textbf{Total energy and closure}:
The total energy is decomposed as
\begin{equation}
E =
\sum_s \rho_s e_{tr,s}(T)
+ E_v
+ \frac{1}{2}\rho (u^2+v^2+w^2)
+ \sum_s \rho_s h_s^0,
\end{equation}
where $e_{tr,s}(T)$ denotes the sensible translational--rotational
internal energy of species $s$, excluding the species formation
enthalpy $h_s^0$ which appears as a separate term. The system is closed by the calorically perfect mixture relation for
each mode and the equation of state $p = \rho R T.$

In nonequilibrium flow simulations, a 5-species model represents air using five chemical species $\mathrm{N_2}$, $\mathrm{O_2}$, $\mathrm{NO}$, $\mathrm{N}$, and $\mathrm{O}$, and is typically used for high-temperature flows where molecular dissociation occurs but ionization remains limited. An 11-species model extends this description by additionally including ionized species and electrons $\mathrm{N_2^+}$, $\mathrm{O_2^+}$, $\mathrm{NO^+}$, $\mathrm{N^+}$, $\mathrm{O^+}$, and $e^-$, allowing the simulation of strongly ionized hypersonic and plasma flows where both chemical and thermal nonequilibrium effects are significant. See Appendix~\ref{appC}.

For clarity and ease of reference, the nomenclature is provided in Appendix~\ref{appD}.

\section{\texttt{MARUT} Solver}
In the present work, the \texttt{MARUT} solver employs a high-order spectral DG method for spatial discretization (see Appendix~\ref{appA} for the entropy-stable strong formulation), while temporal integration is performed using high-order strong-stability-preserving Runge--Kutta (SSP-RK) schemes (see Appendix~\ref{appB} for details). To accurately capture multi-scale flow features, including shocks, boundary layers, and wake structures, an AMR strategy is utilized via the \texttt{P4estMesh} library~\cite{burstedde2011p4est}. The computational domain is represented as a parallel forest-of-trees, where each mesh element is recursively subdivided across multiple refinement levels. Depending on the spatial dimension, this corresponds to a forest-of-quadtrees in two dimensions and a forest-of-octrees in three dimensions, enabling efficient and scalable dynamic mesh adaptation. The initial forest and its inter-tree connectivity are constructed once at start-up using the \texttt{P4estMesh} backend~\cite{burstedde2011p4est}; thereafter, the entire structure is mirrored into a GPU-resident representation, \texttt{GPUForest}, which stores the tree connectivity, space-filling (Morton) ordering, per-element refinement levels, and curvilinear element geometry in GPU memory. All subsequent refinement, coarsening, and solver kernels operate directly on \texttt{GPUForest}, so the mesh data structure remains on the device throughout the simulation.

\begin{figure}[htbp]
    \centering
\includegraphics[width=0.85\textwidth]{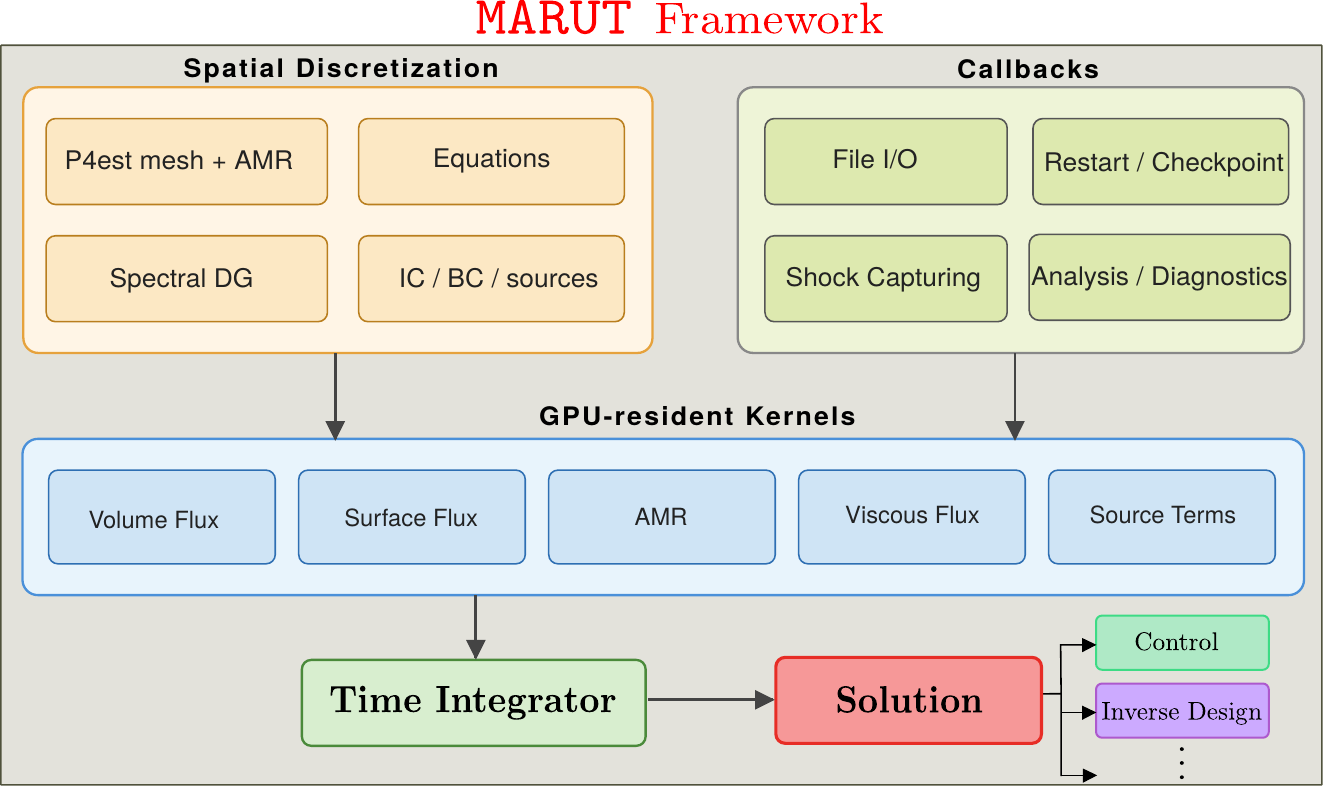}
    \caption{High-level architecture of the \texttt{MARUT} solver. The spatial-discretisation components configure the GPU-resident RHS kernels, which are advanced in time by the SSP-RK integrator under the control of runtime callbacks.}
    \label{fig:marut_framework}
\end{figure}

Mesh refinement is driven by a L\"ohner-type second-difference
indicator~\cite{lohner1987adaptive}, which detects shocks and steep
gradients. 
Figure~\ref{fig:marut_framework} summarises the overall representation of \texttt{MARUT}: the spatial-discretisation layer (mesh, equations, Spectral DG operator, initial/boundary conditions and source terms) feeds the central band of GPU-resident kernels that evaluate volume and surface fluxes, the AMR cycle, viscous fluxes and source terms; the kernels are wrapped by a set of runtime callbacks (file I/O, restart/checkpoint, shock capturing, and analysis/diagnostics) and driven by the SSP-RK integrator that produces the time-advanced solution.

\texttt{MARUT} provides a robust AMR framework based on hierarchical tree-structured meshes, enabling dynamically adaptive high-order DG simulations with excellent conservation and parallel scalability properties.  The initial mesh is imported through the \texttt{P4estMesh} backend, which builds the forest and its inter-tree connectivity at start-up; the adaptive machinery itself then runs on the device-resident \texttt{GPUForest}, which supports localized mesh refinement and coarsening during runtime while preserving nonconforming mesh interfaces subject to a 2:1 balance constraint. By combining physics-based refinement indicators, controller-driven adaptation strategies, and distributed forest-of-octree data structures provided by \texttt{p4est}, \texttt{MARUT} facilitates efficient large-scale simulations of hyperbolic and hyperbolic--parabolic systems on distributed-memory architectures. Although the framework is not specifically intended for fully general unstructured remeshing, its mature quadtree/octree-based AMR capabilities render it highly effective for large-scale scientific computations requiring high accuracy, geometric adaptivity, and computational efficiency.

\subsection{GPU Implementation}
A GPU has many small cores compared to a CPU, but each core is
simpler.  A CPU might have 32 or 64 fast cores with large caches; a
GPU has thousands of cores that are individually slower but run in
parallel.  When a problem can be split into many independent pieces of
roughly the same size, the GPU wins because it processes all the
pieces at once.  When the work is mostly sequential or branches
unpredictably, the CPU wins. Spectral DG falls into the first category.  The volume integral on each element is independent, the face flux at each interface is independent, and the update at each solution point is independent.  A mesh with $N_e$ elements and $(\mathcal{P}+1)^d$ nodes per element gives $N_e(\mathcal{P}+1)^d$ solution points, each of which can be assigned to its own GPU thread.  For typical problem sizes this is $10^4$ to $10^6$ threads, which is enough to keep the GPU completely utilized.

Memory access patterns matter as much as parallelism on a GPU.
Threads that read consecutive addresses in memory get served in a
single transaction (coalesced access); scattered reads waste
bandwidth.  \texttt{MARUT} stores the solution as a four-dimensional array
indexed by $(\text{variable}, \text{node}_i, \text{node}_j,
\text{element})$, here \emph{variable} is the conserved-variable index (e.g.\ $\rho$, $\rho u$, $\rho v$, $\rho w$, $E$), $\text{node}_i$ and $\text{node}_j$ are the tensor-product GLL node indices within an element (extending to $\text{node}_k$ in 3D), and \emph{element} is the global element index in the forest.  Threads working on consecutive elements therefore
read consecutive memory locations, which gives coalesced access for
the most common operations. Each step in the Spectral DG right-hand-side evaluation (volume integral, prolongation to faces, numerical flux, surface integral, Jacobian scaling) is a separate kernel launch.  Keeping kernels small has two benefits: it limits the number of registers each thread needs (which controls how many threads can run simultaneously), and it lets the hardware scheduler overlap independent kernels when possible. All arrays needed during time stepping (solution, metric terms, differentiation matrices, connectivity tables) stay in GPU memory for the entire simulation.  The CPU sets up the mesh and operators at the start, and writes output files periodically, but no data moves between CPU and GPU during the time-stepping loop.

\subsubsection{Kernel Architecture for Hyperbolic Terms}
\label{sec:hyperbolic_kernels}

The hyperbolic right-hand side is built in five steps, listed in
Table~\ref{tab:hyp_kernels}. Every step is a single GPU kernel that
runs over the whole mesh; the steps execute in order inside each
Runge--Kutta substage, and no data is copied back to the host between
them. Throughout this section, $\mathcal{R}_i$ denotes the per-node
contribution to the spatial residual at collocation node~$i$. 
The
assembled residual over all nodes yields the right-hand side operator
$\mathcal{L}(\mathbf{U}^h, t)$ of the semi-discrete system
(Appendix~\ref{appB}).

Each physical element $\Omega_e$ is the image of the reference cube
$\hat{\Omega} = [-1,1]^d$ under an element-wise mapping
$\mathbf{x} = \mathbf{X}(\boldsymbol{\xi})$, where
$\boldsymbol{\xi} = (\xi_1, \dots, \xi_d)$ are the reference coordinates.
The mapping defines a Jacobian matrix $J_{dm} = \partial x_d / \partial \xi_m$,
its determinant $J = \det J_{dm}$, and a set of contravariant basis vectors
$\mathbf{Ja}^{\,l}$ that satisfy the discrete Geometric Conservation Law (GCL) identity
$\sum_l \partial_{\xi_l}(\mathbf{Ja}^{\,l}) = 0$ by construction. The contravariant flux
$\widetilde{\mathbf{F}}^{\,l} \equiv \mathbf{Ja}^{\,l} \cdot
\mathbf{F}_{\mathrm{inv}}$ maps the Cartesian inviscid flux into reference
space. Within each element the solution is collocated at the $(\mathcal{P}+1)^d$
tensor-product Gauss--Lobatto--Legendre (GLL) nodes with associated quadrature
weights $w_i$; differentiation in the $l$-th reference direction is performed
by the one-dimensional GLL differentiation matrix $D$, which includes the
boundary-correction terms required by the Summation-By-Parts (SBP) property of the Lobatto rule.

\begin{table}[htbp]
\centering
\small
\caption{Steps of the hyperbolic RHS evaluation. Each step is one GPU
kernel.}
\label{tab:hyp_kernels}
\begin{tabular}{c l l}
\hline
Step & Operation & Acts on \\
\hline
1 & Shock indicator and smoothing & every element / face \\
2 & Volume integral (weak / split / shock-blended) & every element \\
3 & Interface and mortar fluxes                    & every interior face \\
4 & Boundary flux (one kernel per BC type)         & every boundary face \\
5 & Surface integral and Jacobian rescaling        & every element \\
\hline
\end{tabular}
\end{table}

\paragraph{Shock Indicator and Smoothing:}
Step~1 applies the Hennemann--Gassner indicator~\cite{hennemann2021provably} to the
density--pressure product on the GPU, then smooths the resulting
$\alpha_e$ field across face neighbours so that the blending region
carries no sharp edges. The smoothed coefficient is held in a
per-element scratch array for the remainder of the substage. Performing
this kernel before the volume integral ensures that the blend in
equation~\eqref{eq:vol_blend} below uses the indicator value computed
from the current Runge--Kutta state, not from a stale one carried over
from a previous substage.

\paragraph{Volume Integral:}
Step~2 evaluates the inviscid flux divergence at every solution point
$(i, j[, k])$ of element $e$. Three operators are available and
selected once at solver setup. The standard strong-form nodal SBP Spectral DG
\cite{hesthaven2008nodal} computes the
residual through a single derivative of the contravariant flux
$\widetilde{\mathbf{F}}^{\,l} \equiv \mathbf{Ja}^{\,l}\!\cdot\!\mathbf{F}_{\mathrm{inv}}$
(see~\cite{kopriva2009implementing} for the curvilinear DG-SEM
reference-element setup),
\begin{equation}
    \mathcal{R}^{(\text{strong})}_{i}
    \;=\;
    -\,\frac{1}{J_{i}}\sum_{l=1}^{d}\!\sum_{m=0}^{\mathcal{P}}
        D_{im}\,\widetilde{\mathbf{F}}^{\,l}_{m},
    \label{eq:vol_weak}
\end{equation}
On GLL collocation nodes the strong-form residual above and its weak-form counterpart obtained by integration by parts are algebraically equivalent through the SBP property of the Lobatto quadrature; we use the strong-form expression throughout this section. In the inner sum, $m$
replaces the collocation-node index along reference direction~$l$
in the tensor-product multi-index $(i,j,k)$; the indices in the
remaining directions are held fixed, so the sum runs over the $\mathcal{P}+1$
nodes of direction~$l$ only. The entropy-stable split form
\cite{fisher2013high,gassner2016split} replaces this single
derivative with a symmetric two-point sum,
\begin{equation}
    \mathcal{R}^{(\text{split})}_{i}
    \;=\;
    -\,\frac{2}{J_{i}}\sum_{l=1}^{d}\!\sum_{m=0}^{\mathcal{P}}
        D_{im}\,\mathbf{F}^{\#}\!\Bigl(\mathbf{U}_{i},\,\mathbf{U}_{m};\,\overline{\mathbf{Ja}^{\,l}}_{im}\Bigr),
    \label{eq:vol_split}
\end{equation}
where $\overline{\mathbf{Ja}^{\,l}}_{im} = \tfrac{1}{2}\bigl(\mathbf{Ja}^{\,l}_i + \mathbf{Ja}^{\,l}_m\bigr)$
is the arithmetic average of the contravariant basis at the two
partner nodes and $\mathbf{F}^{\#}$ is the entropy-conservative
two-point flux of Ranocha~\cite{ranocha2018comparison}; the factor
of two compensates the double counting of the symmetric sum and
recovers the standard divergence on smooth solutions through the
consistency $\mathbf{F}^{\#}(\mathbf{U},\mathbf{U}) = \mathbf{F}_{\mathrm{inv}}(\mathbf{U})$.
When shocks are present the volume kernel returns the convex blend
of equation \eqref{eq:vol_split} with a first-order finite-volume residual
built on the GLL sub-cell grid~\cite{hennemann2021provably},
\begin{equation}
    \mathcal{R}^{(\text{blend})}_{i}
    \;=\;
    (1 - \alpha_e)\,\mathcal{R}^{(\text{split})}_{i}
    \;+\; \alpha_e\,\mathcal{R}^{(\text{FV})}_{i}.
    \label{eq:vol_blend}
\end{equation}
The blending coefficient $\alpha_e \in [\alpha_{\min}, \alpha_{\max}]$ controls the per-element mix between the high-order entropy-stable split DG residual and the robust low-order finite-volume residual: $\alpha_e=0$ recovers pure DG (smooth regions), $\alpha_e=1$ recovers pure FV (shock-dominated regions), and intermediate values give a smooth blend. For each element it is computed once per Runge--Kutta substage by the Hennemann--Gassner indicator kernel (Step~1) and is then reused, unchanged, by every subsequent kernel within that substage.

\paragraph{Surface Fluxes:}
At every interior face shared by two same-level elements a numerical
flux $\mathbf{F}^{*}(\mathbf{U}^{-}, \mathbf{U}^{+}; \hat{\mathbf{n}})$
-- either Lax--Friedrichs~\cite{edwards2006dominant} or
HLL~\cite{harten1983upstream}, is evaluated from the interior and
exterior traces $\mathbf{U}^{\pm}$ and the contravariant face normal
$\hat{\mathbf{n}}$.
The two states $\mathbf{U}^\pm$ are read directly from the volume
buffer through precomputed face-indexing tables, which encode the
relative orientation of the two elements that share a face on
unstructured P4est meshes~\cite{burstedde2011p4est}.

Where AMR places a coarse element next to two (in 2D) or four (in 3D)
finer elements, the face is non-conforming and is handled by a
mortar~\cite{kopriva1996conservative}: the solution is projected from
the fine and coarse sides onto the mortar, the Riemann solver is
evaluated on the fine-side quadrature, and the result is projected
back onto both elements through the corresponding $\mathbb{L}_2$ projection
matrices. The reverse projection accumulates contributions from all $2$
(in~2D) or $4$ (in~3D) mortars sharing the coarse face; the
factor of $2$ or $4$ in the projection matrix reflects the ratio
between the coarse face area and a single mortar's area.

Boundary fluxes are dispatched per type. The host keeps a small
dictionary that maps each boundary name in the input mesh to an array
of face indices, and launches one kernel per boundary type
implementing the corresponding boundary functor (slip wall,
supersonic inflow, supersonic outflow, freestream Dirichlet, periodic
with rotation, \dots). One kernel per boundary type lets each functor
be specialised independently.

\paragraph{Surface Integral and Jacobian Rescale:}
Step~5 closes the pipeline. Architecturally, Steps~3 and~4 do
not write directly into the volume residual: each face/boundary kernel
emits its computed Riemann flux into a dedicated, element-indexed
\emph{surface buffer}. Step~5 is then launched as a clean,
massively-parallel per-node volume kernel in which each thread loads its
own volume residual contribution (computed in Step~2) together with the
matching face values from the surface buffer through coalesced reads,
and performs the boundary correction locally with no atomic operations
and no race conditions on shared corner or edge nodes.
For a node $i$ on the element boundary, the local update reads
\begin{equation}
    \frac{d\mathbf{U}_{i}}{dt}
    \;=\;
    \mathcal{R}^{(\text{vol})}_{i}
    \;-\; \frac{1}{J_i\,w_i}\sum_{f\,\ni\, i}\!
        \bigl(\,
        \mathbf{F}^{*}\!\bigl(\mathbf{U}^{-}, \mathbf{U}^{+};\,\mathbf{Ja}^{\,l_f}\bigr)
        \;-\; \widetilde{\mathbf{F}}^{\,l_f}
        \,\bigr)\,\cdot\,\mathbf{n}^{\text{ref}}_{f},
    \label{eq:surf_int}
\end{equation}
where the sum runs over the faces $f$ that contain node $i$;
$l_f \in \{1,\dots,d\}$ is the reference direction normal to face $f$;
$\mathbf{n}^{\text{ref}}_{f} = \pm \hat{\mathbf{e}}_{l_f}$ is the
outward reference-space face normal ($\pm 1$ depending on whether $f$
is the upper or lower face along $l_f$); the numerical Riemann flux
$\mathbf{F}^{*}$ is constructed using the \emph{unnormalised}
contravariant face vector $\mathbf{Ja}^{\,l_f}$ as its normal-direction
argument (so that the surface metric Jacobian is absorbed automatically
and both terms in the bracket live in reference space); and
$\widetilde{\mathbf{F}}^{\,l_f}=\mathbf{Ja}^{\,l_f}\!\cdot\!\mathbf{F}_{\mathrm{inv}}$
is the interior contravariant flux evaluated from the volume state.
The bracketed jump vanishes for interior nodes through the
boundary-interpolation pattern of the GLL basis, so only boundary nodes
of the element receive a non-zero correction. Here
$\mathcal{R}^{(\text{vol})}_{i}$ denotes whichever volume operator is
active, equations~\eqref{eq:vol_weak}, \eqref{eq:vol_split}, or
\eqref{eq:vol_blend}.
A user source term, when present, is added by a final per-node
kernel before the residual is handed to the time integrator.

\subsubsection{Kernel Architecture for Parabolic Terms}
\label{sec:parabolic_kernels}

The viscous right-hand side follows the BR1 mixed formulation of
Bassi and Rebay~\cite{bassi1997high}. Because BR1 introduces an
auxiliary gradient unknown, the viscous residual cannot be assembled
in a single sweep and is built in two steps, listed in
Table~\ref{tab:para_kernels}. When both physics types are active,
the hyperbolic and parabolic pipelines run back-to-back and add
their contributions into a shared residual buffer before the time
integrator advances.

\begin{table}[htbp]
\centering
\small
\caption{Steps of the BR1 viscous RHS evaluation. Each step finishes
on the GPU before the next one starts.}
\label{tab:para_kernels}
\begin{tabular}{c l l}
\hline
Step & Operation & Output \\
\hline
1 & Compute the BR1 gradient $\mathbf{S}^h = \nabla \mathbf{W}^h$ & $\mathbf{S}^h$ \\
2 & Assemble viscous flux and take its divergence              & $\mathcal{R}^{(\text{visc})}_{i}$ \\
\hline
\end{tabular}
\end{table}

\paragraph{Step 1 - BR1 Gradient:}
The auxiliary gradient $\mathbf{S}^h = \nabla \mathbf{W}^h$, where $\mathbf{W} = (\rho, u, v, w, T)^T$ is the vector of primitive gradient variables used in \texttt{MARUT}'s BR1 viscous formulation, is
computed by a strong-form DG operator with the same structure as the
hyperbolic surface pipeline, but with a central interior trace
instead of an upwind numerical flux. At every interior face the
BR1 numerical trace is the arithmetic average,
\begin{equation}
    \widehat{\mathbf{W}}
    \;=\;
    \tfrac{1}{2}\bigl(\mathbf{W}^{h-} + \mathbf{W}^{h+}\bigr),
    \label{eq:br1_avg}
\end{equation}
This central (arithmetic-mean) interface state is the standard BR1 numerical trace introduced in Appendix~\ref{appA}; here it is applied to the primitive gradient variables $\mathbf{W}^h$ rather than to the conserved state
At a no-slip adiabatic wall the boundary kernel sets the wall
velocity to zero and inherits interior $(\rho, T)$; at an
isothermal wall it additionally fixes $T = T_{\text{wall}}$;
at a slip wall it mirrors the velocity across the boundary normal.
Non-conforming faces use the same forward and reverse projections as
the hyperbolic side. After the surface contributions have been
accumulated, the gradient is rescaled by the inverse Jacobian to
express it in physical space.

\paragraph{Step 2 - Viscous Flux and Divergence:}
All quantities in this step are expressed in non-dimensional form
with $R = 1$, so that $T \equiv p/\rho$; the reference values
$\mu_{\text{ref}}$, $T_{\text{ref}}$ and $S$ below are likewise
non-dimensionalised.
A per-node kernel uses the gradient $\mathbf{S}^h$ from Step~1 and
the local state $\mathbf{U}$ to compute the viscosity, the stress
tensor, and the heat flux; the primitive components
$u_i = (\rho u_i)/\rho$ and $T = p/\rho$ needed by these closures are
evaluated on the fly inside this kernel. The viscosity is either constant ($\mu = \mu_0$) or
temperature-dependent through Sutherland's
law~\cite{sutherland1893viscosity},
\begin{equation}
    \mu(T)
    \;=\;
    \mu_{\text{ref}}\,
    \left(\frac{T}{T_{\text{ref}}}\right)^{\!3/2}\!
    \frac{T_{\text{ref}} + S}{T + S}.
    \label{eq:sutherland}
\end{equation}
The Newtonian stress tensor (Stokes hypothesis, zero bulk viscosity)
is
\begin{equation}
    \tau_{ii} \;=\; \tfrac{4}{3}\,\mu\,\partial_{x_i} u_i
    \;-\; \tfrac{2}{3}\,\mu\!\sum_{j\neq i}\!\partial_{x_j} u_j,
    \qquad
    \tau_{ij} \;=\; \mu\,\bigl(\partial_{x_j} u_i + \partial_{x_i} u_j\bigr),
    \quad i \neq j,
    \label{eq:tau}
\end{equation}
the heat flux is $\mathbf{q} = -k\,\nabla T$ with the
non-dimensional thermal conductivity
$k = \gamma\,\mu / [\Pr\,(\gamma-1)]$, and the viscous flux
components $\mathbf{F}_{v,i}$ are assembled from $\boldsymbol{\tau}$
and $\mathbf{q}$ as defined in~Section~\ref{sec:governing}.
A second kernel takes the strong-form divergence of
$\mathbf{F}_{v}$ exactly as in equation~\eqref{eq:vol_weak} applied to
$\mathbf{F}_{v}$ instead of $\mathbf{F}_{\mathrm{inv}}$, and a
central surface pipeline (interface, mortar, and boundary kernels)
supplies the inter-element coupling: the BR1 viscous interface flux is taken as the arithmetic average of the viscous fluxes constructed from the gradients on the two sides,
\begin{equation}
    \mathbf{F}_{v}^{*}
    \;=\;
    \{\!\{\mathbf{F}_v\}\!\}
    \;=\;
    \tfrac{1}{2}\bigl(\mathbf{F}_v(\mathbf{U}^{h-}, \mathbf{S}^{h-})
                       + \mathbf{F}_v(\mathbf{U}^{h+}, \mathbf{S}^{h+})\bigr),
    \label{eq:br1_flux}
\end{equation}
that is, the central trace is taken of the gradient-dependent viscous fluxes rather than of the primitive states alone, which is what makes the BR1 primal discretisation adjoint-consistent and stable.

\paragraph{Time-step Bound:}
A small per-element kernel evaluates the local diffusion stability
number from the per-node viscosity, density and metric, and reduces
over elements to give the parabolic time-step bound
\begin{equation}
    \Delta t_D \;\le\; \tilde c\,\frac{\Delta x_{\min}^2}{(2\mathcal{P}+1)^2\,\nu_{\max}},
    \label{eq:dt_para}
\end{equation}
where $\tilde c$ is the SSP coefficient of the time scheme (Appendix~\ref{appB}), $\Delta x_{\min} = \min_e \Delta x_e$ is the smallest characteristic element size obtained from the device-side reduction, and $\nu_{\max}$ is the maximum of the kinematic viscosity and the thermal diffusivity. This bound is combined with the hyperbolic bound $\Delta t_C$
of Appendix~\ref{appB} through $\Delta t = \min(\Delta t_C, \Delta t_D)$. For completeness, the hyperbolic counterpart is evaluated by an analogous per-element kernel that gives
\begin{equation}
    \Delta t_C \;\le\; \tilde c\,\frac{\Delta x_{\min}}{(2\mathcal{P}+1)\,\lambda_{\max}},
    \label{eq:dt_hyp}
\end{equation}
where $\lambda_{\max} = \max_e \max_i (|u_i| + a)$ is the maximum convective wave speed in the domain and $a$ is the local speed of sound. The full derivation is given in Appendix~\ref{appB}; here, $\Delta t_C$ and $\Delta t_D$ are produced by separate GPU reductions in the same time-step kernel and combined as $\Delta t = \min(\Delta t_C, \Delta t_D)$.

\subsubsection{Kernel Architecture for Source Terms}
\label{sec:source_kernels}
For chemically reacting and thermally non-equilibrium flows the
semi-discrete system carries two source contributions absent in the
inert Navier--Stokes case: the species mass-production rates
$\dot{\omega}_s$ from finite-rate chemistry and the vibrational
energy-exchange term $Q_{T\text{-}v}$ from Landau--Teller relaxation
between the translational--rotational mode at temperature $T$ and the
vibrational mode at temperature $T_v$. Writing the conservative state
as $\mathbf{U}=[\rho_1,\dots,\rho_{N_s},\,\rho u,\rho v,\rho w,\,E,\,E_v]^\top$
in the notation of Sec.~\ref{ssec:tcneq} (with $s=1,\dots,N_s$),
the governing equations read
\begin{equation}
    \frac{\partial \mathbf{U}}{\partial t}
    +\nabla\!\cdot\!\bigl(\mathbf{F}_{\!\mathrm{inv}}(\mathbf{U}) - \mathbf{F}_v(\mathbf{U}, \nabla\mathbf{U})\bigr)
    = \mathbf{S}(\mathbf{U}),
    \label{eq:src_governing}
\end{equation}
where $E_v=\sum_{s\in\mathrm{mol}}\rho_s e_{v,s}(T_v)$ is the total
vibrational energy density summed over molecular species, and the source
vector is
\begin{equation}
    \mathbf{S}(\mathbf{U})
    \;=\;
    \begin{pmatrix}
        \dot\omega_1,\;\dots,\;\dot\omega_{N_s} \\[1pt]
        \mathbf{0} \\[1pt]
        0 \\[1pt]
        Q_{T\text{-}v} \;+\; \sum_{s\in\mathrm{mol}}\dot\omega_s\,e_{v,s}(T_v)
    \end{pmatrix}.
    \label{eq:src_vector}
\end{equation}
Momentum and total energy carry no source contribution: the elementary
reactions of the Park air-5 mechanism \cite{park1989nonequilibrium} conserve atoms and the
Landau--Teller exchange merely redistributes energy between the two
internal modes. The last entry of \eqref{eq:src_vector} contains, in
addition to $Q_{T\text{-}v}$, the convective accompaniment
$\sum_s \dot\omega_s\,e_{v,s}$, which accounts for the vibrational
energy carried by molecules as they are created or destroyed by
chemistry.

The Park air-5 mechanism uses five reactions: three dissociation
reactions
$\mathrm{N}_2+\mathrm{M}\rightleftharpoons 2\mathrm{N}+\mathrm{M}$,
$\mathrm{O}_2+\mathrm{M}\rightleftharpoons 2\mathrm{O}+\mathrm{M}$,
$\mathrm{NO}+\mathrm{M}\rightleftharpoons \mathrm{N}+\mathrm{O}+\mathrm{M}$
in which any of the five species can act as the third body $\mathrm{M}$,
and the two Zeldovich exchange reactions
$\mathrm{N}_2+\mathrm{O}\rightleftharpoons \mathrm{NO}+\mathrm{N}$ and
$\mathrm{N}+\mathrm{O}_2\rightleftharpoons \mathrm{NO}+\mathrm{O}$.
For reaction $r$ with reactant and product stoichiometric coefficients
$\nu'_{sr}$ and $\nu''_{sr}$ (in the notation of
Sec.~\ref{ssec:tcneq}), the net molar rate is
\begin{equation}
    R_r
    \;=\;
    k_{f,r}(T_a)\prod_{s}[X_s]^{\nu'_{sr}}
    \;-\;
    k_{b,r}(T_a)\prod_{s}[X_s]^{\nu''_{sr}},
    \qquad
    [X_s] \equiv \rho_s / W_s,
    \label{eq:src_reaction_rate}
\end{equation}
where $W_s$ is the molar mass and $k_{f,r},\,k_{b,r}$ are evaluated at the
\emph{controlling temperature} $T_a$. Following Park,
$T_a=\sqrt{T\,T_v}$ for the three dissociation reactions, capturing
the fact that molecular dissociation depends jointly on translational
collision energy and vibrational pre-excitation, while $T_a=T$ for the
Zeldovich exchanges, which are dominated by collisional energy alone.
The forward rate uses the modified Arrhenius form
$k_{f,r}(T_a)=A_r\,T_a^{\beta_r}\,\exp\!\bigl(-E_r/(R T_a)\bigr)$ with
tabulated $(A_r,\beta_r,E_r)$, and the backward rate is recovered from
the curve-fit equilibrium constant
$k_{b,r} = k_{f,r} / K_c^r(T_a)$, where $T_a$ takes the per-reaction value defined above ($T_a=\sqrt{T T_v}$ for dissociation reactions, $T_a=T$ for the Zeldovich exchanges); evaluating both $k_{f,r}$ and $K_c^r$ at the same controlling temperature preserves microscopic reversibility and the correct multi-temperature equilibrium state behind strong shocks. The species mass production rate is then
\begin{equation}
    \dot\omega_s
    \;=\;
    W_s\sum_{r=1}^{N_r}\bigl(\nu''_{sr} - \nu'_{sr}\bigr)\,R_r.
    \label{eq:src_omega}
\end{equation}
The atom-conservation property $\sum_s \alpha_s^{(N)}\dot\omega_s = 0$
and $\sum_s \alpha_s^{(O)}\dot\omega_s = 0$, with $\alpha_s^{(N,O)}$ the
number of N and O atoms in species $s$, is enforced algebraically by the
stoichiometry of \eqref{eq:src_reaction_rate}--\eqref{eq:src_omega} and
holds to machine precision in the kernel.

For each molecular species $s\in\mathrm{mol}$, the relaxation toward
local thermal equilibrium between the two modes is modelled by
Landau--Teller,
\begin{equation}
    Q_{T\text{-}v}
    \;=\;
    \sum_{s\in\mathrm{mol}}\rho_s\,
        \frac{e_{v,s}^{eq}(T) - e_{v,s}(T_v)}{\tau_s(T,p)},
    \label{eq:src_landau_teller}
\end{equation}
where $e_{v,s}^{eq}(T)=R_s\theta_{v,s}/[\exp(\theta_{v,s}/T)-1]$ is the
harmonic-oscillator vibrational energy per unit mass at the
translational temperature with characteristic temperature
$\theta_{v,s}$, and the relaxation time
$\tau_s(T,p)$ is the Millikan--White correlation augmented by Park's
high-temperature collision-limiting correction. $Q_{T\text{-}v}>0$
drives $T_v\!\to\!T$ from below (vibrational pumping behind a shock) and
$Q_{T\text{-}v}<0$ from above (vibrational cooling in expansion fans).

The characteristic timescales of \eqref{eq:src_omega} and
\eqref{eq:src_landau_teller} at post-shock conditions can be several
orders of magnitude below the acoustic CFL limit of the hyperbolic
operator, so explicit treatment inside the SSPRK substages would freeze
the global time step at the chemistry rate. \texttt{MARUT} avoids this
through a symmetric second-order Strang split,
\begin{equation}
    \mathbf{U}^{n+1}
    \;=\;
    \mathcal{S}_{\Delta t/2}\!\circ\!\mathcal{T}_{\Delta t}\!\circ\!\mathcal{S}_{\Delta t/2}\bigl(\mathbf{U}^{n}\bigr),
    \label{eq:strang_split}
\end{equation}
where $\mathcal{T}_{\Delta t}$ is the hyperbolic--parabolic update of
Sections~\ref{sec:hyperbolic_kernels}--\ref{sec:parabolic_kernels} taken
explicitly with SSPRK54, and $\mathcal{S}_{\Delta t/2}$ is the local
source update taken implicitly.

Because $\mathbf{S}$ in \eqref{eq:src_vector} is purely local,
depending only on $(\rho_s, T, T_v)$ at a single node and not on
neighbours, the entire source substep collapses to one per-node GPU
kernel with no inter-thread communication and no ghost data. At each
node the kernel advances the reduced state
$\mathbf{Y}=(\rho_1,\dots,\rho_{N_s},\,E_v)\in\mathbb{R}^{N_s+1}$ by
Newton iteration on the backward-Euler residual. Because $\sum_s \dot\omega_s = 0$ holds exactly by stoichiometry and the momentum and total-energy entries of $\mathbf{S}$ in~\eqref{eq:src_vector} are identically zero, the total mixture density $\rho = \sum_s \rho_s$, the momentum components $(\rho u, \rho v, \rho w)$, and the total energy $E$ remain invariant under this step and are treated as fixed parameters; only the reduced state $\mathbf{Y}$ is evolved.
\begin{equation}
    \mathbf{R}(\mathbf{Y}^{\star})
    \;=\;
    \mathbf{Y}^{\star} - \mathbf{Y}^{n} - \tfrac{\Delta t}{2}\,\mathbf{S}_{\!Y}(\mathbf{Y}^{\star}),
    \qquad
    \mathbf{S}_{\!Y}
    \;=\;
    \bigl(\dot\omega_1,\,\dots,\,\dot\omega_{N_s},\,
          Q_{T\text{-}v} + \textstyle\sum_{s\in\mathcal{M}}\dot\omega_s\,e_{v,s}(T_v)\bigr).
    \label{eq:source_residual}
\end{equation}
One Newton sweep consists of four operations executed by a single fused
device function: (i)~temperature recovery, where an inner Newton solves
the caloric equation of state with NASA-9 polynomial $c_v(T)$ to give
$(T, T_v, p)$ from $\mathbf{Y}^{\star}$ and the fixed momentum and total
energy; (ii)~rate evaluation, where $k_f^r(T_a^r)$,
$\mathcal{R}_r$, $\dot\omega_s$, and $Q_{T\text{-}v}$ are computed from
\eqref{eq:src_reaction_rate}--\eqref{eq:src_landau_teller};
(iii)~Jacobian assembly, where
$\mathbf{J}=\partial\mathbf{R}/\partial\mathbf{Y}^{\star}\in\mathbb{R}^{(N_s+1)\times(N_s+1)}$
is built column by column from forward (one-sided) finite differences in $\mathbf{Y}^{\star}$ with relative perturbation $\varepsilon = \max(10^{-7}|Y_k^{\star}|,\,10^{-12})$
-- analytical Jacobians of the NASA-9 thermodynamics
and Park rates are avoided to keep the kernel general across mechanisms, and the one-sided rule halves the number of full chemistry+caloric rate evaluations per Newton step from $2(N_s+1)$ to $(N_s+1)$; and
(iv)~the linear update
$\mathbf{Y}^{\star}\!\leftarrow\!\mathbf{Y}^{\star} - \mathbf{J}^{-1}\mathbf{R}$
is taken with Gauss--Jordan elimination directly in thread-local
registers. The state of one node never leaves registers between rate
evaluations, so the only global-memory traffic per Newton step is one
read and one write of the nine-variable nodal state.

\subsubsection{GPU-resident Adaptive Mesh Refinement}
\label{sec:amr_kernels}

A defining feature of \texttt{MARUT} is that the adaptive mesh refinement
cycle runs entirely on the GPU. Because AMR is invoked every few
time steps, a host round-trip would dominate the wall time;
\texttt{MARUT} therefore keeps the forest, (\texttt{GPUForest}), the connectivity, and
the projection operators on the device and treats the AMR cycle as
one further sequence of GPU kernels alongside the hyperbolic and
parabolic right-hand sides. The cycle is built in five steps,
listed in Table~\ref{tab:amr_kernels}.

\begin{table}[htbp]
\centering
\small
\caption{Steps of the GPU-resident AMR cycle. Each step is a small
group of GPU kernels; the only routine GPU-to-host traffic across
the whole cycle is the per-boundary index download in step~5.}
\label{tab:amr_kernels}
\begin{tabular}{c l l}
\hline
Step & Operation & Output \\
\hline
1 & Indicator and flagging                                           & per-element flag $\in \{-1, 0, +1\}$ \\
2 & 2:1 level-balance enforcement                                    & balanced flag field \\
3 & Solution transfer ($\mathbb{L}_{2}$ projection, Jacobian-aware)           & new solution buffer \\
4 & Geometry recompute (GCL-preserving contravariant basis)          & new metric $J,\;\mathbf{Ja}^{\,l}$ \\
5 & Connectivity, boundary matching and buffer resize                & new interface/mortar/BC arrays \\
\hline
\end{tabular}
\end{table}

\paragraph{Forest Data Structure:}
The adapted mesh is represented by \texttt{GPUForest}, a custom GPU-resident forest-of-trees whose fields are all GPU arrays. For every active leaf $e$ of the
octree we store a sorted triple
\begin{equation}
    \bigl(\,T_e,\;m_e,\;\ell_e\,\bigr),
    \label{eq:forest_keys}
\end{equation}
where $T_e$ is the coarse-tree identifier, $m_e$ the local Morton
code~\cite{morton1966computer}, and $\ell_e$ the refinement level.
The coarse mesh contributes the fixed tree-to-tree and
tree-to-face adjacency tables of \texttt{p4est}~\cite{burstedde2011p4est}
together with the corner-coordinate array of every coarse element.
Connectivity arrays (interface, mortar and boundary index lists)
are rebuilt from scratch at every cycle, sized to the freshly
adapted forest. A pool of reusable scratch buffers (refine flags,
new keys, prefix-scan buffers, indicator buffers) holds the
intermediate state without any further allocations.

The five steps of Table~\ref{tab:amr_kernels} execute as a unified, device-resident pipeline, minimizing host-device synchronization to a single index synchronization phase at the conclusion of the cycle. 
The pipeline begins with a per-element indicator kernel that evaluates local solution gradients, typically based on key thermodynamic or kinematic quantitie, to classify elements for refinement, retention, or coarsening. Following this, a parallel balance kernel enforces the standard 2:1 octree connectivity constraint across element faces, iteratively resolving refinement levels across neighboring trees via device-optimized adjacency lookups. Once the new mesh topology is established, a parallel scan and allocation sequence configures the updated forest structure. Element data is then transferred using a conservative, Jacobian-weighted $L_2$ projection operator designed to maintain conservation properties on curved geometries. The geometric metrics and contravariant bases for the newly formed elements are subsequently recomputed using conservative discretization forms that satisfy the discrete GCL, ensuring consistency with the underlying entropy-stable volume operators. Finally, the cycle concludes with a parallel connectivity-rebuild phase. This kernel evaluates element faces to dynamically reconstruct interior interfaces, non-conforming mortars, and physical domain boundaries. After updating the relevant boundary indexing arrays, scratch memory is dynamically adjusted to the new element count, and control is returned to the time integration loop.

\section{Results}
In this study, the \texttt{MARUT} solver is tested on a range of two- and three-dimensional benchmark problems for both the Euler and Navier--Stokes equations. The considered cases span multiple flow regimes, including subsonic, transonic, supersonic, and hypersonic conditions, thereby demonstrating the robustness and versatility of the framework across a wide spectrum of compressible flow physics. In addition, we present performance and scalability results for both single- and multi-GPU configurations. These results highlight the efficiency of the parallel implementation and its ability to maintain strong scaling across increasing computational resources. 

Among all test cases, the mesh for the two-dimensional cylinder case was adopted from the \texttt{Trixi.jl} \cite{schlottkelakemper2025trixi} example suite, although all simulations were performed exclusively within \texttt{MARUT}. For the other test cases, including the Taylor–Green vortex, ONERA M6 wing, non-equilibrium blast wave, and RAE 2822 airfoil cases, the meshes were generated using \texttt{Gmsh} \cite{geuzaine2009gmsh}.

\subsection{Supersonic Inviscid Flow over a Two-Dimensional Cylinder}
\label{sec:cylinder}
As a first validation case we consider transient supersonic inviscid flow past a
circular cylinder confined in a channel. The configuration combines a detached
bow shock, an expansion around the cylinder shoulders, and an unsteady wake, and
therefore exercises the shock capturing, non-conforming mortar, and
adaptive-refinement machinery of \texttt{MARUT} simultaneously. The mesh is the unstructured quadrilateral grid with the cylinder centerd in a rectangular
channel and five labeled boundaries: \texttt{Left}, \texttt{Right},
\texttt{Top}, \texttt{Bottom}, and \texttt{Circle}. We solve the two-dimensional compressible Euler equations for an ideal gas
with ratio of specific heats $\gamma = 1.4$, closed by the equation of state
$p = (\gamma-1)\bigl(\rho e - \tfrac{1}{2}\rho\|\mathbf{u}\|^2\bigr)$. The domain is initialised with the uniform free-stream state
\begin{equation}
    \rho_\infty = 1.4, \quad u_\infty = 3.0, \quad v_\infty = 0, \quad p_\infty = 1.0,
    \label{eq:cyl_freestream}
\end{equation}
corresponding to a free-stream Mach number $M_\infty = u_\infty/\sqrt{\gamma p_\infty/\rho_\infty} = 3$.
The \texttt{Left} boundary is a supersonic inflow at which the exterior state is
fixed to~\eqref{eq:cyl_freestream} and the flux is evaluated directly from that
state; the \texttt{Right} boundary is a supersonic outflow that uses the interior
state; and the \texttt{Top}, \texttt{Bottom}, and \texttt{Circle} boundaries are
treated as inviscid slip walls. All boundary conditions are enforced weakly
through the numerical surface flux.

The solution is discretised with a nodal DG-SEM operator on a Lobatto--Legendre
basis of polynomial degree $\mathcal{P} = 3$. The volume integral uses the entropy-stable
split form of Ranocha flux, while a local Lax--Friedrichs flux provides the inter-element and surface dissipation. Shocks are captured by the
Hennemann--Gassner indicator applied to the density--pressure product
$\rho\,p$. Time integration uses the five-stage, fourth-order SSPRK54 and
a per-stage positivity floor enforces $\rho \ge 10^{-6}$ and $p \ge 5\times 10^{-7}$. AMR is driven by a L\"ohner-type second-derivative indicator
evaluated on the density field. A three-level controller is used with base
level $\ell_{\text{base}} = 0$, an intermediate level $\ell_{\text{med}} = 3$
activated at an indicator threshold of $0.02$, and a maximum level
$\ell_{\max} = 5$ activated at $0.05$. The AMR step is invoked every
$50$ time steps. The initial condition is pre-adapted with up to five AMR
sweeps at $t = 0$ so that the bow shock is already resolved at the correct
level before time integration begins.

\begin{figure}[htpb]
    \centering
    \includegraphics[scale=1.2,clip=true]{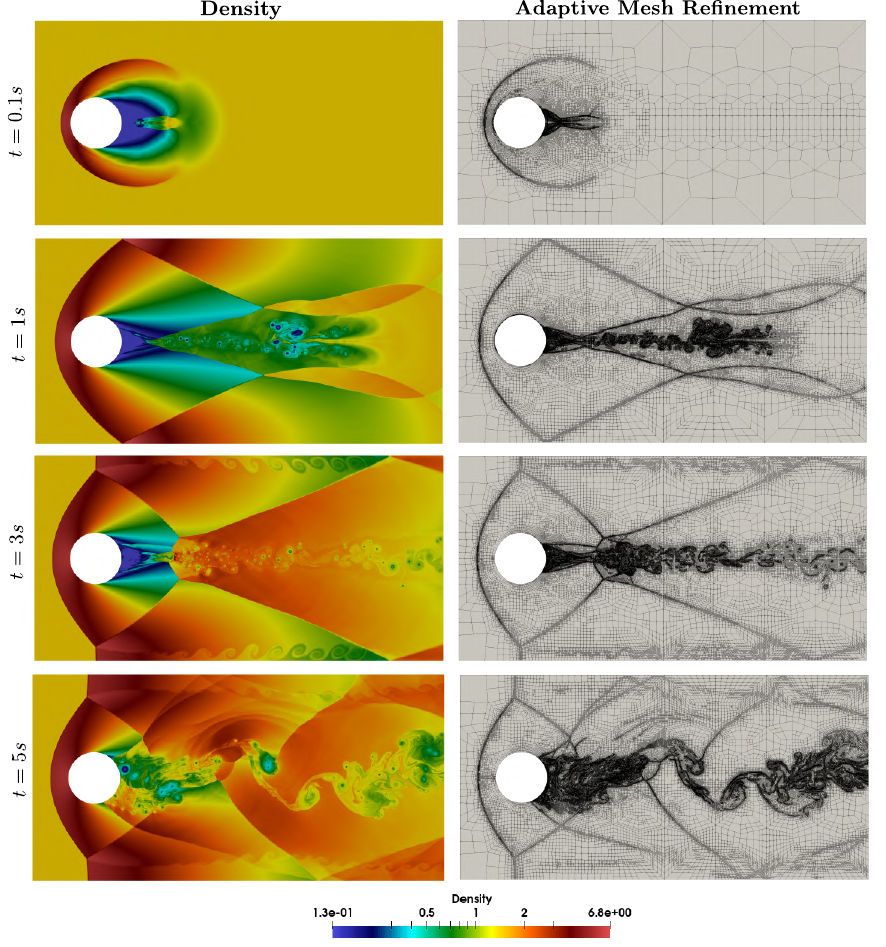}
    \caption{Mach~3 supersonic Euler flow past a cylinder in a
    channel: density field (left column) and AMR mesh (right column)
    at $t = 0.1,\,1,\,3,$ and $5$\,s. Final adapted mesh:
    $147{,}801$ elements ($\approx 9.46\times 10^{6}$ DOFs), with
    two-thirds at the maximum level $\ell_{\max} = 5$; density
    colour scale shared across all panels.}
    \label{fig:cyl_density}
\end{figure}
The simulation is advanced to $t_{\text{end}} = 5$ time units on a single
NVIDIA L40S GPU and completes in approximately $2$ hours of wall-clock
time. By the final time the adapted mesh contains $147{,}801$ active
elements, of which $\approx 66.5\%$ reside at the maximum refinement
level $\ell_{\max} = 5$ and a further $\approx 26.5\%$ at $\ell = 4$;
with $(\mathcal{P}+1)^2 = 16$ Lobatto--Legendre nodes per element this corresponds
to $\approx 2.36\times 10^{6}$ grid points and
$\approx 9.46\times 10^{6}$ conserved-variable degrees of freedom.
Figure~\ref{fig:cyl_density} reports the evolution of the density
field together with the corresponding adapted mesh at four
representative instants $t = 0.1,\,1,\,3,$ and $5$ s, with the left
column showing the density contours and the right column shows
the GPU-resident AMR element grid. By $t = 1$\,s the
detached bow shock is formed upstream of the cylinder, with
characteristic expansion fans emanating from the shoulders and an
early recirculation pocket in the near wake. Between $t = 3$\,s and
$t = 5$\,s the wake develops into a fully unsteady vortex-shedding
regime, with shed structures convecting downstream and producing the
broad-band density signature visible in the bottom-left panel. The
right column shows that the GPU-resident AMR tracks
each of these features as they emerge.
The same refinement pattern visibly migrates downstream as the wake
sheds and reorganises between $t = 3$ and $t = 5$\,s, demonstrating
that the indicator and the $\mathbb{L}_2$ projection reorganise the active
element set every fifty steps without disturbing the resolved flow. The simulation video is available \href{https://www.youtube.com/watch?v=JO8kq3bqk8s}{Supersonic Flow Over a 2D Cylinder}.


\subsection{Three-Dimensional Viscous Compressible Taylor–Green Vortex}
\label{sec:tgv}
As a three-dimensional viscous validation case we simulate the classical
Taylor--Green vortex (TGV) at Reynolds number $\mathrm{Re} = 1600$. The
flow is initially laminar, develops small scales through non-linear
vortex stretching, and eventually decays to an approximately isotropic
turbulent state. Because the exact initial condition is smooth and
periodic, any spurious dissipation produced by the numerical scheme
appears directly in the kinetic-energy and dissipation histories, making
the TGV a standard probe of dispersion, dissipation, and
shock-free AMR behaviour for high-order methods. We solve the three-dimensional compressible Navier--Stokes equations for
a Newtonian, calorically perfect gas with $\gamma = 1.4$, dynamic
viscosity $\mu = 6.25\times 10^{-4}$, and Prandtl number $\mathrm{Pr} = 0.72$.
With the reference velocity, density, and length scales implied by the
initial condition below, these parameters correspond to
$\mathrm{Re} = 1/\mu = 1600$. The viscous stress tensor and heat flux
are taken in standard form, and the pressure closure is
$p = (\gamma-1)\bigl(E - \tfrac{1}{2}\rho\|\mathbf{u}\|^2\bigr)$.
Viscous terms are discretised with a BR1 mixed formulation using the
primitive variables as the gradient unknowns. The computational domain is the triply-periodic cube
$\Omega = [-\pi, \pi]^3$, with periodic boundary conditions imposed on
all six faces through the interior-interface treatment of the DG
operator. No physical walls are present.
The solution is initialized using
$A = 1$ and reference the Mach number $M_s$:
\begin{equation}
\begin{aligned}
    \rho(\mathbf{x},0) &= 1, \\
    u(\mathbf{x},0) &=  A\,\sin x_1 \cos x_2 \cos x_3, \\
    v(\mathbf{x},0) &= -A\,\cos x_1 \sin x_2 \cos x_3, \\
    w(\mathbf{x},0) &= 0, \\
    p(\mathbf{x},0)   &= \frac{A^2\,\rho}{\gamma\,M_s^2}
        + \frac{A^2\,\rho}{16}\!\left[\cos 2x_1 \cos 2x_3 + 2\cos 2x_2
        + 2\cos 2x_1 + \cos 2x_2 \cos 2x_3\right].
\end{aligned}
\label{eq:tgv_ic}
\end{equation}
The additive pressure term is the hydrostatic correction consistent
with the divergence-free velocity field in the incompressible limit.

The governing equations are advanced with a nodal DG-SEM operator on a
Lobatto--Legendre basis of polynomial degree $\mathcal{P} = 6$. The inviscid
volume integral is computed in entropy-stable split form using the
Ranocha two-point flux, and the inter-element surface coupling uses the
local Lax--Friedrichs flux. Time integration uses
the SSPRK54, and a per-stage positivity floor
enforces $\rho, p \ge 10^{-6}$. The simulation is advanced to
$t_{\text{end}} = 20$ time units. Three mesh configurations are considered, all using the same DG-SEM
polynomial degree $\mathcal{P} = 6$ so that each element carries $(\mathcal{P}+1)^3 = 343$
nodal degrees of freedom per conserved variable:
\begin{itemize}
    \item \textbf{Case~1} — $12\times 12\times 12$ uniform Cartesian mesh,
          no AMR, yielding $\approx 0.59$\,Million grid points.
    \item \textbf{Case~2} — $24\times 24\times 24$ uniform Cartesian mesh,
          no AMR, yielding $\approx 4.74$\,Million grid points.
    \item \textbf{Case~3} — $8\times 8\times 8$ base mesh with GPU-resident
          AMR up to level $\ell_{\max} = 3$, reaching
          $\approx 170,000$ active elements and
          $\approx 58.31$\,Million grid points.
\end{itemize}

In Case~3, refinement is driven by a L\"ohner-type second-derivative
indicator evaluated on the local velocity magnitude
$|\mathbf{u}| = \sqrt{u^2 + v^2 + w^2}$, which tracks the
vortex-stretching regions and the small-scale structures that emerge
during transition. A three-level controller is used with base level
$\ell_{\text{base}} = 0$, an intermediate level
$\ell_{\text{med}} = 1$ activated at an indicator threshold of $0.1$, and
a maximum level $\ell_{\max} = 3$ activated at $0.2$. The AMR step is
invoked every $5$ time steps so that the adapted mesh tracks the rapidly
evolving small scales during the transition phase. Two reference
Mach numbers are considered to assess the solver in both nearly
incompressible and clearly compressible regimes: a subsonic
configuration with $M_s = 0.1$ and a supersonic configuration with
$M_s = 1.25$. The geometry, polynomial degree, AMR controller and
time-stepping method are identical between the two regimes; only
the reference Mach number $M_s$ in the initial-condition pressure
field~\eqref{eq:tgv_ic} changes.

\subsubsection{Subsonic regime ($M_s = 0.1$):}
The vortex topology is visualised through iso-surfaces of the second
invariant of the velocity-gradient tensor,
\begin{equation}
    Q \;=\; \tfrac{1}{2}\bigl(\Omega_{ij}\Omega_{ij} \;-\; S_{ij}S_{ij}\bigr),
    \qquad
    S_{ij} = \tfrac{1}{2}(\partial_j u_i + \partial_i u_j),
    \quad
    \Omega_{ij} = \tfrac{1}{2}(\partial_j u_i - \partial_i u_j),
    \label{eq:q_criterion}
\end{equation}
where $S_{ij}$ and $\Omega_{ij}$ are the symmetric and
antisymmetric parts of $\partial_j u_i$, and $u_i = (u,v,w)$. Positive iso-surfaces of
$Q$ enclose regions in which rotation dominates strain, and are a
standard descriptor of post-transition coherent vortex structures.
\begin{figure}[htpb]
    \centering
\includegraphics[scale=0.9,clip=true]{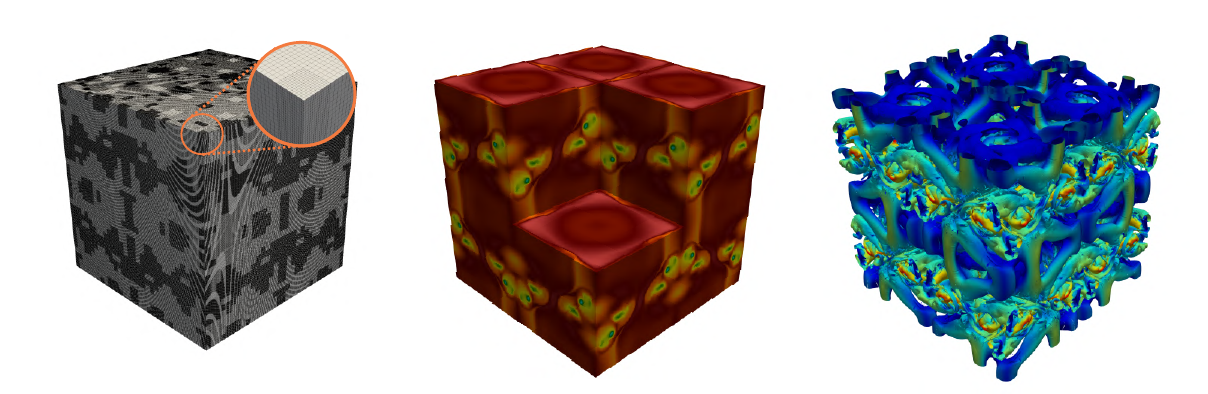}
    \caption{Subsonic TGV (Case~3, $M_s = 0.1$, $\mathrm{Re} = 1600$)
    at $t = 10$, $\approx 120{,}000$ active elements: AMR mesh
    (left), density slice (middle), and iso-surface of $Q = 0.01$
    coloured by velocity magnitude (right).}
    \label{fig:tgv_sub}
\end{figure}

Figure~\ref{fig:tgv_sub} reports the adapted mesh, a density slice
through the periodic cube, and an iso-surface of $Q = 0.01$ for the
subsonic case at $t = 10$. The velocity-magnitude indicator
concentrates the highest refinement level on the vortex-stretching
cores and on the freshly nucleated small-scale structures of the
transition phase; the unrefined elements visible in the corners of
the mesh panel correspond to regions in which the laminar
large-scale field has already broken down into low-amplitude
broadband motion that the indicator does not flag. The density
slice is essentially uniform at the imposed reference value, as
expected for $M_s = 0.1$, and the $Q = 0.01$ iso-surface confirms
the expected post-transition structure of densely packed,
finite-thickness vortex tubes filling the periodic cube.

\begin{figure}
    \centering
    \includegraphics[scale=0.48,clip=true, trim=0.15cm 0.2cm 0.35cm 0cm]{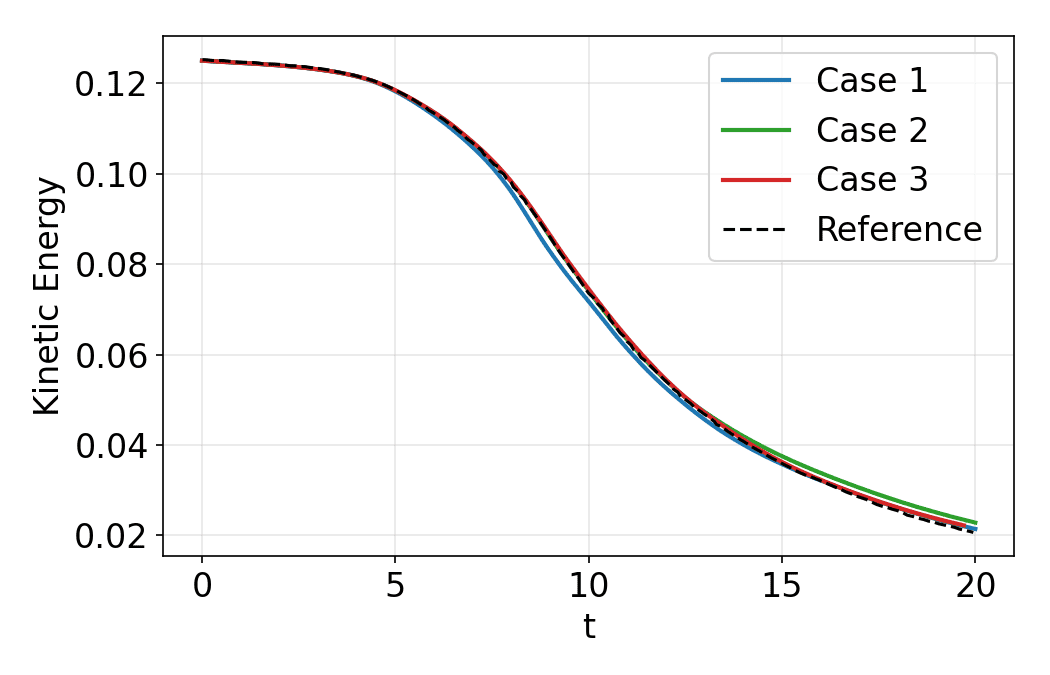}
    \includegraphics[scale=0.48,clip=true, trim=0.35cm 0.2cm 0.15cm 0cm]{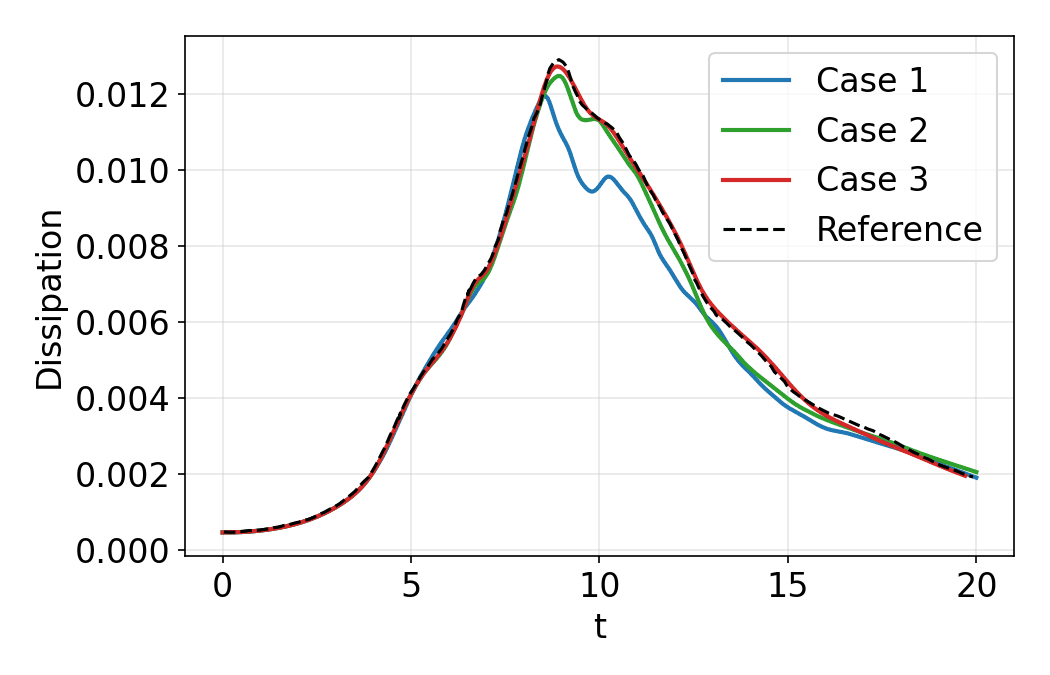}
    \caption{Subsonic TGV: kinetic energy $E_k(t)$ (left) and
    dissipation rate $\varepsilon(t) = -dE_k/dt$ (right) for the
    three grids of Sec.~\ref{sec:tgv} (Case~1: $12^3$, Case~2: $24^3$, Case~3:
    $8^3$ with AMR up to $\ell_{\max} = 3$), against pseudo-spectral DNS reference data from van Rees et al.~\cite{vanrees2011comparison}.}
    \label{fig:sub_ke}
\end{figure}
The numerical validation is performed using the volume-averaged
kinetic energy and its dissipation rate, defined respectively as
\begin{equation}
    E_k(t) \;=\; \frac{1}{2\,\rho_0\,U_0^2\,|\Omega|}\!\int_{\Omega} \rho\,\mathbf{u}\!\cdot\!\mathbf{u}\,d\Omega,
    \qquad
    \varepsilon(t) \;=\; -\,\frac{dE_k}{dt},
    \label{eq:tgv_ke_eps}
\end{equation}
where $|\Omega| = (2\pi)^3$ is the volume of the periodic domain,
$\rho_0$ is the reference density and $U_0 = M\,c_0$ is the reference
velocity built from the reference Mach number $M$ and the reference
sound speed $c_0 = \sqrt{\gamma\,p_0/\rho_0}$. The non-dimensional
normalisation makes $E_k$ independent of the absolute pressure and
velocity scales, so that the subsonic ($M = 0.1$) and supersonic
($M = 1.25$) histories can be compared against the same reference
DNS curves.
Figure~\ref{fig:sub_ke} plots $E_k(t)$ and $\varepsilon(t)$ for the
three grid cases against the established
reference data curve of \cite{vanrees2011comparison}. All three GPU runs follow the reference $E_k$ curve essentially perfectly through the
laminar phase, and the dissipation peak around $t \approx 9$ is
recovered to within graphical accuracy by both the fully resolved
$24^3$ uniform mesh (Case~2) and the AMR run starting from $8^3$
(Case~3); the under-resolved $12^3$ uniform mesh (Case~1) exhibits
the expected damped peak with a slight phase delay. The agreement
between Cases~2 and~3 is the central validation result: the AMR
machinery reproduces the dissipation history of the fully resolved
Cartesian mesh while deploying refined elements only where the
indicator detects them, and the on-device $\mathbb{L}_2$ projection
introduces no measurable spurious dissipation across the hundreds
of AMR cycles required to reach $t_{\text{end}} = 20$.

\subsubsection{Supersonic regime ($M_s = 1.25$):}
The same geometric and numerical setup is now run with the
reference Mach number raised to $M_s = 1.25$, which generates
finite-amplitude density and pressure fluctuations and weak
shocklets that the subsonic limit suppresses.
\begin{figure}
    \centering
\includegraphics[scale=0.9,clip=true]{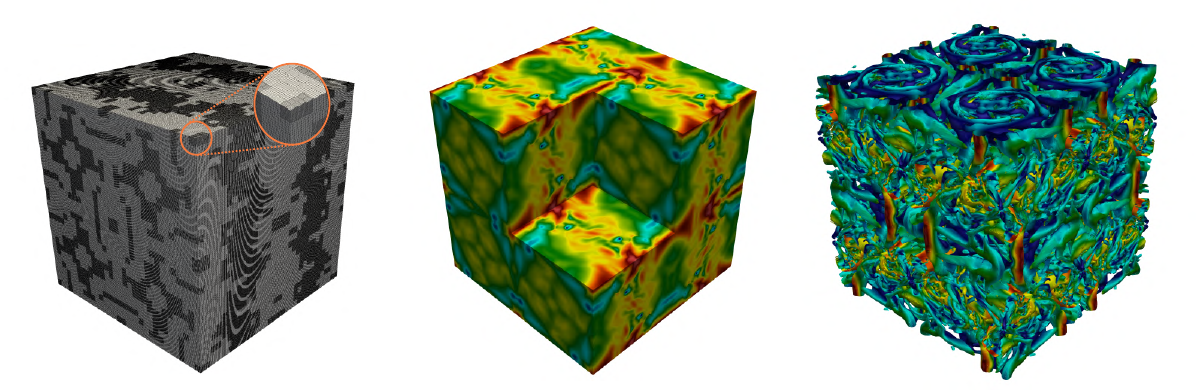}
    \caption{Supersonic TGV ($M_s = 1.25$, $\mathrm{Re} = 1600$) at
    $t = 11$, approximately $ 140{,}000$ active elements: same panel
    layout as Fig.~\ref{fig:tgv_sub}. Weak shocklets are visible
    in the density slice, and the $Q$-iso-surface shows finer
    vortex tubes than the subsonic case.}
    \label{fig:tgv_sup}
\end{figure}
Figure~\ref{fig:tgv_sup} reports the same three diagnostics
(adapted mesh, density slice, and iso-surface of $Q = 0.01$ as
defined in~\eqref{eq:q_criterion}) for the supersonic case at
$t = 11$. The AMR mesh has converged to $\approx 140{,}000$ active
elements, slightly more than in the subsonic case at a comparable
time, reflecting the additional small scales generated by
compressibility. The density slice now exhibits clearly visible
weak shocklets and finite-amplitude density gradients that are
absent at $M_s = 0.1$, but the same indicator and $\mathbb{L}_2$ projection handles both regimes without modification. The
$Q$-iso-surface shows finer and more numerous vortex tubes than
the subsonic case at the same nominal time, consistent with the
stronger non-linear cascade triggered by the compressible modes.

For the supersonic regime the total dissipation rate is further
split into its rotational (solenoidal) and acoustic (dilatational)
contributions, defined as in \cite{peyvan2025h3pc} by
\begin{equation}
    \varepsilon_s(t) \;=\; \frac{L^2}{\mathrm{Re}\,U_0^2\,|\Omega|}\!\int_{\Omega} \frac{\mu(T)}{\mu_0}\,\boldsymbol{\omega}\!\cdot\!\boldsymbol{\omega}\,d\Omega,
    \qquad
    \varepsilon_d(t) \;=\; \frac{L^2}{3\,\mathrm{Re}\,U_0^2\,|\Omega|}\!\int_{\Omega} \frac{\mu(T)}{\mu_0}\,(\nabla\!\cdot\!\mathbf{u})^2\,d\Omega,
    \label{eq:eps_split}
\end{equation}
where $L = 2\pi$ is the box side, $\mathrm{Re} = \rho_0\,U_0\,L/\mu_0$
is the reference Reynolds number, $\mu_0$ the reference viscosity,
and $\boldsymbol{\omega} = \nabla\!\times\!\mathbf{u}$ the vorticity
field. The local viscosity $\mu(T)$ varies in space through
Sutherland's law in the supersonic configuration (so the ratio
$\mu(T)/\mu_0$ stays inside the integral), while in the subsonic
configuration $\mu(T) = \mu_0$ and the ratio reduces to unity. In
the incompressible limit $\nabla\!\cdot\!\mathbf{u} \to 0$ the
dilatational contribution vanishes and the total dissipation
reduces to its solenoidal component; finite values of
$\varepsilon_d$ therefore quantify the acoustic content of the
dissipation budget and provide a stringent test of the
entropy-stable split-form volume operator and the BR1 viscous
treatment in the compressible regime.

\begin{figure}[htpb]
    \centering
    \includegraphics[scale=0.45,clip=true, trim=0.15cm 0.2cm 0.35cm 0cm]{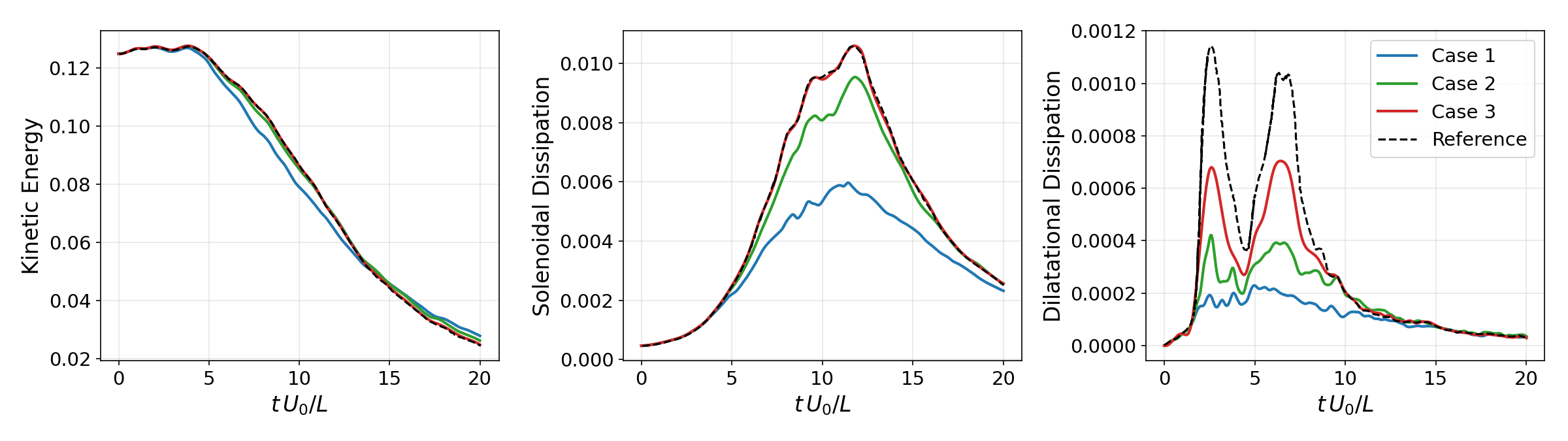}
    \caption{Supersonic TGV: $E_k(t)$ (left), solenoidal
    dissipation $\varepsilon_s(t)$ (middle), and dilatational
    dissipation $\varepsilon_d(t)$ (right) for the same three grids
    as in Fig.~\ref{fig:sub_ke}, against TENO reference data from Chapelier et al.~\cite{chapelier2024comparison}.}
    \label{fig:sup_ke}
\end{figure}

Figure~\ref{fig:sup_ke} plots $E_k(t)$, $\varepsilon_s(t)$ and
$\varepsilon_d(t)$ as defined by \eqref{eq:tgv_ke_eps} and
\eqref{eq:eps_split} against the corresponding reference data
curves. The kinetic-energy decay is captured by all three
configurations and is essentially indistinguishable between
Case~2 and Case~3, as in the subsonic regime. The solenoidal
dissipation closely matches the reference across the three cases,
showing that the rotational cascade is correctly resolved by both
the fully resolved Cartesian mesh and the AMR run. The
dilatational component captures the additional acoustic
dissipation arising from compressibility at $M_s = 1.25$; the AMR
Case~3 and the fully resolved Case~2 both track the reference
curve through the peak and into the decay phase, while the
under-resolved Case~1 systematically under-predicts
$\varepsilon_d$. Taken together, the agreement of both
$\varepsilon_s$ and $\varepsilon_d$ with the reference confirms
that the entropy-stable split-form volume operator and the BR1
viscous treatment carry over correctly to the compressible
regime, and that the GPU-resident AMR cycle preserves these
dissipation budgets across the hundreds of refinement and
coarsening passes required to reach $t_{\text{end}} = 20$.

\subsection{Transonic Viscous Flow over ONERA M6 Wing}
As an external-aerodynamics validation of the three-dimensional
viscous solver we simulate the canonical transonic flow over the
ONERA M6 swept wing at the wind-tunnel conditions~\cite{schmitt1979pressure}. The configuration combines a
detached supercritical region on the suction side, a
$\lambda$-shock structure across the span, and a tip-vortex roll-up;
together these features exercise the curvilinear hexahedral
geometry, the wall boundary-condition stack, and the combined
hyperbolic--parabolic GPU pipeline of \texttt{MARUT} on a realistic complex
geometry. The mesh carries $217{,}088$ hex elements with five
labelled boundaries: \texttt{Symmetry} (the root plane $z = 0$),
\texttt{FarField} (the upstream, top, and outflow faces of the
domain box), \texttt{TopWing} and \texttt{BottomWing} (the suction
and pressure surfaces of the wing). The wing
itself has root chord $c_{\text{root}} = 0.6837$, tip chord
$c_{\text{tip}} = 0.385$, half-span $b/2 = 1.0418$, leading-edge
sweep of $30^\circ$; the computational
domain extends to $x \in [-6.4, 7.4]$, $y \in [-6.4, 6.4]$, and
$z \in [0, 7.4]$.

We solve the three-dimensional compressible Navier--Stokes
equations for a Newtonian, perfect gas with
$\gamma = 1.4$, Prandtl number $\mathrm{Pr} = 0.72$, and constant
dynamic viscosity $\mu = \rho_\infty\,u_\infty\,L_{\text{ref}} / \mathrm{Re} = 1.003\times 10^{-7}$
obtained from the reference state below with $L_{\text{ref}} = 1$.
The pressure closure is
$p = (\gamma-1)\bigl(E - \tfrac{1}{2}\rho\|\mathbf{u}\|^2\bigr)$
and the viscous fluxes are discretised with the same BR1 mixed
formulation used for the Taylor--Green vortex. The domain is initialized uniformly at the free-stream state,
\begin{equation}
    \rho_\infty = 1.4, \quad u_\infty = 0.84, \quad p_\infty = 1.0,
    \quad
    \alpha = 3.06^\circ,
    \label{eq:onera_freestream}
\end{equation}
corresponding to a free-stream Mach number
$M_\infty = u_\infty/\sqrt{\gamma\,p_\infty/\rho_\infty} = 0.84$ and
a chord Reynolds number based on $L_{\text{ref}} = 1$ of
$\mathrm{Re} = 1.172\times 10^{7}$, matching the
test condition. The free-stream velocity is rotated into the
geometric frame by the angle of attack $\alpha$ as
$u_{\infty} = u_\infty\cos\alpha,$
$v_{\infty} = u_\infty\sin\alpha,$
$w_{\infty} = 0$, so that positive $\alpha$ tilts the inflow
toward the suction surface and produces positive lift. The
hyperbolic and parabolic boundary conditions are applied
separately, in the standard split-form treatment of the
compressible Navier--Stokes equations: the \texttt{Symmetry} face
is an inviscid slip wall for the Euler part and a
slip-plus-adiabatic wall for the viscous part; the \texttt{FarField}
face is a Dirichlet boundary fixed to~\eqref{eq:onera_freestream}
in both parts; and the \texttt{TopWing} and \texttt{BottomWing}
surfaces are inviscid slip walls for the Euler flux and no-slip
adiabatic walls for the viscous flux, so that the inviscid surface
flux sees only the wall pressure and the viscous gradient operator
enforces no-slip and zero heat flux. All boundary conditions are
imposed weakly through the numerical surface flux.

The solution is discretised with a nodal DG-SEM operator on a
Lobatto--Legendre basis of polynomial degree $\mathcal{P} = 3$, giving
$(\mathcal{P}+1)^3 = 64$ nodes per element and a total of
$\approx 13.89\times 10^{6}$ grid points
on the static mesh. The inviscid volume integral is computed in
entropy-stable split form using the Ranocha two-point flux, and the
inter-element surface coupling uses the local Lax--Friedrichs flux.
Time integration uses the five-stage, fourth-order SSPRK54 with a
fixed CFL number of $0.5$ and a combined hyperbolic/parabolic
time-step estimator; a per-stage positivity floor enforces
$\rho, p \ge 10^{-6}$. AMR is disabled and the simulation runs on
the static mesh throughout, so that the present case isolates the
performance and accuracy of the curvilinear viscous operator from
the AMR. 
The full run is performed on a single NVIDIA H100/H200 GPU.

\begin{figure}[htpb]
    \centering
    \includegraphics[scale=0.82,clip=true, trim=0.15cm 0.2cm 0.35cm 0cm]{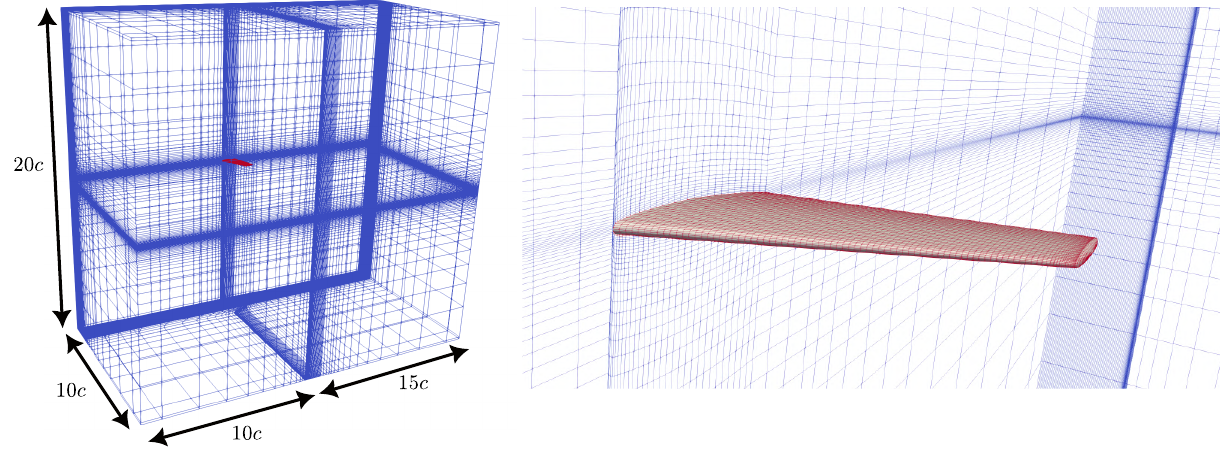}
    \caption{(Left) Three-dimensional computational domain bounding the
    ONERA M6 wing, with extents expressed in units of the root chord $c$:
    $10c$ upstream of the leading edge, $15c$ downstream of the trailing
    edge, $10c$ in the cross-stream (spanwise) direction, and $20c$ in the
    vertical direction; the wing is shown in red at the centre of the box.
    (Right) Surface mesh of the wing geometry.}
    \label{fig:onera_msh}
\end{figure}

\begin{figure}[htpb]
    \centering
    \includegraphics[width=1.0\textwidth]{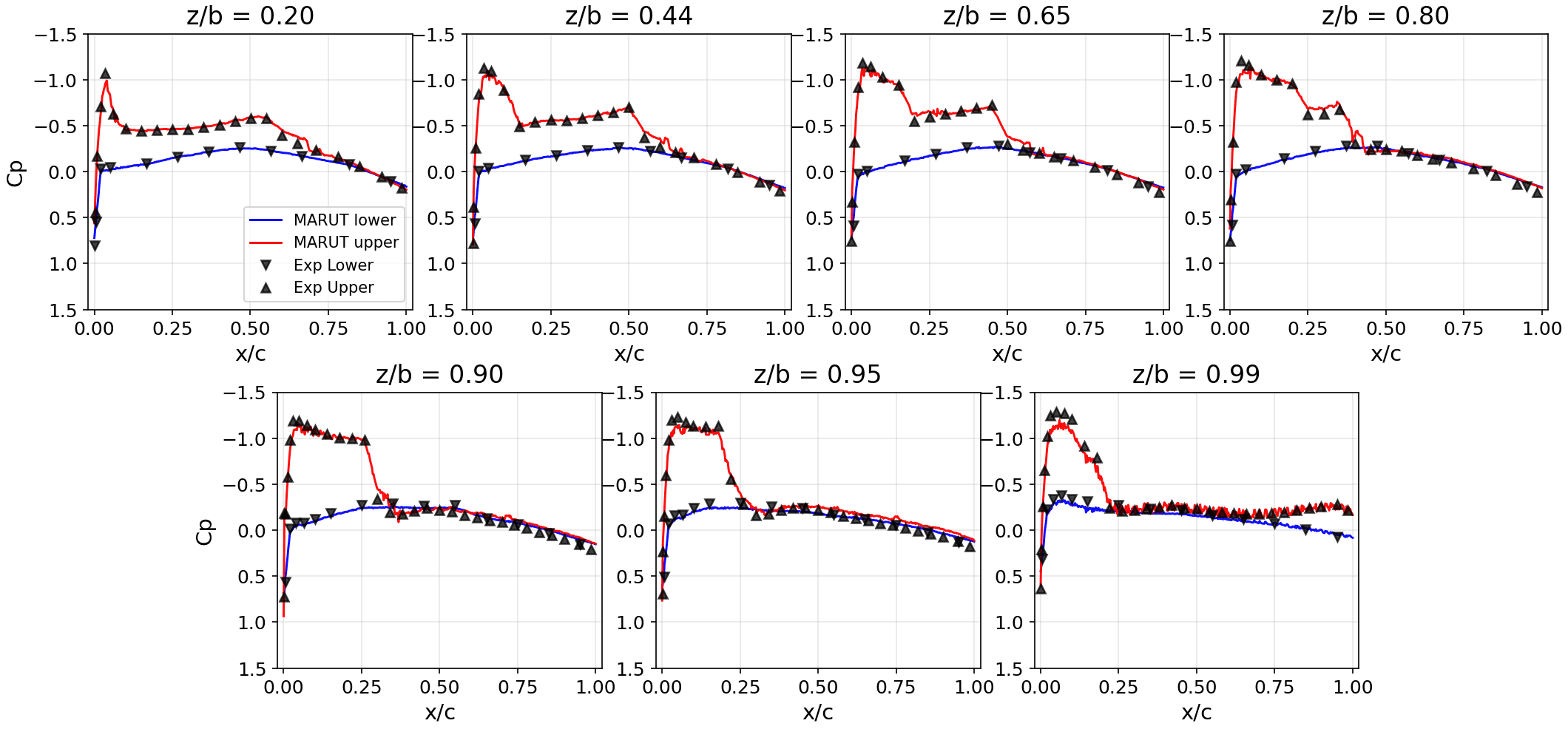}
    \caption{Surface pressure coefficient $C_p$ distributions on the ONERA M6 wing at $M_\infty = 0.84$ for the seven spanwise stations $(z/b) \in \{0.20,\,0.44,\,0.65,\,0.80,\,0.90,\,0.95,\,0.99\}$.
    Solid lines: \texttt{MARUT} solution at $t=9.0$ convective time units
    on the static $217{,}088$-element $\mathcal{P}=3$ mesh; black markers:
    Schmitt and Charpin wind-tunnel data~\cite{schmitt1979pressure}
    (downward triangles: lower/pressure surface, upward triangles:
    upper/suction surface).}
    \label{fig:onera_cp}
\end{figure}

\begin{figure}[htbp]
    \centering
    \includegraphics[width=0.85\textwidth]{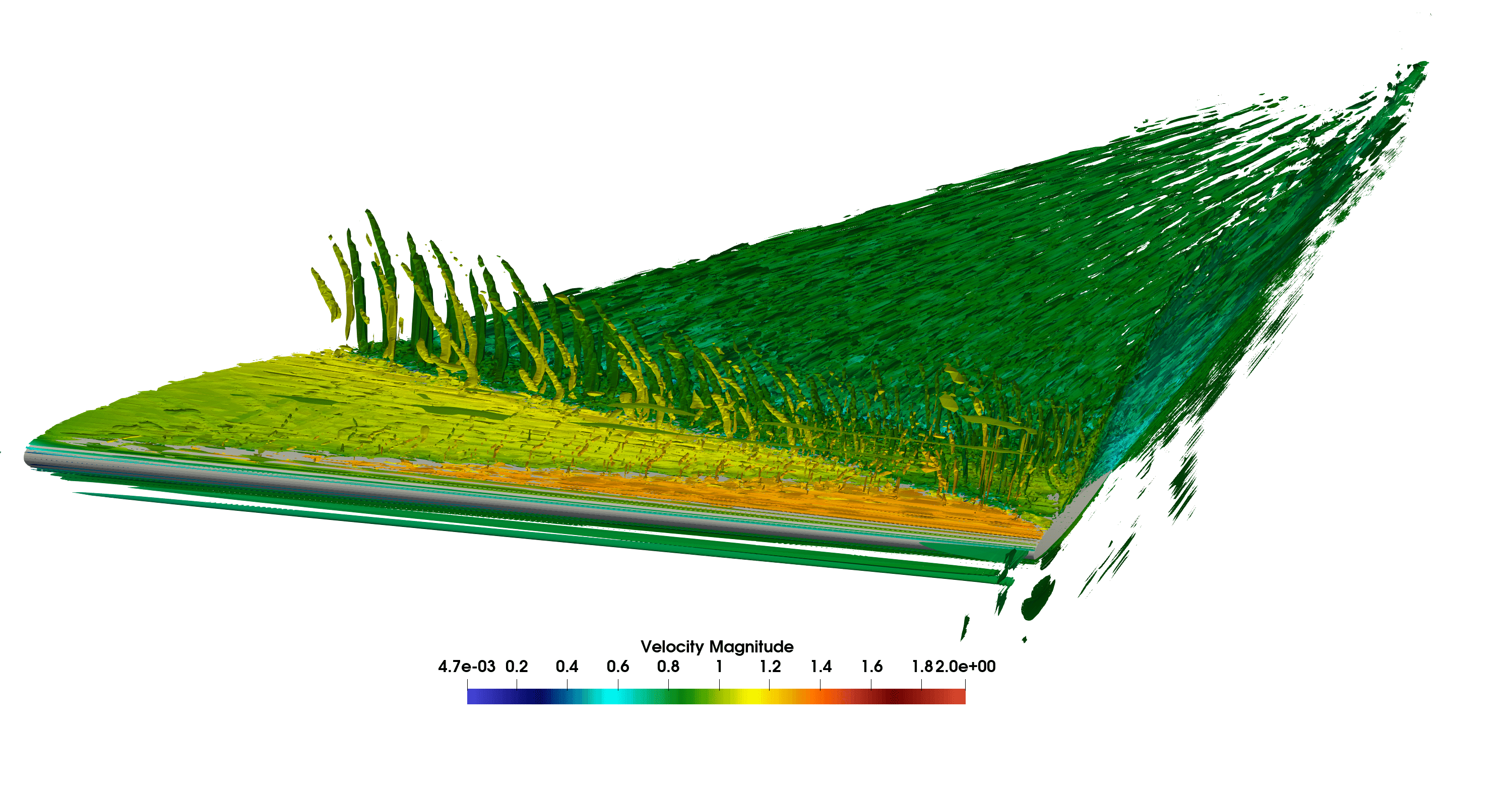}
    \caption{Iso-surfaces of the $Q$-criterion coloured by velocity
    magnitude for the transonic ONERA M6 wing at $M_\infty = 0.84$,
    showing the tip-vortex and the trailing-edge shear layer.}
    \label{fig:onera_qcrit}
\end{figure}

Figure~\ref{fig:onera_cp} compares the computed surface pressure
coefficient with the Schmitt and Charpin
experiment~\cite{schmitt1979pressure} at the seven canonical spanwise
sections. Across the inboard stations
($z/b = 0.20$--$0.65$), the suction-side $\lambda$-shock is captured at
the correct chordwise location, with the leading shock branch near
$x/c \approx 0.2$--$0.3$ and the recompression branch near
$x/c \approx 0.6$--$0.7$; the predicted pre-shock plateau and post-shock
pressure recovery follow the measured trend, and the pressure-side
distribution coincides with the experimental data over the entire chord.
Toward the tip the two shock branches coalesce into a single oblique
shock, in agreement with experiment, and \texttt{MARUT} reproduces this
behaviour at $z/b = 0.80$ and $0.90$. At the outermost station
$z/b = 0.99$, the simulation captures the strong upper-surface
suction peak associated with the tip-vortex roll-up. The overall
agreement with experiment demonstrates that the combined
hyperbolic--parabolic GPU pipeline, together with the curvilinear
hexahedral operator and the weakly imposed no-slip adiabatic wall,
reproduces the canonical transonic ONERA M6 solution on the
wall-resolved static mesh. The associated vortical structure is shown in
Figure~\ref{fig:onera_qcrit} through iso-surfaces of the
$Q$-criterion coloured by velocity magnitude. The roll-up of the
tip vortex, the shear layer shed from the trailing edge, and the
streamwise vortices that develop downstream of the suction-side
$\lambda$-shock are all resolved on the curvilinear hexahedral mesh,
consistent with the strong tip suction peak observed at the outermost
$C_p$ station.

\subsection{Non-Equilibrium Blast Reaction}
We consider a two-dimensional, chemically and
thermally reacting circular blast wave from Grossman et al. \cite{grossman1990flux}.
Unlike the inviscid Sedov problem, this configuration couples a strong
cylindrical shock to finite-rate air-5 chemistry \cite{park1989nonequilibrium} and a
two-temperature Landau--Teller vibrational relaxation model, and therefore
exercises the shock-capturing operator, the Zhang--Shu positivity-preserving
limiter, the implicit chemistry kernel, and the two-temperature state-recovery
machinery of \texttt{MARUT} simultaneously. The five-species mixture
$\mathcal{S}=\{\text{N},\,\text{O},\,\text{NO},\,\text{N}_2,\,\text{O}_2\}$
is initialised with a hot, partially dissociated driver inside a cold,
diatomic-dominated ambient driven gas separated by a circular interface at
$r=0.5$ on the square domain
$(x,y)\in[-1,1]^2$:
\begin{equation}
    \bigl(T,\,T_v,\,p,\,\mathbf{u}\bigr) =
    \begin{cases}
        \bigl(9000\,\text{K},\,9000\,\text{K},\,195256\,\text{Pa},\,\mathbf{0}\bigr) & r<0.5,\\[2pt]
        \bigl(300\,\text{K},\,300\,\text{K},\,10000\,\text{Pa},\,\mathbf{0}\bigr) & r>0.5,
    \end{cases}
    \label{eq:blast_ic}
\end{equation}
with chemical-equilibrium species composition:
$\bigl(\rho_{\text{N}},\,\rho_{\text{O}},\,\rho_{\text{NO}},\,\rho_{\text{N}_2},\,\rho_{\text{O}_2}\bigr)
=(2.79\!\times\!10^{-2},\,8.94\!\times\!10^{-3},\,3.49\!\times\!10^{-5},\,1.58\!\times\!10^{-3},\,5.02\!\times\!10^{-7})$~kg/m$^3$
(predominantly atomic at $9000$\,K), whereas the driven gas is ambient air
$(\rho_{\text{N}_2},\,\rho_{\text{O}_2}) = (0.0887,\,0.0269)$~kg/m$^3$ with
the three minor species at trace levels below $10^{-17}$~kg/m$^3$. To
prevent Gibbs oscillations from the radial discontinuity contaminating the
positivity bounds, the interface is regularised by a $\tanh$ profile of
width $\delta = \Delta x/2$. All four boundaries are inviscid slip walls,
which is the closed-domain configuration used in the reference; the shock
does not reach the boundary before the final time $t_{\text{end}} =
2\!\times\!10^{-4}$\,s.

The domain is discretised on a $64\times 64$ Cartesian grid of $4096$
elements with a nodal DG-SEM operator of polynomial degree $\mathcal{P}=7$, yielding
$262\,144$ Lobatto--Legendre nodes and roughly $2.36\!\times\!10^{6}$
conserved-variable degrees of freedom (nine variables per node: five
species densities, two momentum components, total energy, and vibrational
energy). The volume integral uses the kinetic-energy preserving central
two-point flux in flux-differencing split form, blended against a
subcell Lax--Friedrichs finite-volume operator through the
Hennemann--Gassner indicator evaluated on the smoothness of the
$\rho p$ field. Surface fluxes use local Lax--Friedrichs with a
strong-form boundary correction that subtracts $(1-\alpha_e)\,f(\mathbf{U}^-)\!\cdot\!\mathbf{n}$
from the Riemann flux on each face; this correction is essential at
shock-flagged element interfaces, where neighbouring elements carry
different blending coefficients and the strong-form DG volume operator
otherwise produces an uncanceled boundary contribution that drifts mass
conservation. The Zhang--Shu limiter enforces $\rho_s \ge 0$ per species. Time integration
of the hyperbolic substep uses the five-stage, fourth-order SSPRK54
scheme at $\mathrm{CFL}=0.05$.
\begin{figure}
    \centering
    \includegraphics[scale=0.95,clip=true]{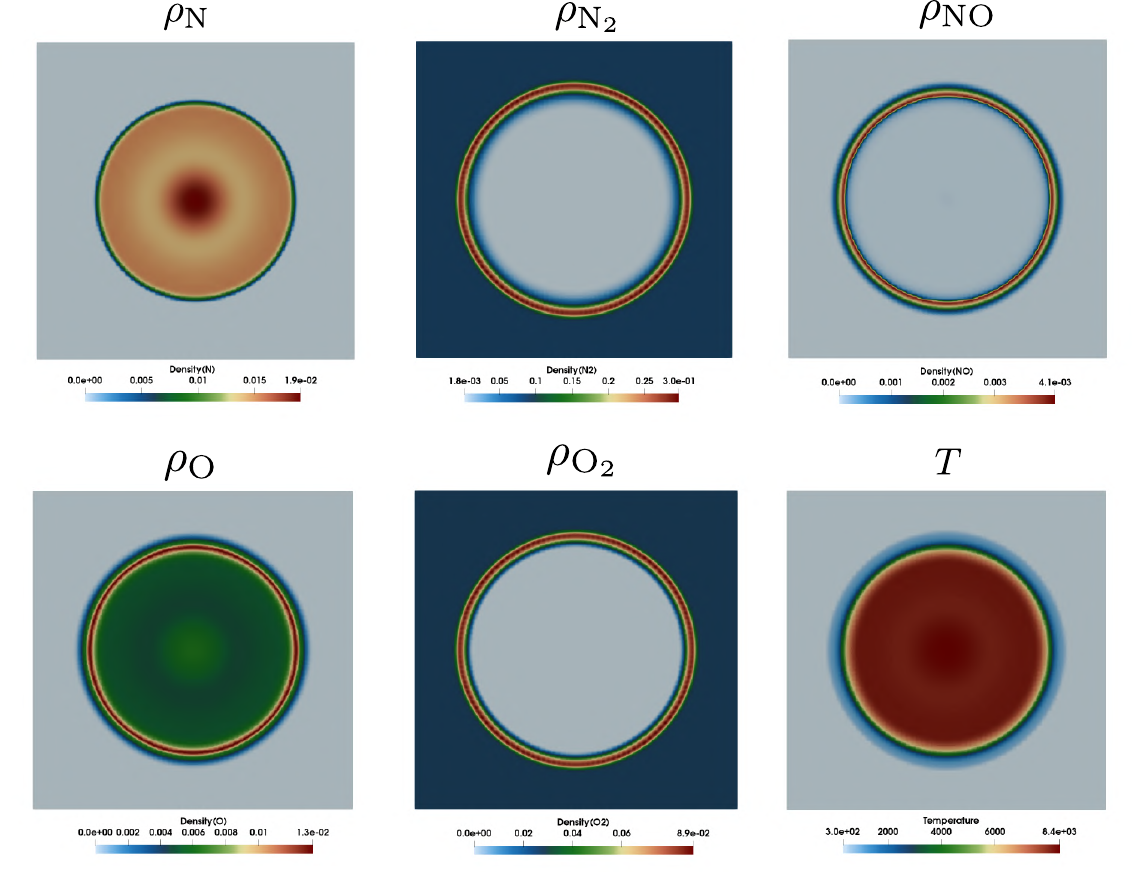}
    \caption{Non-equilibrium reactive blast wave at $t_{\text{end}}=2\times 10^{-4}$\,s
    on a $64\times 64$ grid with $\mathcal{P}=7$ spectral DG and SSPRK54. Top row,
    left to right: atomic nitrogen $\rho_{\text{N}}$, molecular nitrogen
    $\rho_{\text{N}_2}$, nitric oxide $\rho_{\text{NO}}$. Bottom row:
    atomic oxygen $\rho_{\text{O}}$, molecular oxygen $\rho_{\text{O}_2}$,
    and translational--rotational temperature $T$ (in K). Color scales are
    individual to each panel. The cylindrical shock has propagated from
    the initial interface at $r=0.5$ to $r\approx 0.7$.}
    \label{fig:blast}
\end{figure}
The simulation is completed on a single NVIDIA H100
GPU in $\sim\!90$ minutes.
Figure~\ref{fig:blast} shows three distinct regions at the final time: a
hot atomic core ($r\lesssim 0.55$, $T\approx 8.4\!\times\!10^{3}$\,K) where
$\rho_{\text{N}}$ peaks at the centre and $\rho_{\text{O}}$ exhibits a
characteristic annular maximum from faster oxygen recombination; a
cylindrical shock at $r\approx 0.7$ compressing the ambient diatomics by
$\approx 3.4\times$ (well below the Rankine--Hugoniot bound of $6$); and a
thin nitric-oxide ring at the shock front from the Zeldovich exchanges
$\text{N}_2+\text{O}\!\rightleftharpoons\!\text{NO}+\text{N}$ and
$\text{NO}+\text{O}\!\rightleftharpoons\!\text{O}_2+\text{N}$. 

\subsection{Performance Analysis}

\begin{figure}[htpb]
    \centering
    \includegraphics[trim=0.2cm 0cm 0.2cm 0cm, scale=0.32,clip=true]{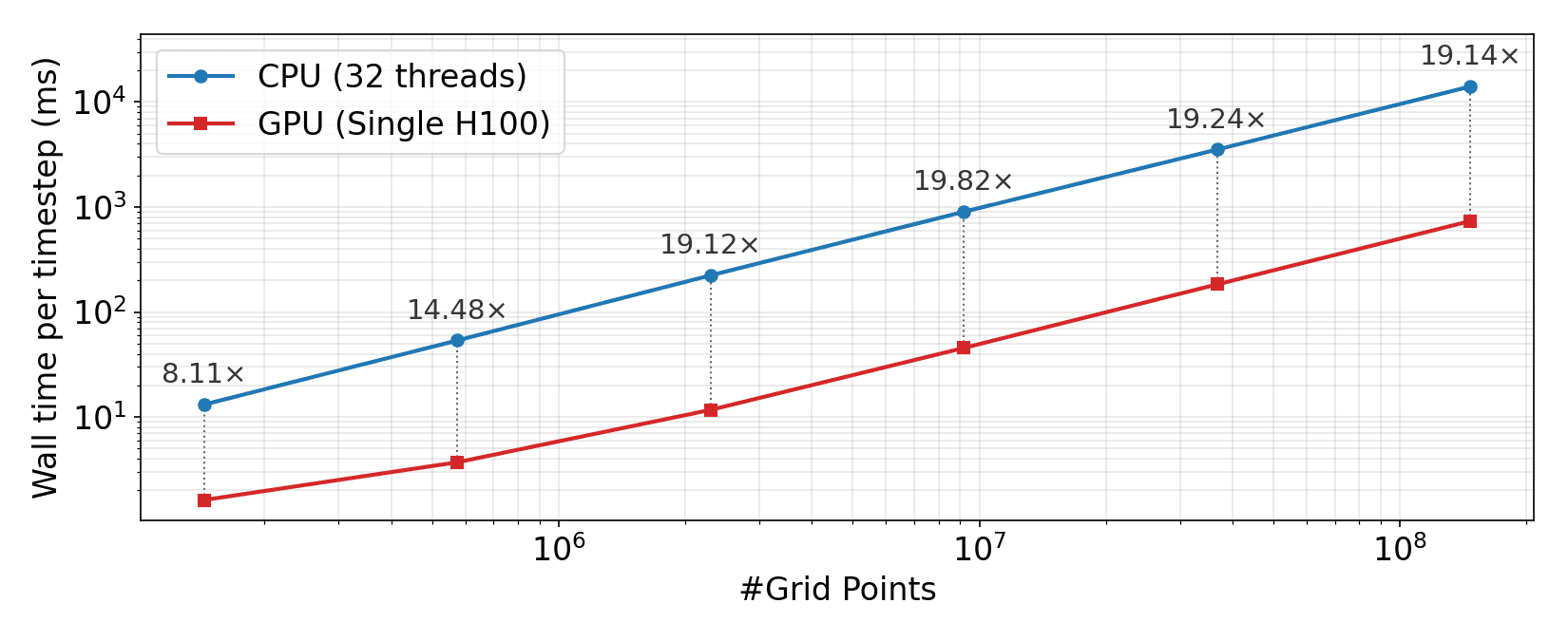}
        \includegraphics[trim=0.2cm 0cm 0.2cm 0cm, scale=0.32,clip=true]{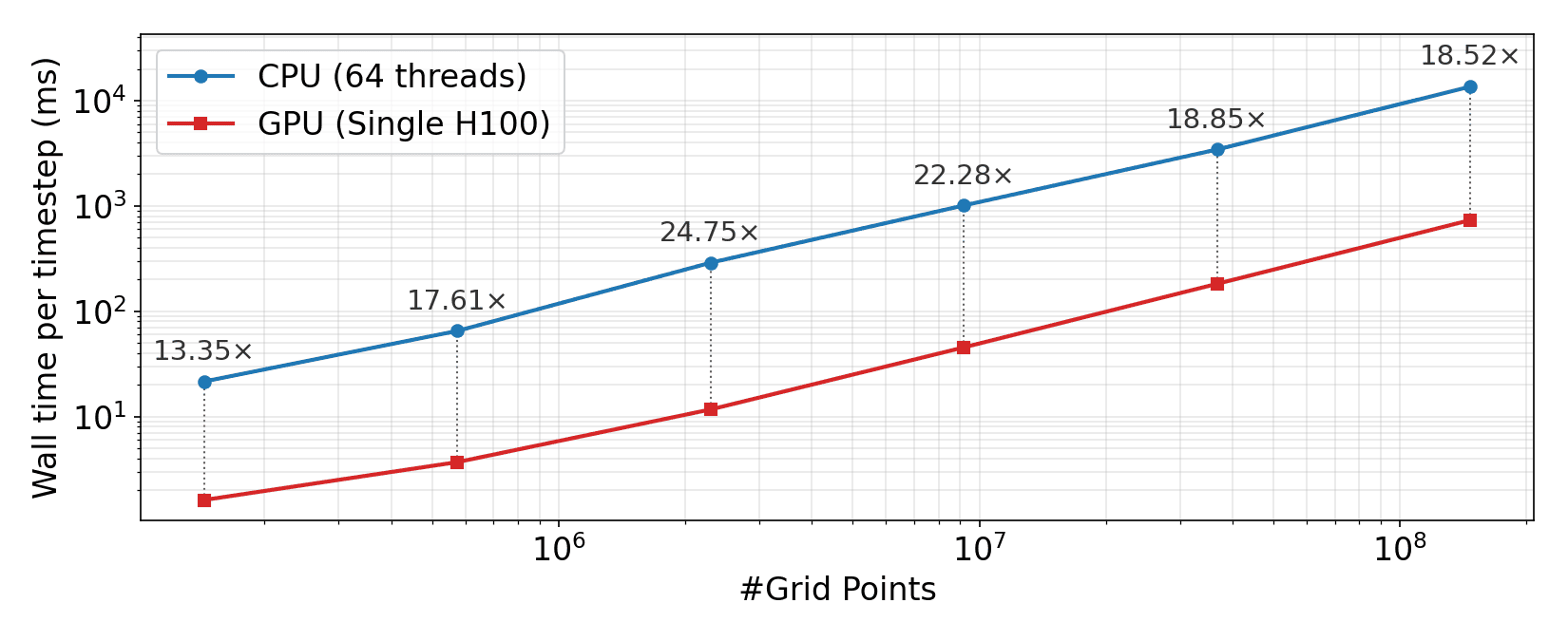}
    \caption{2D Euler cylinder benchmark: wall-time per SSPRK54 step vs.\ number of grid points, comparing CPU baselines with 32 (left) and 64 (right) threads against \texttt{MARUT} on a single NVIDIA H100 GPU at $\mathcal{P}= 3$. Annotations indicate the achieved GPU speed-up for each configuration.}
    \label{fig:bench_cyl_32}
\end{figure}

\noindent \textbf{2D Euler (Cylinder)}:  The performance of the GPU implementation on the supersonic cylinder
(Sec.~\ref{sec:cylinder}) is measured by running each candidate mesh
for a fixed number of SSPRK54 time steps and reporting the wall-time
per step as a function of the total number of grid points
$N_{\text{dof}} = N_e\,(\mathcal{P}+1)^2$. Figures~\ref{fig:bench_cyl_32} show this curve for two CPU baselines 
(\texttt{Trixi.jl} \cite{schlottkelakemper2025trixi}: 32-thread and 64-thread shared-memory runs of the corresponding
reference DG-SEM library on the same problem) against \texttt{MARUT} on a
single NVIDIA H100 GPU.

Both baselines exhibit the expected linear dependence of wall time on
$N_{\text{dof}}$ in log--log coordinates, but with substantially
different slopes. Against the 32-thread CPU run
(Fig.~\ref{fig:bench_cyl_32}), the GPU advantage grows from
$\approx 8.1\times$ at the smallest mesh ($\sim 10^6$ grid points) to
$\approx 19.1\times$--$19.8\times$ in the $10^7$--$10^8$ regime, where
the GPU has enough work to saturate its streaming multiprocessors and
the constant per-kernel launch overhead is amortised over a much
larger element pool. Doubling the CPU baseline to 64 threads does not, counterintuitively, reduce
the GPU advantage at small and intermediate problem sizes: the
reported speed-up actually rises to $\approx 13.4\times$ at
$\sim 10^6$ grid points and peaks at $\approx 24.7\times$ near
$10^7$. This is because the 64-thread CPU run does not scale
linearly from the 32-thread baseline at these sizes, the smaller
per-thread workload incurs proportionally more synchronisation,
cache-line contention, and NUMA traffic, so doubling the threads
yields well below a $2\times$ wall-time reduction and the GPU/CPU
ratio correspondingly grows. Only beyond $\sim 10^7$ grid points,
where the CPU operator becomes memory-bandwidth bound and 64 threads
finally saturate the available bandwidth, does the 64-thread
speed-up curve fall back onto the 32-thread plateau at
$\approx 18.5\times$. The two figures therefore converge in the
asymptotic large-mesh regime, as one would physically expect, while
their offset at small sizes is a direct fingerprint of CPU thread
scaling rather than of any property of the GPU implementation.

\vspace{0.2cm}
\noindent \textbf{3D Navier-Stokes (TGV)}:
\begin{figure}[htpb]
    \centering
    \includegraphics[scale=0.6,clip=true]{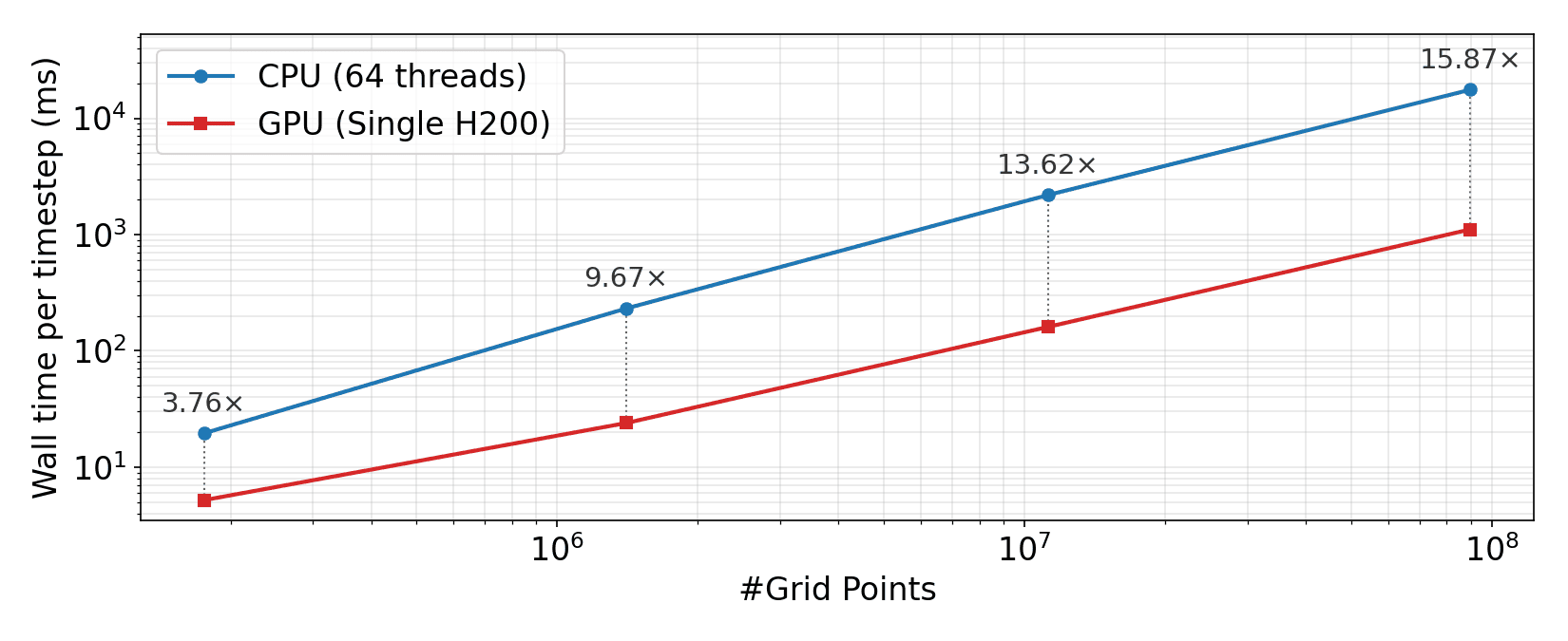}
    \caption{3D Navier--Stokes TGV benchmark: wall-time per SSPRK54
    step vs.\ number of grid points, for a 64-thread CPU baseline
    and \texttt{MARUT} on a single NVIDIA H200 GPU at $\mathcal{P} = 6$, averaged over
    5000 steps. Annotations report the GPU speed-up.}
    \label{fig:bench_tgv}
\end{figure}
The corresponding three-dimensional benchmark uses the
Taylor--Green vortex configuration (Sec.~\ref{sec:tgv}) and is shown in
Figure~\ref{fig:bench_tgv}, which compares \texttt{MARUT} on a single NVIDIA
H200 GPU against a 64-thread CPU baseline (\texttt{Trixi.jl} \cite{schlottkelakemper2025trixi}) running the same operator
at $\mathcal{P} = 6$ for 5000 SSPRK54 steps. The GPU advantage starts at
$\approx 3.8\times$ at the smallest mesh ($\sim 10^5$ grid points) and
climbs steadily to $\approx 9.7\times,\ 13.6\times,$ and finally
$\approx 15.9\times$ at $\sim 10^8$ grid points. The lower speed-up
floor relative to the two-dimensional benchmark is consistent with
the higher arithmetic intensity of the 3D operator: at $\mathcal{P} = 6$ each
element carries $(\mathcal{P}+1)^3 = 343$ Lobatto--Legendre nodes, so a 3D element exposes
$\approx 21\times$ more work per element than its 2D counterpart, and
problem sizes of $\sim 10^5$ grid points correspond to only a few
hundred elements -- not enough to saturate the GPU. Once the mesh
crosses $\sim 10^6$ grid points the H200 reaches steady occupancy and
the $\approx 13$--$16\times$ speed-up regime that obtains in the
$10^7$--$10^8$ range mirrors the trend seen for the cylinder. 

\section{Multi-GPU Scaling}
\begin{figure}[htbp]
    \centering
    \includegraphics[width=0.95\textwidth]{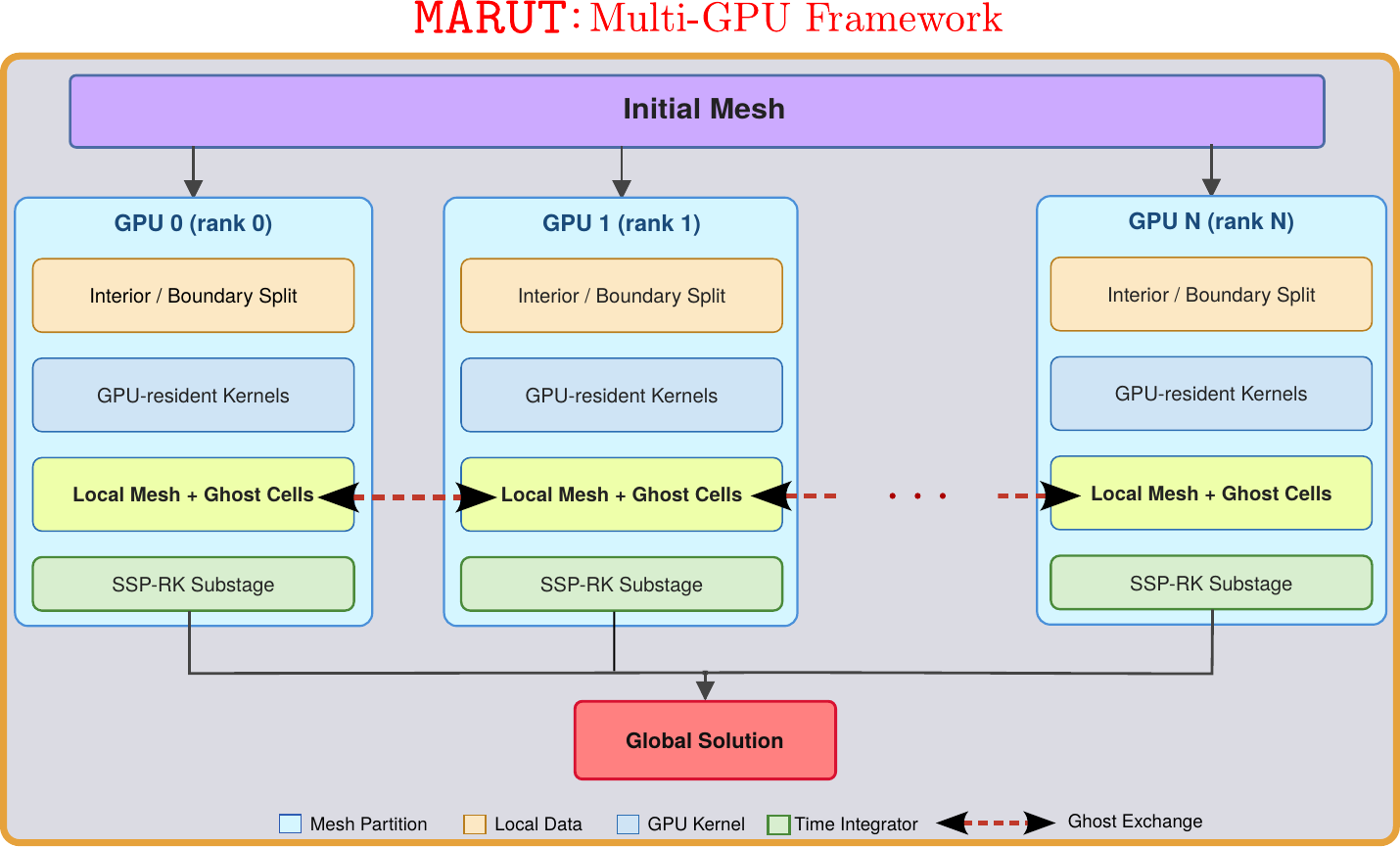}
    \caption{\texttt{MARUT} multi-GPU representation. The mesh is partitioned across ranks, and each GPU advances its own partition through an identical pipeline; ghost-cell exchanges overlap with the interior compute on each rank.}
    \label{fig:marut_multigpu}
\end{figure}

In distributed GPU implementations of DG solvers, performance is governed not only by arithmetic throughput but by the balance between computation and data movement. Although modern GPUs achieve high efficiency for local flux evaluations, scalability is often limited by communication overhead associated with inter-device data exchange. A primary bottleneck arises from \textit{halo exchanges} between elements residing on different GPUs. These transfers incur latency from the underlying interconnect (e.g., PCIe, NVLink, or InfiniBand), which can exceed the cost of local compute kernels \cite{li2019evaluating}. As a result, computation is periodically stalled, reducing overall device utilization. This effect is further amplified by global synchronization, where progress is constrained by the slowest communication path, introducing system-wide idle time. To address these limitations, \texttt{MARUT} combines latency-hiding, communication-avoiding, and kernel-level optimizations. Communication is overlapped with computation by prioritizing interior element updates while asynchronously issuing non-blocking halo exchanges. This reduces idle time by allowing data transfer and arithmetic execution to proceed concurrently. Figure~\ref{fig:marut_multigpu} sketches this distributed execution model: the initial mesh is partitioned across ranks along a space-filling curve, and each GPU runs an identical pipeline, interior/boundary split, GPU-resident kernels, local mesh with ghost cells, and SSP-RK substage, with ghost-cell data exchanged asynchronously between neighbouring ranks at every substage before the per-step solutions are gathered into the global solution.

\subsection{Scalability Challenges in Multi-GPU Adaptive Mesh Refinement}
Efficient multi-GPU scaling with AMR requires a hierarchy of numerical and communication operations to preserve accuracy, conservation, and scalability across dynamically evolving meshes. Central to AMR are the \textit{prolongation} and \textit{restriction} operators, which transfer solution data between coarse and fine refinement levels. Prolongation interpolates coarse-grid data onto finer grids during refinement and ghost-cell reconstruction, whereas restriction projects fine-grid solutions back onto coarser levels to maintain multilevel consistency and conservation;  see, \cite{berger1989local,berger1984adaptive}. Most AMR implementations additionally enforce a 2:1 refinement balance constraint, ensuring that adjacent cells differ by at most one refinement level; this substantially simplifies stencil construction, coarse-fine synchronization, and timestep stability in distributed octree- and block-structured meshes \cite{zhang2021amrex}. Scalable multi-GPU execution further depends on efficient halo and ghost-region exchange across refinement levels and device boundaries. In conservative finite-volume formulations, this is coupled with refluxing procedures that reconcile flux mismatches at coarse–fine interfaces to preserve global conservation laws \cite{berger1989local}. Many AMR algorithms also employ temporal subcycling, in which finer refinement levels evolve with smaller timesteps than coarse levels, requiring recursive synchronization and multilevel communication \cite{ berger1984adaptive}. Because adaptive refinement dynamically redistributes computational work, scalable implementations additionally require load-balancing strategies, often based on space-filling curves or hierarchical patch migration, to minimize communication overhead while maintaining balanced GPU utilization \cite{zhang2021amrex}.

\subsection{Strong and Weak Scaling}
The parallel efficiency of \texttt{MARUT} is assessed through both strong and weak scaling experiments on up to four NVIDIA L40S GPUs interconnected via InfiniBand, using CPU-staged MPI communication. The test configuration is a two-dimensional transonic viscous flow over the RAE~2822 airfoil at $M=0.85$, $Re=50{,}000$, discretized with polynomial degree $\mathcal{P}=5$ and advanced with the SSPRK54 time integrator for 1000 steps. Inter-GPU data exchange employs the face-only communication strategy as discussed before, in which only the $(\mathcal{P}+1)\times n_v$ surface degrees of freedom per shared interface are communicated, rather than the full $(\mathcal{P}+1)^2\times n_v$ volume data of each ghost element.

\begin{figure}[htbp]
    \centering
    \begin{subfigure}[b]{0.48\textwidth}
        \centering
        \includegraphics[width=\textwidth]{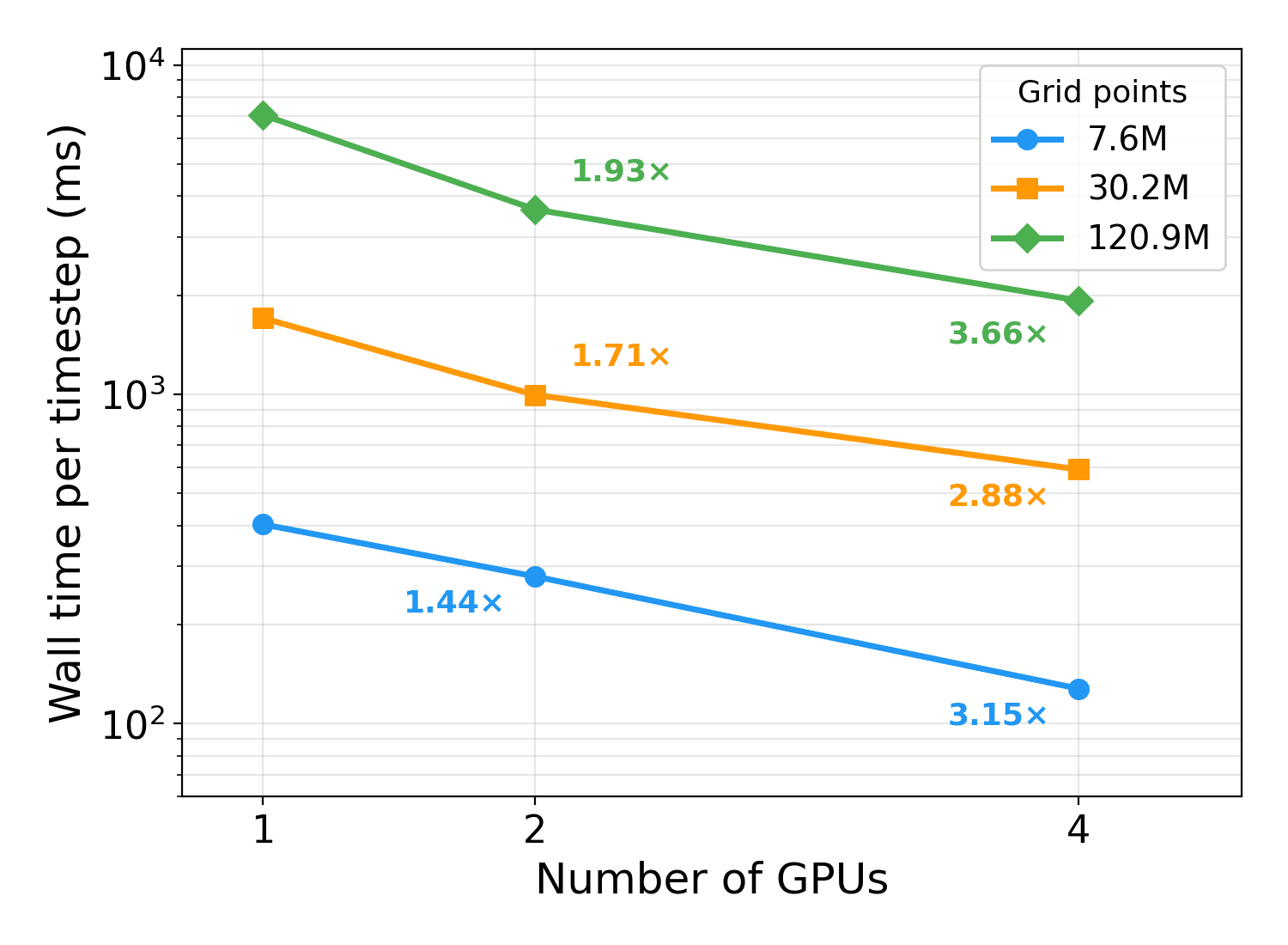}
    \end{subfigure}
    \hfill
    \begin{subfigure}[b]{0.48\textwidth}
        \centering
        \includegraphics[width=\textwidth]{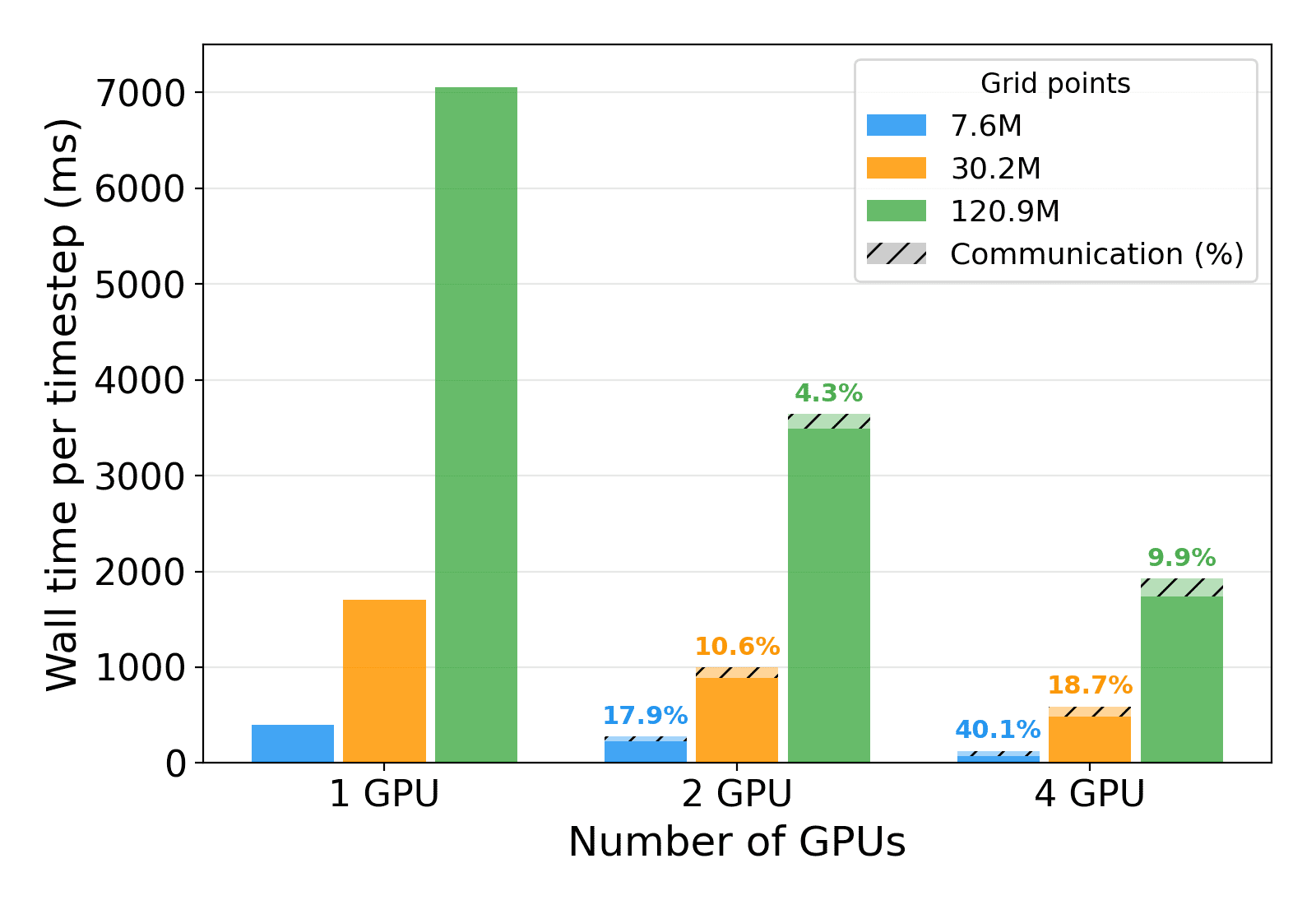}
    \end{subfigure}
    \caption{Strong scaling on up to four L40S GPUs. (Left) Wall time per SSPRK54 step versus GPU count for three problem sizes; annotations indicate the measured speedup. (Right) Per-step compute (solid) and communication (hatched) time grouped by GPU count; percentages indicate the communication fraction.}
    \label{fig:strong_scaling}
\end{figure}

\paragraph{Strong scaling:}
In a strong scaling study the total problem size is held fixed while the number of GPUs is increased. The ideal speed-up on $N$ GPUs is $S_N ~ (\triangleq  T_1 / T_N) = N$, and the parallel efficiency is defined as
\begin{equation}
\eta_{\text{strong}} = \frac{S_N}{N} = \frac{T_1}{N\,T_N},
\label{eq:strong_eff}
\end{equation}
where $T_1$ and $T_N$ denote the wall time per step on one and $N$ GPUs, respectively. Three mesh sizes are considered: $2.1\times10^5$ elements ($7.6\times10^6$ grid points), $8.4\times10^5$ elements ($3.0\times10^7$ grid points), and $3.4\times10^6$ elements ($1.2\times10^8$ grid points). At the largest problem size, a single GPU cannot accommodate the mesh ($>44$\,GB required); the single-GPU baseline is therefore extrapolated from the measured per-element cost at smaller sizes. Figure~\ref{fig:strong_scaling} (Left) reports the wall time per SSPRK54 step as a function of GPU count for each problem size, with annotations indicating the achieved speedup. At the smallest mesh ($2.1\times10^5$ elements) the communication fraction reaches $40\%$ on four GPUs, limiting the speedup to $3.15\times$ ($78.8\%$ efficiency). Increasing the problem size to $8.4\times10^5$ elements improves the four-GPU efficiency to $72.0\%$ ($2.88\times$ speedup), while the largest mesh ($3.4\times10^6$ elements) achieves $3.66\times$ speedup at $91.5\%$ efficiency, with communication accounting for only $9.9\%$ of the wall time. This trend reflects the favorable surface-to-volume ratio of the DG operator: the communication cost scales with the number of shared inter-partition faces ($\sim N_e^{1/2}$ in 2D), whereas the compute cost grows linearly with the total number of elements. Figure~\ref{fig:strong_scaling} (Right) decomposes each configuration into compute and communication time per step. At every GPU count, the hatched region representing communication remains approximately constant across problem sizes, confirming that the absolute communication overhead is governed by the partition interface topology rather than by the total element count. The compute portion, by contrast, decreases proportionally with the number of GPUs, and this asymmetry is the mechanism by which larger problems sustain higher parallel efficiency.

\begin{figure}[htbp]
    \centering
    \begin{subfigure}[b]{0.48\textwidth}
        \centering
        \includegraphics[width=\textwidth]{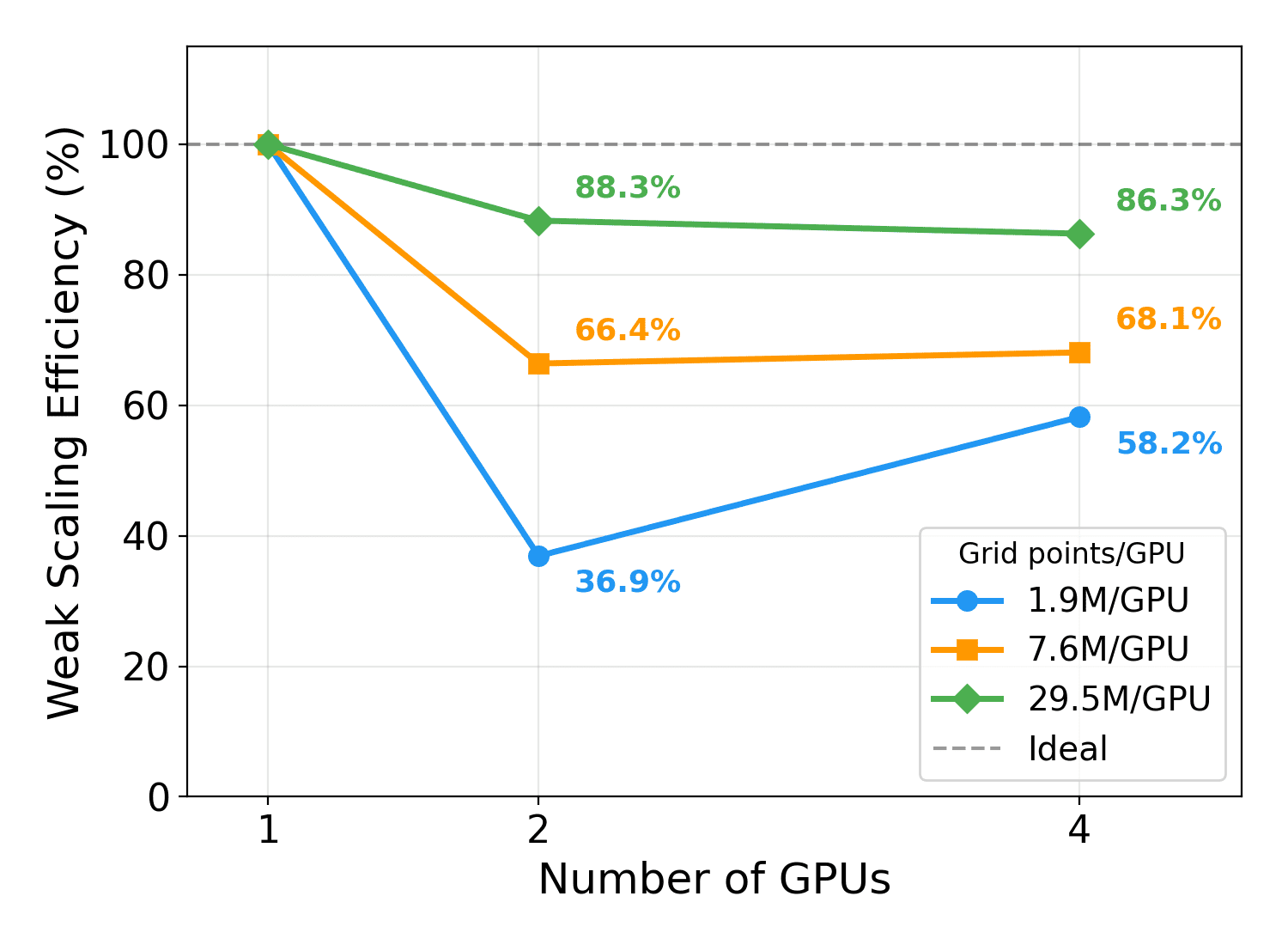}
    \end{subfigure}
    \hfill
    \begin{subfigure}[b]{0.48\textwidth}
        \centering
        \includegraphics[width=\textwidth]{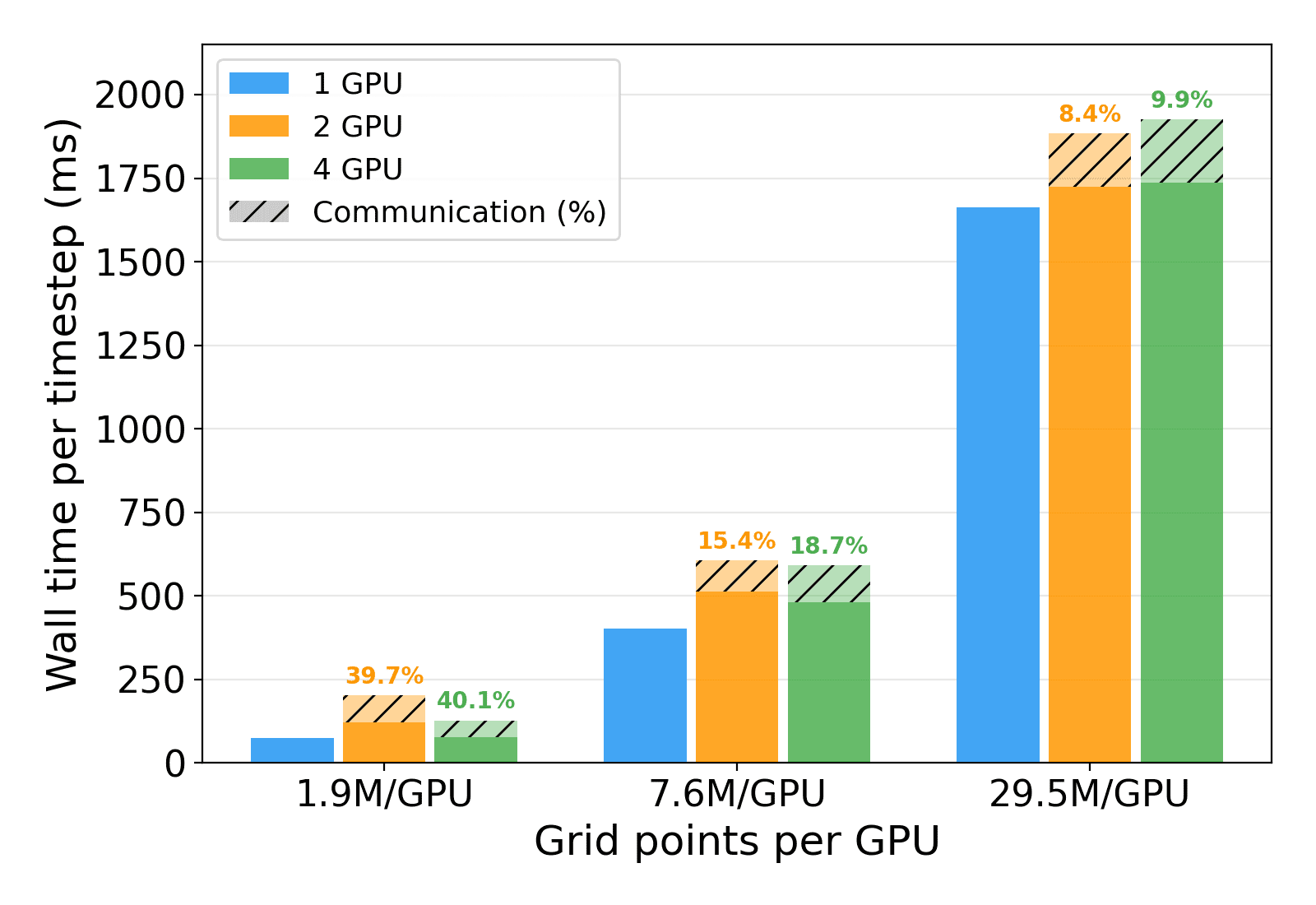}
    \end{subfigure}
    \caption{Weak scaling on up to four L40S GPUs. (Left) Weak scaling efficiency versus GPU count for three per-GPU problem sizes; the dashed line indicates ideal scaling ($\eta=100\%$). (Right) Per-step compute (solid) and communication (hatched) time grouped by per-GPU problem size; the three bars in each group correspond to 1, 2, and 4~GPUs.}
    \label{fig:weak_scaling}
\end{figure}

\paragraph{Weak scaling:}
In a weak scaling study the problem size per GPU is held approximately constant while both the total problem size and the GPU count are increased simultaneously. The ideal behavior is constant wall time per step regardless of the number of GPUs, and the weak scaling efficiency is defined as
\begin{equation}
\eta_{\text{weak}} = \frac{T_1}{T_N},
\label{eq:weak_eff}
\end{equation}
where $T_1$ is the wall time per step on a single GPU with its local problem size, and $T_N$ is the wall time per step on $N$ GPUs with the scaled problem. Any deviation from $\eta_{\text{weak}}=1$ reflects communication overhead introduced by the partitioning. Three weak scaling series are constructed at approximately $5.2\times10^4$, $2.1\times10^5$, and $8.2\times10^5$ elements per GPU, using purpose-built meshes to ensure consistent element quality across GPU counts. At the smallest per-GPU load ($\sim52$K elements), the communication fraction dominates and the four-GPU efficiency drops to $58.2\%$. At $\sim210$K elements per GPU the efficiency improves to $68.1\%$. The most practically relevant series at $\sim820$K elements per GPU achieves $88.3\%$ efficiency on two GPUs and $86.3\%$ on four GPUs, with the communication fraction remaining below $10\%$. Figure~\ref{fig:weak_scaling} (Left) plots the weak scaling efficiency as a function of GPU count for all three series. The clear separation between the curves demonstrates that the per-GPU problem size is the primary determinant of parallel efficiency: once each GPU has sufficient work to amortize the communication latency, the solver sustains high throughput with minimal degradation. Figure~\ref{fig:weak_scaling} (Right) decomposes the per-step wall time into compute and communication contributions at each problem size. In the $\sim820$K series, the three bars are nearly equal in height, confirming that the compute cost per GPU remains constant while the communication overhead grows modestly from $0\%$ (single GPU) to $10\%$ (four GPUs). The face-only exchange strategy achieves a consistent $\sim23\times$ reduction in communication volume compared to a full ghost-element exchange, which is the key enabler of the observed scaling behavior.

\section{Conclusions}
In this work, we present \texttt{MARUT}, a GPU-native high-order computational fluid dynamics (CFD) framework developed for simulations of compressible reacting and non-reacting flows with finite-rate chemistry on modern heterogeneous architectures. Implemented entirely on NVIDIA GPUs with adaptive mesh refinement (AMR), the framework combines high-order numerical accuracy and scalable many-core performance within a unified solver infrastructure, thereby addressing one of the major challenges in contemporary scientific computing: the efficient and accurate simulation of strongly multiscale, multiphysics flow phenomena. The development of such methodologies has become increasingly important as advances in aerospace propulsion, hypersonics, and high-enthalpy flow physics continue to outpace the capabilities of conventional CFD frameworks, particularly in regimes where shocks, turbulence and chemically reacting interfaces interact across widely disparate spatial and temporal scales. By coupling high-order discretizations with fully GPU-resident AMR algorithms, \texttt{MARUT} is designed to minimize memory-transfer overheads and maximize hardware utilization while preserving the numerical fidelity required for predictive simulations of nonlinear compressible flows. Beyond its computational performance, the framework also reflects the broader transition of the scientific computing community towards modular, AI-enabled and differentiable simulation environments. Owing to its implementation in \texttt{Julia}, \texttt{MARUT} can be naturally integrated with modern machine-learning and scientific-AI ecosystems, enabling future extensions involving surrogate modeling, reinforcement-learning-based flow control, inverse problems and data-driven optimization. In this sense, \texttt{MARUT} represents not only a high-performance CFD solver, but also a step towards next-generation autonomous scientific computing platforms in which high-fidelity numerical simulation, adaptive algorithms and artificial intelligence operate within a tightly coupled computational paradigm. 

The main contributions of this study can be summarized as follows:
\begin{enumerate}
  
  \item \textbf{Fully GPU-resident high-order solver:} The \texttt{MARUT} solver employs a high-order spectral discontinuous Galerkin (DG) method for spatial discretization and high-order strong-stability-preserving Runge--Kutta (SSP-RK) schemes for time integration. Both the hyperbolic (inviscid) and parabolic (viscous) operators are evaluated entirely on the GPU (for curvilinear, unstructured \texttt{P4estMesh} geometries) as a sequence of element-local kernels with no CPU data transfer during the time-stepping loop. This includes the on-device execution of entropy-stable split-form volume integrals, shock-capturing indicators, and interface fluxes. This fully GPU-resident formulation eliminates communication bottlenecks, enabling low-dissipation, high-fidelity resolution of steep gradients, shocks, and boundary 
layers in compressible and reacting flows.


   \item \textbf{GPU-resident conservative AMR:} We introduce \texttt{GPUForest}, a forest-of-quadtree/octree data structure whose connectivity, space-filling (Morton) ordering, refinement levels, and curvilinear element geometry reside entirely in GPU memory, on which a conservative AMR cycle runs as a sequence of GPU kernels. The framework dynamically refines regions containing strong solution gradients and multiscale flow structures, including shock waves, shock/boundary-layer interactions, shear layers, and reaction fronts, while eliminating the major computational bottlenecks associated with CPU--GPU data movement. This delivers substantially improved computational efficiency, scalability, and memory locality compared with refined meshes, making large-scale high-fidelity simulations feasible on modern multi-GPU architectures.


\item \textbf{GPU-resident finite-rate chemistry and thermal non-equilibrium:}  The solver integrates finite-rate chemical kinetics 
and Landau--Teller vibrational--translational energy relaxation of a two-temperature model directly on the GPU through per-node source-term kernels, advanced together with the hyperbolic and parabolic operators in a single device-resident pipeline.

\item \textbf{Multi-GPU scalability and exascale readiness:} The solver demonstrates excellent strong and weak scaling performance across multiple NVIDIA GPUs, achieving high parallel efficiency for large-scale simulations on modern high-performance computing systems. The GPU-native implementation is carefully designed to maximize data locality, parallel throughput, and inter-GPU communication efficiency while minimizing computational overhead. The resulting framework is exascale-ready and capable of supporting production-scale, high-order simulations with sustained numerical fidelity for complex compressible and reactive-flow applications on next-generation heterogeneous supercomputing platforms.

\item \textbf{Broad validation across compressible flow regimes:} The framework is validated over a wide range of compressible flow regimes, including subsonic, transonic, supersonic, and hypersonic flows with finite-rate chemical reactions, demonstrating robustness in capturing strongly nonlinear phenomena such as shock dynamics, turbulence, and chemically reacting flows with stiff source terms. These capabilities are relevant to a broad spectrum of aerospace applications, ranging from commercial passenger aircraft operating in the transonic regime to high-speed aerospace systems such as missiles, military aircraft, and atmospheric re-entry vehicles.


\end{enumerate}
The proposed \texttt{MARUT} framework provides a unified high-order, adaptive, and GPU-efficient approach for simulating compressible flows across multiple regimes, including hypersonic flows with real-chemistry. The results indicate that combining high-order DG methods with AMR and GPU acceleration is a promising pathway toward resolving complex flow physics at scale, particularly for applications in high-speed aerodynamics, propulsion systems, and reactive flow modeling. The \texttt{Julia}-based implementation of \texttt{MARUT} also offers a natural route towards the incorporation of and machine-learning (ML) methodologies. Owing to the rapidly growing scientific ML ecosystem in \texttt{Julia}, including libraries for differentiable programming, neural networks and scientific machine learning, the framework can be readily coupled with data-driven models and optimization algorithms. Such integration enables the development of hybrid computational strategies in which physics-based solvers are augmented by neural-network surrogates, reinforcement-learning agents or physics-informed learning approaches, thereby enhancing predictive capability and computational efficiency for complex flow problems.
On the application side, \texttt{MARUT} has already been employed in flow-control studies \cite{mondal2025shocks,Mondal2026Hypersonic} and can be further extended to inverse problems, parameter estimation, and reduced-order modeling.

\appendix
\section{Finite-rate chemical kinetics}
\label{appC}
The five-species air model considers only the major neutral species
$$\mathrm{N_2},~ \mathrm{O_2},~ \mathrm{NO},~ \mathrm{N},~ \mathrm{O},$$
and is generally adequate for moderately high-temperature hypersonic
flows where ionization effects remain weak. The model captures the
dominant dissociation and recombination reactions of air while
maintaining relatively low computational cost. In contrast, the eleven-species air model additionally includes charged
particles and ionized species,
\[
\mathrm{N_2},\;
\mathrm{O_2},\;
\mathrm{NO},\;
\mathrm{N},\;
\mathrm{O},\;
\mathrm{NO^+},\;
\mathrm{N^+},\;
\mathrm{O^+},\;
\mathrm{N_2^+},\;
\mathrm{O_2^+},\;
e^-,
\]
thereby accounting for ionization, electron-impact reactions, and
plasma effects that become important at very high temperatures,
typically encountered in atmospheric entry flows and strong shock layers.
The eleven-species model therefore provides improved physical fidelity
for high-enthalpy nonequilibrium flows, at the expense of increased
stiffness and computational complexity due to the larger number of
species and reactions.


\vspace{0.1cm}
\noindent \textbf{For the five-species air model}:
the five-species air model \cite{park1989nonequilibrium} $(\mathrm{N_2}, \mathrm{O_2}, \mathrm{NO}, \mathrm{N}, \mathrm{O})$,
the dominant dissociation and exchange reactions are commonly written as
\begin{equation*}
\mathrm{N_2 + M \rightleftharpoons N + N + M},
\end{equation*}

\begin{equation*}
\mathrm{O_2 + M \rightleftharpoons O + O + M},
\end{equation*}

\begin{equation*}
\mathrm{NO + M \rightleftharpoons N + O + M},
\end{equation*}
where $M$ denotes a third-body collision partner that facilitates
energy transfer during dissociation and recombination events.

\vspace{0.1cm}
\noindent \textbf{For the eleven-species air model}: 
The eleven-species air model \cite{gupta1990review} includes dissociation, recombination,
exchange, ionization, and electron-impact reactions. Representative
reactions are

\begin{equation*}
\mathrm{N_2 + M \rightleftharpoons N + N + M},
\end{equation*}

\begin{equation*}
\mathrm{O_2 + M \rightleftharpoons O + O + M},
\end{equation*}

\begin{equation*}
\mathrm{NO + M \rightleftharpoons N + O + M},
\end{equation*}

\begin{equation*}
\mathrm{N + O \rightleftharpoons NO^+ + e^-},
\end{equation*}

\begin{equation*}
\mathrm{O + e^- \rightleftharpoons O^+ + 2e^-},
\end{equation*}

\begin{equation*}
\mathrm{N + e^- \rightleftharpoons N^+ + 2e^-},
\end{equation*}

\begin{equation*}
\mathrm{N_2 + e^- \rightleftharpoons N_2^+ + 2e^-},
\end{equation*}

\begin{equation*}
\mathrm{O_2 + e^- \rightleftharpoons O_2^+ + 2e^-}.
\end{equation*}
The presence of charged particles introduces additional coupling
between chemical kinetics and thermodynamic non-equilibrium through the
electron temperature and electron-impact reaction rates.

\section{Strong-Form Entropy-Stable Spectral DG Formulation}
\label{appA}

Let the computational domain be partitioned as $\Omega = \bigcup_{e=1}^{N_e} \Omega_e,$ where $\Omega_e$ denotes the $e$-th element and $N_e$ is the total number of elements. Let $\mathcal{P} \geq 0$ be the polynomial approximation degree. In the context of the discontinuous Galerkin spectral element method (DGSEM), the discrete trial and test space is the broken, discontinuous piecewise polynomial space defined via the direct sum of the local element spaces:
\[
\mathcal{V}^h := \bigoplus_{e=1}^{N_e} \mathcal{V}_e = \bigoplus_{e=1}^{N_e} \left[ \mathbb{P}^\mathcal{P}(\Omega_e) \right]^m,
\]
where $m$ is the number of conserved variables and $\mathbb{P}^\mathcal{P}(\Omega_e)$ denotes the space of polynomials spanning up to degree $\mathcal{P}$ in each spatial coordinate direction. 

The global approximate solution $\mathbf{U}^h \in \mathcal{V}^h$ and the corresponding Cartesian flux fields are represented as the direct sum of their independent elemental components:
\[
\mathbf{U}^h = \bigoplus_{e=1}^{N_e} \mathbf{U}_e^h, \quad 
\mathbf{F}^h = \bigoplus_{e=1}^{N_e} \mathbf{F}_e^h, \quad 
\mathbf{G}^h = \bigoplus_{e=1}^{N_e} \mathbf{G}_e^h, \quad 
\mathbf{H}^h = \bigoplus_{e=1}^{N_e} \mathbf{H}_e^h.
\]
On a specific hexahedral element $\Omega_e$, the local solution field $\mathbf{U}_e^h \in \mathcal{V}_e$ is expanded using a 3D tensor-product Lagrange polynomial basis collocated directly at the physical Cartesian coordinate positions of the Gauss--Lobatto--Legendre (GLL) nodes:
\[
\mathbf{U}_e^h(x,y,z,t) = \sum_{i=0}^{\mathcal{P}}\sum_{j=0}^{\mathcal{P}}\sum_{k=0}^{\mathcal{P}} \mathbf{U}_{ijk,e}^h(t)\,\ell_i(x)\ell_j(y)\ell_k(z),
\]
where $\ell_i(x)$, $\ell_j(y)$, and $\ell_k(z)$ represent the 1D Lagrange polynomials interpolating the GLL nodal distribution along the $x$-, $y$-, and $z$-directions of the element, respectively, and $\mathbf{U}_{ijk,e}^h(t)$ are the time-dependent nodal degrees of freedom.

Similarly, the local Cartesian fluxes are approximated using the matching collocated group finite-element representation:
\[
\mathbf{F}_e^h(x,y,z,t)=\sum_{i=0}^{\mathcal{P}}\sum_{j=0}^{\mathcal{P}}\sum_{k=0}^{\mathcal{P}} \mathbf{F}_{ijk,e}^h(t)\ell_i(x)\ell_j(y)\ell_k(z),
\]
with identical tensor-product expansions applied to the remaining local spatial flux distributions $\mathbf{G}_e^h(x,y,z,t)$ and $\mathbf{H}_e^h(x,y,z,t)$. The inviscid fluxes are computed point-wise as functions of the local state vector, i.e., $\mathbf{F}_{\mathrm{inv},e}^h=\mathbf{F}_{\mathrm{inv}}(\mathbf{U}_e^h)$, $\mathbf{G}_{\mathrm{inv},e}^h=\mathbf{G}_{\mathrm{inv}}(\mathbf{U}_e^h)$, and $\mathbf{H}_{\mathrm{inv},e}^h=\mathbf{H}_{\mathrm{inv}}(\mathbf{U}_e^h)$.

Multiplying the governing equations by a local test function $\mathbf{v}_e^h \in \mathcal{V}_e$ and integrating over the element $\Omega_e$ yields the local variational formulation:
\begin{equation}
\int_{\Omega_e}\mathbf{v}_e^h\frac{\partial \mathbf{U}_e^h}{\partial t}\,d\Omega
+
\int_{\Omega_e}\mathbf{v}_e^h
\left(
\frac{\partial \mathbf{F}_{\mathrm{inv},e}^h}{\partial x}
+
\frac{\partial \mathbf{G}_{\mathrm{inv},e}^h}{\partial y}
+
\frac{\partial \mathbf{H}_{\mathrm{inv},e}^h}{\partial z}
\right)d\Omega
=
\int_{\Omega_e}\mathbf{v}_e^h
\left(
\frac{\partial \mathbf{F}_{v,e}^h}{\partial x}
+
\frac{\partial \mathbf{G}_{v,e}^h}{\partial y}
+
\frac{\partial \mathbf{H}_{v,e}^h}{\partial z}
\right)d\Omega.
\label{eq:strong_start}
\end{equation}

Because the global space $\mathcal{V}^h$ is constructed via a direct sum, functions are permitted to be fully discontinuous across element boundaries, requiring interface coupling to be handled through surface penalty terms. On a tensor-product GLL grid, the standard derivative discrete matrix operators naturally satisfy the summation-by-parts (SBP) property, allowing the spatial derivatives to remain on the volume fluxes while preserving conservation and stability. 

Replacing the multi-valued physical boundary flux values with consistent, unique numerical interface fluxes $\mathbf{F}_e^{h,*}$, $\mathbf{G}_e^{h,*}$, and $\mathbf{H}_e^{h,*}$ yields the strong-form nodal semi-discrete system for a single element $\Omega_e$:
\begin{align}
\int_{\Omega_e}\mathbf{v}_e^h &\frac{\partial \mathbf{U}_e^h}{\partial t}\,d\Omega 
+ \int_{\Omega_e}\mathbf{v}_e^h
\left(
\frac{\partial \mathbf{F}_{\mathrm{inv},e}^h}{\partial x}
+
\frac{\partial \mathbf{G}_{\mathrm{inv},e}^h}{\partial y}
+
\frac{\partial \mathbf{H}_{\mathrm{inv},e}^h}{\partial z}
\right)d\Omega \nonumber \\
&
+ \int_{\partial\Omega_e}\mathbf{v}_e^h
\left( (\mathbf{F}_{\mathrm{inv},e}^{h,*} - \mathbf{F}_{\mathrm{inv},e}^h ) n_x
+ (\mathbf{G}_{\mathrm{inv},e}^{h,*} - \mathbf{G}_{\mathrm{inv},e}^h ) n_y
+ (\mathbf{H}_{\mathrm{inv},e}^{h,*} - \mathbf{H}_{\mathrm{inv},e}^h ) n_z \right) dS \nonumber \\
&= \int_{\Omega_e}\mathbf{v}_e^h
\left(
\frac{\partial \mathbf{F}_{v,e}^h}{\partial x}
+
\frac{\partial \mathbf{G}_{v,e}^h}{\partial y}
+
\frac{\partial \mathbf{H}_{v,e}^h}{\partial z}
\right)d\Omega
+ \int_{\partial\Omega_e}\mathbf{v}_e^h
\left( (\mathbf{F}_{v,e}^{h,*} - \mathbf{F}_{v,e}^h ) n_x
+ (\mathbf{G}_{v,e}^{h,*} - \mathbf{G}_{v,e}^h ) n_y
+ (\mathbf{H}_{v,e}^{h,*} - \mathbf{H}_{v,e}^h ) n_z \right) dS,
\label{eq:strong_element_form}
\end{align}
where $\partial\Omega_e$ denotes the boundary of element $\Omega_e$, and $\hat{\mathbf{n}}=(n_x,n_y,n_z)^T$ is the outward unit normal vector. The surface integrals penalize the jump between the interior element traces and the common numerical interface fluxes. These interface fluxes depend on the interior and exterior traces of the solution, $\mathbf{U}_e^{h-}$ and $\mathbf{U}_e^{h+}$, thereby coupling adjacent elements and weakly enforcing boundary consistency.

Finally, the global strong-form semi-discrete system is obtained by summing the independent contributions across the direct sum space:
\begin{align}
\sum_{e=1}^{N_e}
\Bigg[
&
\int_{\Omega_e}\mathbf{v}_e^h\frac{\partial \mathbf{U}_e^h}{\partial t}\,d\Omega
+ \int_{\Omega_e}\mathbf{v}_e^h
\left(
\frac{\partial \mathbf{F}_{\mathrm{inv},e}^h}{\partial x}
+
\frac{\partial \mathbf{G}_{\mathrm{inv},e}^h}{\partial y}
+
\frac{\partial \mathbf{H}_{\mathrm{inv},e}^h}{\partial z}
\right)d\Omega  \notag\\
&
+ \int_{\partial\Omega_e}\mathbf{v}_e^h
\left( (\mathbf{F}_{\mathrm{inv},e}^{h,*} - \mathbf{F}_{\mathrm{inv},e}^h ) n_x
+ (\mathbf{G}_{\mathrm{inv},e}^{h,*} - \mathbf{G}_{\mathrm{inv},e}^h ) n_y
+ (\mathbf{H}_{\mathrm{inv},e}^{h,*} - \mathbf{H}_{\mathrm{inv},e}^h ) n_z \right) dS \notag\\
&
- \int_{\Omega_e}\mathbf{v}_e^h
\left(
\frac{\partial \mathbf{F}_{v,e}^h}{\partial x}
+
\frac{\partial \mathbf{G}_{v,e}^h}{\partial y}
+
\frac{\partial \mathbf{H}_{v,e}^h}{\partial z}
\right)d\Omega
- \int_{\partial\Omega_e}\mathbf{v}_e^h
\left( (\mathbf{F}_{v,e}^{h,*} - \mathbf{F}_{v,e}^h ) n_x
+ (\mathbf{G}_{v,e}^{h,*} - \mathbf{G}_{v,e}^h ) n_y
+ (\mathbf{H}_{v,e}^{h,*} - \mathbf{H}_{v,e}^h ) n_z \right) dS
\Bigg]=0.
\end{align}

\vspace{0.2cm}
\noindent\textbf{Inviscid Volume Flux (Entropy-Conservative Split Form)}:
For high-Mach-number, under-resolved flows and long-time integrations, the standard strong-form inviscid volume derivatives inside the element-level integral $\int_{\Omega_e}\mathbf{v}_e^h \left( \nabla \cdot \mathbf{F}_{\mathrm{inv},e}^h \right)d\Omega$ in \eqref{eq:strong_element_form} are evaluated using an entropy-conservative two-point split formulation \cite{fisher2013high,gassner2016split}. 

By utilizing collocated GLL quadrature, the volume integrals reduce to pointwise nodal evaluations. At a specific node $(i,j,k)$ within element $\Omega_e$, the strong spatial derivatives are directly approximated via the symmetric two-point split-flux operator:
\begin{equation}
\begin{aligned}
\left.\frac{\partial \mathbf{F}_{\mathrm{inv},e}^h}{\partial x}\right|_{ijk}
&\approx
2 \sum_{m=0}^{\mathcal{P}} D_{im}\,
\mathbf{F}^{\#}(\mathbf{U}_{ijk,e}^h, \mathbf{U}_{mjk,e}^h), \\
\left.\frac{\partial \mathbf{G}_{\mathrm{inv},e}^h}{\partial y}\right|_{ijk}
&\approx
2 \sum_{m=0}^{\mathcal{P}} D_{jm}\,
\mathbf{G}^{\#}(\mathbf{U}_{ijk,e}^h, \mathbf{U}_{imk,e}^h), \\
\left.\frac{\partial \mathbf{H}_{\mathrm{inv},e}^h}{\partial z}\right|_{ijk}
&\approx
2 \sum_{m=0}^{\mathcal{P}} D_{km}\,
\mathbf{H}^{\#}(\mathbf{U}_{ijk,e}^h, \mathbf{U}_{ijm,e}^h),
\end{aligned}
\label{eq:split_form_3D}
\end{equation}
where $D$ denotes the standard one-dimensional strong differentiation matrix at the GLL nodes. The factor of two is a consequence of the symmetric two-point flux construction; it ensures that for a spatially uniform state, the split operator simplifies back to the standard strong-form derivative (i.e., $2\sum D_{im}\mathbf{F} = \sum D_{im}\mathbf{F} + \mathbf{F}\sum D_{im} = \partial_x \mathbf{F}$), thereby maintaining consistency for smooth solutions.

The two-point numerical volume fluxes $\mathbf{F}^{\#}$, $\mathbf{G}^{\#}$, and $\mathbf{H}^{\#}$ are symmetric and consistent, meaning:
\begin{equation}
\mathbf{F}^{\#}(\mathbf{U},\mathbf{U}) = \mathbf{F}_{\mathrm{inv}}(\mathbf{U}),
\quad
\mathbf{G}^{\#}(\mathbf{U},\mathbf{U}) = \mathbf{G}_{\mathrm{inv}}(\mathbf{U}),
\quad
\mathbf{H}^{\#}(\mathbf{U},\mathbf{U}) = \mathbf{H}_{\mathrm{inv}}(\mathbf{U}).
\end{equation}
In this work, we employ the kinetic energy preserving and entropy-conservative flux of Ranocha \cite{ranocha2018comparison}. In the $x$-direction, this split-flux vector is given by:
\begin{equation}
\begin{aligned}
F^{\#}_{\rho}    &= \{\!\!\{\rho\}\!\!\}^{\ln} \{\!\!\{u\}\!\!\}, \\
F^{\#}_{\rho u}  &= \{\!\!\{\rho\}\!\!\}^{\ln} \{\!\!\{u\}\!\!\}^2 + \{\!\!\{p\}\!\!\}, \\
F^{\#}_{\rho v}  &= \{\!\!\{\rho\}\!\!\}^{\ln} \{\!\!\{u\}\!\!\}\{\!\!\{v\}\!\!\}, \\
F^{\#}_{\rho w}  &= \{\!\!\{\rho\}\!\!\}^{\ln} \{\!\!\{u\}\!\!\}\{\!\!\{w\}\!\!\}, \\
F^{\#}_{\rho E}  &= \left(
\frac{1}{2(\gamma-1)\{\!\!\{T^{-1}\}\!\!\}^{\ln}}
+ \tfrac{1}{2}\bigl(\{\!\!\{u\}\!\!\}^2
+ \{\!\!\{v\}\!\!\}^2
+ \{\!\!\{w\}\!\!\}^2\bigr)
\right)
\{\!\!\{\rho\}\!\!\}^{\ln} \{\!\!\{u\}\!\!\}
+ \{\!\!\{p u\}\!\!\},
\end{aligned}
\end{equation}
with the corresponding $y$- and $z$-direction volume fluxes obtained through a cyclic permutation of the velocity components. Here, $\{\!\!\{a\}\!\!\}=\tfrac{1}{2}(a_L+a_R)$ represents the standard arithmetic mean, $\{\!\!\{a\}\!\!\}^{\ln} = (a_L - a_R)/(\ln a_L - \ln a_R)$ denotes the logarithmic mean, and $T=p/\rho$ is the fluid temperature. When paired with a matching dissipative numerical interface flux in \eqref{eq:strong_element_form}, this strong-form volume configuration guarantees a globally nonlinearly entropy-stable semi-discretization.

\vspace{0.2cm}
\noindent \textbf{Shock Capturing}: 
The sub-cell shock capturing strategy follows the provably robust DG/finite-volume (FV) blending framework of Hennemann and Gassner \cite{hennemann2021provably}. The core idea is to locally blend the high-order, entropy-stable strong-form DGSEM discretization with a robust first-order FV method on an embedded sub-cell grid. For each element $\Omega_e$, a dimensionless smoothness indicator is constructed via a modal decomposition of a chosen indicator field, such as the pressure or density. The local solution is projected onto an orthogonal tensor-product Legendre basis with element-local modal coefficients $\hat{m}_{ijk,e}$. For an approximation of polynomial degree $\mathcal{P}$, the total modal energy up to a given degree $K$ is defined as:
\begin{equation*}
E_{\mathrm{tot}}^{(K)} =
\sum_{i,j,k=0}^{K} \hat{m}_{ijk,e}^2.
\end{equation*}

\noindent The energy contained within the highest modes at the outer boundaries of the tensor-product spectrum is measured by:
\begin{equation*}
E_{\mathrm{high}}^{(\mathcal{P})} =
\sum_{\max(i,j,k)=\mathcal{P}} \hat{m}_{ijk,e}^2,
\qquad
E_{\mathrm{high}}^{(\mathcal{P}-1)} =
\sum_{\max(i,j,k)=\mathcal{P}-1} \hat{m}_{ijk,e}^2.
\end{equation*}

\noindent The dimensionless smoothness indicator for element $\Omega_e$ is then given by:
\begin{equation*}
\mathcal{E}_e =
\max\!\left(
\frac{E_{\mathrm{high}}^{(\mathcal{P})}}{E_{\mathrm{tot}}^{(\mathcal{P})}},
\quad
\frac{E_{\mathrm{high}}^{(\mathcal{P}-1)}}{E_{\mathrm{tot}}^{(\mathcal{P}-1)}}
\right).
\end{equation*}
This indicator is mapped to a local blending factor $\alpha_e \in [0,1]$ via a smooth sigmoid function:
\begin{equation*}
\alpha_e =
\frac{1}{1 + \exp\!\left(
-\dfrac{s}{\mathcal{T}}\left(\mathcal{E}_e - \mathcal{T}\right)
\right)},
\end{equation*}
where the activation threshold $\mathcal{T}$ is dynamically set according to the polynomial degree as:
\begin{equation*}
\mathcal{T} = 0.5 \cdot 10^{-1.8(\mathcal{P}+1)^{1/4}},
\end{equation*}
and $s>0$ is a steepness parameter controlling the sharpness of the transition between pure DGSEM and sub-cell FV behavior.

To prevent isolated element switches and enforce spatial coherence across discontinuities, the blending factor is regularized using a face-neighbourhood maximum:
\begin{equation*}
\alpha_e \leftarrow
\max\!\bigl(\alpha_e,\{\alpha_{e'}\}_{e' \in \mathcal{N}(e)}\bigr),
\end{equation*}
where $\mathcal{N}(e)$ denotes the set of elements sharing a face with $\Omega_e$. The final semi-discrete residual is then constructed as a local convex combination of the high-order strong-form DGSEM residual and the robust sub-cell finite-volume residual:
\begin{equation}
\mathcal{R}_e =
(1 - \alpha_e)\,\mathcal{R}_e^{\mathrm{DG}}
+
\alpha_e\,\mathcal{R}_e^{\mathrm{FV}}.
\end{equation}

\noindent Consistent with our strong-form nodal formulation, $\mathcal{R}_e^{\mathrm{DG}}$ represents the discrete operator form of equation \eqref{eq:strong_element_form}:
\begin{align*}
\mathcal{R}_e^{\mathrm{DG}} =
&\int_{\Omega_e} \mathbf{v}_e^h \frac{\partial \mathbf{U}_e^h}{\partial t}\, d\Omega
+ \int_{\Omega_e} \mathbf{v}_e^h \left(
\frac{\partial \mathbf{F}_{\mathrm{inv},e}^h}{\partial x} + \frac{\partial \mathbf{G}_{\mathrm{inv},e}^h}{\partial y} + \frac{\partial \mathbf{H}_{\mathrm{inv},e}^h}{\partial z}
\right) d\Omega \\
&+ \int_{\partial\Omega_e} \mathbf{v}_e^h \left(
(\mathbf{F}_{\mathrm{inv},e}^{h,*} - \mathbf{F}_{\mathrm{inv},e}^h) n_x + (\mathbf{G}_{\mathrm{inv},e}^{h,*} - \mathbf{G}_{\mathrm{inv},e}^h) n_y + (\mathbf{H}_{\mathrm{inv},e}^{h,*} - \mathbf{H}_{\mathrm{inv},e}^h) n_z
\right) dS \\
&- \int_{\Omega_e} \mathbf{v}_e^h \left(
\frac{\partial \mathbf{F}_{v,e}^h}{\partial x} + \frac{\partial \mathbf{G}_{v,e}^h}{\partial y} + \frac{\partial \mathbf{H}_{v,e}^h}{\partial z}
\right) d\Omega \\
&- \int_{\partial\Omega_e} \mathbf{v}_e^h \left(
(\mathbf{F}_{v,e}^{h,*} - \mathbf{F}_{v,e}^h) n_x + (\mathbf{G}_{v,e}^{h,*} - \mathbf{G}_{v,e}^h) n_y + (\mathbf{H}_{v,e}^{h,*} - \mathbf{H}_{v,e}^h) n_z
\right) dS.
\end{align*}

\noindent The term $\mathcal{R}_e^{\mathrm{FV}}$ denotes a first-order finite-volume residual computed on a tensor-product sub-cell grid embedded within $\Omega_e$, utilizing the matching numerical surface flux functions for interface consistency. 

This blending strategy guarantees that the scheme preserves high-order spectral accuracy in smooth regions ($\alpha_e \approx 0$), while automatically activating the dissipative, shock-capturing sub-cell finite-volume scheme near sharp gradients and discontinuities ($\alpha_e \approx 1$), providing rigorous nonlinear stability.

\vspace{0.2cm}
\noindent\textbf{Viscous Volume and Interface Fluxes}: Unlike the inviscid terms, the viscous fluxes depend on both the local solution state and its gradients, i.e.,
\[
\mathbf{F}_{v,e}^h = \mathbf{F}_v(\mathbf{U}_e^h, \nabla \mathbf{U}_e^h), \quad
\mathbf{G}_{v,e}^h = \mathbf{G}_v(\mathbf{U}_e^h, \nabla \mathbf{U}_e^h), \quad
\mathbf{H}_{v,e}^h = \mathbf{H}_v(\mathbf{U}_e^h, \nabla \mathbf{U}_e^h).
\]

\noindent\textit{Volume contribution:}
The viscous volume terms are evaluated directly using the nodal strong-form operator and do not require a split-form treatment. To compute these terms, an auxiliary gradient variable tensor is introduced locally within each element:
\[
\mathbf{S}_e^h = \nabla \mathbf{W}_e^h,
\]
where $\mathbf{W}_e^h = (\rho, u, v, w, T)_e^T$ is the local vector of primitive variables and $\mathbf{S}_e^h \in \mathbb{R}^{m \times 3}$ is its corresponding gradient tensor. Taking the gradient of the primitive variables (rather than of the conserved state $\mathbf{U}_e^h$) is the standard choice for the Bassi--Rebay 1 (BR1) formulation, since physical viscous closures (such as shear stress and heat flux) act on velocity and temperature gradients directly. 

Consistent with our strong-form nodal framework, the auxiliary gradient tensor $\mathbf{S}_e^h$ is computed locally within each element using a collocated SBP derivative construction rather than a global weak integration:
\[
\int_{\Omega_e} \mathbf{v}_e^h \, \mathbf{S}_e^h \, d\Omega
=
\int_{\Omega_e} \mathbf{v}_e^h \, \nabla \mathbf{W}_e^h \, d\Omega
+ \int_{\partial \Omega_e} \mathbf{v}_e^h \, \left( \widehat{\mathbf{W}}_e - \mathbf{W}_e^h \right) \otimes \hat{\mathbf{n}} \, dS,
\]
where $\mathbf{v}_e^h \in \mathcal{V}_e$ is the local test function. On collocated GLL nodes, the first integral simplifies to applying the strong differentiation matrix $D$ directly to the nodal values of $\mathbf{W}_e^h$, while the surface integral penalizes the boundary discrepancy using the arithmetic average as the unique numerical interface trace:
\[
\widehat{\mathbf{W}}_e = \{\!\!\{\mathbf{W}_e\}\!\!\}
= \tfrac{1}{2}(\mathbf{W}_e^{h-} + \mathbf{W}_e^{h+}),
\]
where $\mathbf{W}_e^{h-}$ and $\mathbf{W}_e^{h+}$ denote the interior and exterior traces of the primitive variables across the interface of element $\Omega_e$, respectively. Once $\mathbf{S}_e^h$ is determined at the GLL nodes, the viscous fluxes $\mathbf{F}_{v,e}^h, \mathbf{G}_{v,e}^h, \mathbf{H}_{v,e}^h$ are evaluated pointwise as functions of $\mathbf{U}_e^h$ and $\mathbf{S}_e^h$.

\noindent\textit{Interface flux:}
At element interfaces, the unique numerical viscous fluxes required by the boundary penalty terms in \eqref{eq:strong_element_form} are evaluated using the BR1 scheme \cite{bassi1997high}. The interface flux is defined by a symmetric arithmetic average of the interior and exterior flux states:
\[
\mathbf{F}_{v,e}^{h,*} =
\frac{1}{2} \left[
\mathbf{F}_v(\mathbf{U}_e^{h-}, \mathbf{S}_e^{h-}) +
\mathbf{F}_v(\mathbf{U}_e^{h+}, \mathbf{S}_e^{h+})
\right],
\]
with analogous symmetric expressions applied to define $\mathbf{G}_{v,e}^{h,*}$ and $\mathbf{H}_{v,e}^{h,*}$.

Due to the mathematical equivalence established by the SBP property, this strong-form approach retains the strict conservation, geometric flexibility, and weak boundary consistency of traditional weak-form DG methods while unlocking sub-cell split-flux evaluation and achieving spectral (exponential) convergence for smooth solutions.

\section{SSP-RK Time Discretization}
\label{appB}

The semi-discrete DGSEM formulation results in a global system of ordinary differential equations evolving in time:
\begin{equation}
\frac{d\mathbf{U}^h}{dt} = \mathcal{L}(\mathbf{U}^h, t)
\end{equation}
where $\mathbf{U}^h \in \mathcal{V}^h$ denotes the global vector of discrete solution degrees of freedom, and the global spatial semi-discrete operator $\mathcal{L}$ is assembled via the direct sum of the local element residuals $\mathcal{R}_e$:
\[
\mathcal{L}(\mathbf{U}^h, t) = \bigoplus_{e=1}^{N_e} \mathcal{R}_e(\mathbf{U}_e^h, t).
\]
To advance this global system in time, we apply the general form of an $s$-stage strong stability-preserving Runge--Kutta (SSP-RK) method \cite{spiteri2002new}:
\begin{align}
\mathbf{U}^{h,(0)} &= \mathbf{U}^{h,n}, \\
\mathbf{U}^{h,(i)} &= \sum_{k=0}^{i-1} \left[\alpha_{ik} \mathbf{U}^{h,(k)} + \Delta t \, \beta_{ik} \, \mathcal{L}\left(\mathbf{U}^{h,(k)}\right)\right], \quad i = 1, 2, \ldots, s, \\
\mathbf{U}^{h,n+1} &= \mathbf{U}^{h,(s)},
\end{align}
where $\mathbf{U}^{h,(i)}$ represents the global discrete solution vector at the intermediate stage $i$. Consistency and stability are ensured by requiring the coefficients to satisfy $\alpha_{ik} \geq 0$, $\beta_{ik} \geq 0$, and $\sum_{k=0}^{i-1} \alpha_{ik} = 1$. Since each stage is a convex combination of forward-Euler steps, any convexity-based bound (such as positivity or the entropy inequality) satisfied by the spatial operator under a forward-Euler constraint $\Delta t \leq \Delta t_{\text{FE}}$ is strictly inherited by the full SSP method under the relaxed constraint $\Delta t \leq \tilde{c}\,\Delta t_{\text{FE}}$, where $\tilde{c}$ is the SSP coefficient of the scheme.

The \texttt{MARUT} solver implements two distinct SSP-RK schemes. The three-stage, third-order method (SSP-RK33) of Shu and Osher \cite{shu1988efficient} has an SSP coefficient of $\tilde{c} = 1$ and is the most economical choice for moderately stiff hyperbolic problems where third-order temporal accuracy is sufficient. The five-stage, fourth-order method (SSP-RK54) of Spiteri and Ruuth \cite{spiteri2002new} achieves fourth-order accuracy with an SSP coefficient $\tilde{c} \approx 1.508$, permitting roughly $50\%$ larger time steps than SSPRK33 for the same spatial operator. The additional two stages carry a modest per-step computational cost but are heavily offset by the larger admissible $\Delta t$, making SSPRK54 the preferred integrator for production runs in the present work. 

The global time-step constraint for the convective (hyperbolic) contributions is limited by the worst-case element across the entire direct sum space:
\begin{equation}
\Delta t_C \leq \tilde{c} \, \min_{e=1,\dots,N_e} \left( \frac{\Delta x_e}{(2\mathcal{P}+1)\lambda_{\max,e}} \right),
\end{equation}
where $\Delta x_e$ represents the characteristic grid size of element $\Omega_e$ and $\lambda_{\max,e}$ denotes the maximum local wave speed within that element. 
Similarly, the stability constraint associated with the diffusion or viscous terms is defined as:
\begin{equation}
\Delta t_D \leq \tilde{c} \, \min_{e=1,\dots,N_e} \left( \frac{\Delta x^2_e}{(2\mathcal{P}+1)^2\nu_{\max,e}} \right),
\end{equation}
where $\nu_{\max,e}$ corresponds to the maximum local value comparing the kinematic viscosity and the thermal diffusivity within element $\Omega_e$.

Consequently, the global time step $\Delta t$ is selected as the minimum of the convective and diffusive constraints across all elements: 
\[
\Delta t = \min(\Delta t_C, \Delta t_D).
\]
The SSPRK54 method ensures fourth-order temporal accuracy while maintaining strong stability properties, which is critical for accurately resolving shocks encountered in compressible transonic and supersonic flows without inducing unphysical oscillations.

\section{Nomenclature}
\label{appD}


\renewcommand{\arraystretch}{1.18}
\begin{longtable}{@{}p{0.22\textwidth} p{0.74\textwidth}@{}}
\caption{Nomenclature of symbols used in the paper.}\label{tab:nomenclature}\\
\hline
\textbf{Symbol} & \textbf{Definition} \\
\hline
\endfirsthead
\multicolumn{2}{l}{\textit{(Table~\ref{tab:nomenclature} continued)}}\\
\hline
\textbf{Symbol} & \textbf{Definition} \\
\hline
\endhead
\hline
\endfoot

\multicolumn{2}{@{}l}{\textbf{\textit{Compressible Navier--Stokes Equations}}}\\
$\Omega \subset \mathbb{R}^3$ & Physical domain \\
$\mathbf{x}=(x,y,z)$ & Physical coordinates \\
$t \in \mathbb{R}^+$ & Time \\
$\rho$ & Fluid (mixture) density \\
$(u,v,w)$ ~\text{and}~ $\mathbf{u}$ & Velocity components and vector \\
$p$ & Static pressure \\
$T$ & Translational (--rotational) temperature \\
$E$ & Total energy per unit volume \\
$\gamma$ & Ratio of specific heats \\
$\mu$ & Dynamic viscosity \\
$k$ & Thermal conductivity \\
$\tau_{ij}$ & Component of viscous stress tensor \\
$q_j$ & Heat-flux vector component \\
$\delta_{ij}$ & Kronecker delta \\
$\mathbf{U}$ & Vector of conserved variables \\
$\mathbf{W}$ & Vector of primitive variables \\
$\mathbf{F}_{\mathrm{inv}},\mathbf{G}_{\mathrm{inv}},\mathbf{H}_{\mathrm{inv}}$ & Inviscid Cartesian fluxes in x-, y-, and z-directions, respectively\\
$\mathbf{F}_{v},\mathbf{G}_{v},\mathbf{H}_{v}$ & Viscous Cartesian fluxes in x-, y-, and z-directions, respectively \\

\multicolumn{2}{@{}l}{}\\
\multicolumn{2}{@{}l}{\textbf{\textit{Thermochemical non-equilibrium and finite-rate chemistry}}}\\
$N_s$ & Number of species \\
$\rho_s$ & Partial density of species $s$ \\
$Y_s = \rho_s/\rho$ & Mass fraction of species $s$ \\
$R$ & Mixture specific gas constant \\
$R_s$ & Specific gas constant of species $s$ \\
$W_s$ & Molar mass of species $s$ \\
$T_v$ & Vibrational--electronic temperature \\
$T_a$ & Two-temperature controlling temperature \\
$E_v$ & Vibrational--electronic energy per unit volume \\
$e_{v,s}(T_v)$ & Vibrational energy per unit mass of species $s$ \\
$e_{v,s}^{eq}(T)$ & Equilibrium vibrational energy at $T$ \\
$\theta_{v,s}$ & Characteristic vibrational temperature of species $s$ \\
$h_s$, $h_s^0$ & Sensible and formation enthalpies of species $s$ \\
$c_{p,s}$ & Specific heat at constant pressure of species $s$ \\
$J_{s,i}$ & Diffusive mass flux of species $s$ in direction $i$ \\
$D_s$ & Mixture-averaged diffusion coefficient of species $s$ \\
$k_{tr},k_v$ & Translational and vibrational thermal conductivities \\
$q_i, q_{v,i}$ & Translational and vibrational heat fluxes \\
$\dot\omega_s$ & net chemical production rate of species $s$ \\
$R_r$ & Net molar rate of reaction $r$ \\
$k_{f,r}, k_{b,r}$ & Forward / backward rate coefficients \\
$K_c^r$ & Equilibrium constant of reaction $r$ \\
$A_r,\beta_r,E_r$ & Modified-Arrhenius constants \\
$\nu'_{sr},\nu''_{sr}$ & Reactant and product stoichiometric coefficients \\
$[X_s]$ & Molar concentration of species $s$, $\rho_s/W_s$ \\
$\mathrm{mol}$ & Set of molecular species (subscript) \\
$\tau_s(T,p)$ & Vibrational relaxation time of species $s$ \\
$Q_{T\text{-}v}$ & Landau--Teller vibrational relaxation source \\
$\mathbf{S}(\mathbf{U})$ & Source vector \\
$\alpha_s^{(N,O)}$ & N- and O-atom counts in species $s$ \\

\multicolumn{2}{@{}l}{}\\
\multicolumn{2}{@{}l}{\textbf{\textit{Spectral DG (geometry and basis)}}}\\
$\Omega_e$ & Physical element \\
$\mathcal{P}$ & Polynomial approximation degree \\
$\mathbf{v}^h$ & Discrete test function \\
$d$ & Spatial dimension \\
$\hat\Omega=[-1,1]^d$ & Reference element \\
$\boldsymbol\xi=(\xi_1,\dots,\xi_d)$ & Reference coordinates \\
$\mathbf{X}(\boldsymbol\xi)$ & Element-wise reference-to-physical map \\
$J_{dm}=\partial x_d/\partial\xi_m$ & Jacobian matrix \\
$J=\det J_{dm}$ & Jacobian determinant \\
$\mathbf{Ja}^{\,l}$ & Contravariant basis vector \\
$\widetilde{\mathbf{F}}^{\,l}$ & Contravariant inviscid flux, $\mathbf{Ja}^{\,l}\!\cdot\!\mathbf{F}_{\mathrm{inv}}$ \\
$w_i$ & GLL quadrature weight at node $i$ \\
$D$ & 1D GLL differentiation matrix \\
$\mathcal{V}^h$ & Discrete trial/test space \\
$l(x)$ & 1D Lagrange polynomials \\
$\mathbf{U}^h$ & Approximate solution  \\
$\mathbf{W} = (\rho,u,v,w,T)^T$ & Vector of primitive (gradient) variables used in the BR1 viscous formulation \\
$\mathbf{S}^h = \nabla\mathbf{W}^h$ & BR1 auxiliary gradient \\
$\hat{\mathbf{n}}$ & Outward face unit normal in reference space \\
$\mathbf{U}^{\pm}$ & Interior/exterior face traces \\
$\mathbf{F}^{*}$ & Numerical surface flux (LF, HLL) \\

\multicolumn{2}{@{}l}{}\\
\multicolumn{2}{@{}l}{\textbf{\textit{Residuals, shock capturing, and AMR}}}\\
$\mathcal{R}_i$ & Per-node spatial residual at node $i$ \\
$\mathcal{R}_e$ & Per-element weak-form residual \\
$\mathcal{L}(\mathbf{U}^h,t)$ & Semi-discrete RHS operator \\
$\alpha_e$ & Hennemann--Gassner DG/FV blending coefficient \\
$\alpha_{\min},\alpha_{\max}$ & Blending coefficient bounds \\
$\eta_e$ & L\"ohner second-difference indicator value \\
$\varphi$ & Scalar field used by L\"ohner indicator \\
$f_w$ & L\"ohner filter weight \\
$\mathrm{flag}_e \in \{-1,0,+1\}$ & Per-element AMR coarsen/keep/refine flag \\
$m_e$ & Morton (Z-order) code of element $e$ \\
$\mathcal{F},\mathcal{R}$ & $\mathbb{L}_2$ Forward / Reverse projection matrices \\
$\nu_e$ & Per-element effective viscosity (max of kinematic, thermal) \\

\multicolumn{2}{@{}l}{}\\
\multicolumn{2}{@{}l}{\textbf{\textit{Time integration (SSP-RK)}}}\\
$\Delta t$ & Global time step \\
$\Delta t_C,\Delta t_D$ & Convective / diffusive (parabolic) bounds \\
$\Delta t_{\text{FE}}$ & Forward-Euler stability bound \\
$\Delta x_{\min}$ & Minimum characteristic element size, $\min_e \Delta x_e$ \\
$\lambda_{\max}$ & Maximum wave speed in the domain \\
$\nu_{\max}$ & Maximum diffusion coefficient \\
$\tilde c$ & SSP coefficient of the time scheme \\
$s$ & Number of SSP-RK stages \\
$\mathbf{U}^{h,(i)}$ & Solution at intermediate stage $i$ \\
$\alpha_{ik},\beta_{ik}$ & Shu--Osher form coefficients \\
$\mathcal{S}_{\Delta t},\mathcal{T}_{\Delta t}$ & Strang-split source and transport operators \\

\multicolumn{2}{@{}l}{}\\
\multicolumn{2}{@{}l}{\textbf{\textit{Multi-GPU performance}}}\\
$N$ & Number of GPUs / MPI ranks \\
$T_1,T_N$ & Wall time per step on 1 and $N$ GPUs \\
$S_N=T_1/T_N$ & Parallel speedup on $N$ GPUs \\
$\eta_{\text{strong}}$ & Strong-scaling efficiency, $S_N/N$ \\
$\eta_{\text{weak}}$ & Weak-scaling efficiency, $T_1/T_N$ \\
$n_v$ & Number of conserved variables (face-only exchange) \\

\hline
\end{longtable}
\renewcommand{\arraystretch}{1.0}

\section*{Acknowledgments} 
The authors gratefully acknowledge the Turing HPC Cluster at Worcester Polytechnic Institute, USA, for providing the computational infrastructure and GPU resources used in this work.

\bibliographystyle{unsrt}
\bibliography{ref}

\end{document}